\documentclass[prX,aps,amssymb,shownopacs,twocolumn,10pt]{revtex4-1}
\usepackage{amsmath}
\usepackage{amssymb}
\usepackage{amsthm}
\usepackage{amsfonts}
\usepackage{listings}
\usepackage{easybmat}
\usepackage{enumerate}
\usepackage{latexsym}
\usepackage{hyperref}

\usepackage{psfrag}
\usepackage{color}
\usepackage{bm}
\usepackage[pdftex]{graphicx}
\usepackage{subfigure}

\newcommand{\beq}{\begin{equation}}
\newcommand{\eneq}{\end{equation}}
\newcommand{\bra}[1]{\left\langle#1\right|}
\newcommand{\ket}[1]{\left|#1\right\rangle}

\newcommand{\pref}[1]{(\ref{#1})}

\def\be{\begin{equation}}
\def\ee{\end{equation}}
\def\ba{\begin{eqnarray}}
\def\ea{\end{eqnarray}}

\def\mbf{\mathbf}

\def\a{\alpha}

\def\e{\epsilon}

\def\k{\kappa}

\def\n{\nu}

\input{epsf}

\begin{document}

\tolerance 10000

\newcommand{\vk}{{\bf k}}

\title{$\mathbb Z_2$ fractional topological insulators in two dimensions}

\author{C. Repellin$^1$}
\author{B. Andrei Bernevig$^2$}
\author{N. Regnault$^{2,1}$}
\affiliation{$^1$ Laboratoire Pierre Aigrain, ENS and CNRS, 24 rue Lhomond, 75005 Paris, France\\
$^2$ Department of Physics, Princeton University, Princeton, NJ 08544}

\begin{abstract}
We propose a simple microscopic model to numerically investigate the stability of a two dimensional fractional topological insulator (FTI). The simplest example of a FTI consists of two decoupled copies of a Laughlin state with opposite chiralities. We focus on bosons at half filling. We study the stability of the FTI phase upon addition of two coupling terms of different nature: an interspin interaction term, and an inversion symmetry breaking term that couples the copies at the single particle level. Using exact diagonalization and entanglement spectra, we numerically show that the FTI phase is  stable against both perturbations. We compare our system to a similar bilayer fractional Chern insulator. We show evidence that the time reversal invariant system survives the introduction of interaction coupling on a larger scale than the time reversal symmetry breaking one,  stressing the importance of time reversal symmetry in the FTI phase stability. 
We also discuss possible fractional phases beyond $\nu = 1/2$.
\end{abstract}

\date{\today}
\maketitle
\section{Introduction}
The field of topological insulators has exploded in recent years, fueled by the theoretical prediction~\cite{kane-PhysRevLett.95.226801, Bernevig15122006} and experimental realization~\cite{Koenig02112007, hsieh-nature2008452} of the topological insulators that preserve time reversal symmetry. Recently, the role of strong interactions in topological insulators has received much attention, especially in time reversal breaking systems~\cite{neupert-PhysRevLett.106.236804,sheng-natcommun.2.389,regnault-PhysRevX.1.021014} (see also Refs.~\cite{Bergholtz-2013IJMPB..2730017B,Parameswaran-2013CRPhy..14..816P} and references therein).
It has also been noticed that adding symmetry to topological order might create even more diverse quantum phases~\cite{Hung-PhysRevB.87.195103, Wen-PhysRevB.87.155114}. Such phases have been dubbed symmetry enriched topological phases, and include $\mathbb Z_2$ topological order. The existence of $\mathbb Z_2$ topological order is supported by the numerical observation of $\mathbb Z_2$ spin liquids~\cite{Yan-Science03062011}. The generalization of the FQH effect to time reversal invariant (TRI) systems considered in Refs.~\cite{PhysRevLett.96.106802,levin-PhysRevLett.103.196803} is another example of  $\mathbb Z_2$ topological order. 

The simplest example of a two dimensional TRI topological insulator is built from two decoupled copies of a Chern insulator. Chern insulators~\cite{haldane-1988PhRvL..61.2015H} are band insulators exhibiting a non-zero quantized Hall conductance (or integer quantum Hall effect) in the absence of a magnetic field. 
Chern insulators provide the simplest microscopic models of time reversal symmetry breaking topological insulators. They can also be paired to create a TRI topological insulator: in the system of two decoupled copies of a Chern insulator with opposite chiralities, one for spin-up and one for spin-down, the counterpropagating spin 
currents add up to a zero Hall conductance~\cite{kane-PhysRevLett.95.226801}. Such a quantum spin Hall system is characterized by a $\mathbb Z_2$ topological invariant~\cite{PhysRevLett.95.146802}, which takes value $0$ for a trivial insulator and $1$ for a topological insulator.  Two copies of a Chern insulator with an odd Chern number $C$ form a system with an odd number of Kramer's pairs. Following the argument of Kane and Mele~\cite{PhysRevLett.95.146802}, the composite system is non-trivial when $C$ is odd.

In the presence of strong interactions, fractionally filled Chern bands host a physics similar to that of the FQH effect, with excitations that carry fractional charge and statistics. The existence of these fractional Chern insulators (FCI) is supported by strong numerical~\cite{neupert-PhysRevLett.106.236804, sheng-natcommun.2.389, Bernevig-2012PhysRevB.85.075128, wang-PhysRevLett.107.146803, wang-PhysRevLett.108.126805,
Wu-2012PhysRevB.85.075116, Venderbos-PhysRevLett.108.126405, Kourtis-PhysRevB.86.235118, Scaffidi-PhysRevLett.109.246805,Liu-2013PhysRevB.87.205136, Lauchli-PhysRevLett.111.126802, qi-PhysRevLett.107.126803,Wu-2012arXiv1206.5773W,  Liu-PhysRevB.87.035306, Jian-PhysRevB.88.165134} and analytical~\cite{Roy-2012arXiv1208.2055R, goerbig-2012epjb, parameswaran-PhysRevB.85.241308, Dobardzic-PhysRevB.88.115117} evidence. 
Two copies of fractional Chern insulators with opposite chiralities for spin up and down can realize a TRI fractional topological insulator (FTI)~\cite{PhysRevLett.96.106802}. However, the existence of a fractional topological phase in a Chern band does not guarantee the stability of the composite system, just as a non-zero Chern number does not guarantee the existence of a quantum spin Hall effect in the non-interacting case. Levin and Stern~\cite{levin-PhysRevLett.103.196803} developed a criterion to determine if a FCI phase yields a stable FTI when doubled into a time 
reversal invariant system. They proved that a state with abelian quasiparticles of charge $e^*$ is a FTI when $\sigma_{SH} / e^*$ is odd, with $\sigma_{SH}$ the spin-hall conductance. Following this argument, any Laughlin-like FCI phase is a suitable candidate for the realization of an FTI. Subsequently, Refs.~\cite{levin-PhysRevB.86.115131, PhysRevB.84.165107, Koch-2013arXiv1311.6507K,Cappelli-2013JHEP...12..101C} also discussed a criterion to predict the stability of a TRI topological insulator in the presence of strong interactions.

 An earlier work~\cite{PhysRevB.84.165107} studied the stability of a fermionic FTI phase made of two FCI copies coupled by an interlayer interaction. The filling factor is $\nu = 2/3$ in each copy, which should lead to an unstable FTI phase according to Levin and Stern's criterion. However, one should not use this argument to explain the rather narrow range of stability of the FTI phase in Ref.~\cite{PhysRevB.84.165107}. Indeed, the numerical evidence presented in this work only uses the bulk properties of the system, while Levin and Stern's criterion concerns the stability of the edge states. In this article, we choose a filling fraction that respects Levin and Stern's stability rule , even though we only consider the bulk properties of our system. Additionally, the work of Kane and Mele~\cite{PhysRevLett.95.146802,kane-PhysRevLett.95.226801} proves that the topology of the band structure does not require the conservation of spin. Consequently, 
we 
study the stability of the FTI phase when the two FCI layers are coupled at the band structure level. This type of coupling is particularly interesting as in realistic systems, the FCI copies are coupled by a Rashba term that emerges from spin-orbit coupling.

The purpose of this paper is to propose and study a simple and stable microscopic model for a FTI. Our model is based on the kagome lattice model with interacting bosons at half filling.
The bulk stability of the FTI phase is probed through the ground state degeneracy and the particle entanglement spectrum (PES)~\cite{li-08prl010504, sterdyniak-PhysRevLett.106.100405}. This allows us to obtain the phase diagram indicating the 
stability of the FTI state with respect to the amplitude of the various coupling terms. The most salient feature is that the stability region is larger than that of a usual bilayer FCI system at the same filling fraction. When decoupled, these two systems are indistinguishable from the energy spectrum or PES perspective. As a result, time reversal symmetry seems to be a crucial ingredient of the stability of the FTI.

This article is organized as follows. In Sec.~II, we present the one-body lattice model used throughout the paper, and draw the phase diagram that gives the non-interacting topology of the two coupled Chern insulator copies. We also describe the interaction Hamiltonian. In Sec.~III we show the phase diagram when the two layers are only coupled by an interaction term. We compare our system to the time reversal breaking system of two FCI copies, and show that the TRI FTI is significantly more stable. In Sec.~IV, we discuss the stability of the FTI phase when an inversion symmetry breaking term couples the two layers at the band structure level. In Sec.~V, we discuss the other filling factors, with a particular attention to the bosonic $\nu = 1/3$ case. 

\section{Model Hamiltonian}
\label{sec:ModelHamiltonian}
\subsection{One-body model}
\label{sec:OneBodyHamiltonian}
\begin{figure}
\includegraphics[width=0.99\linewidth]{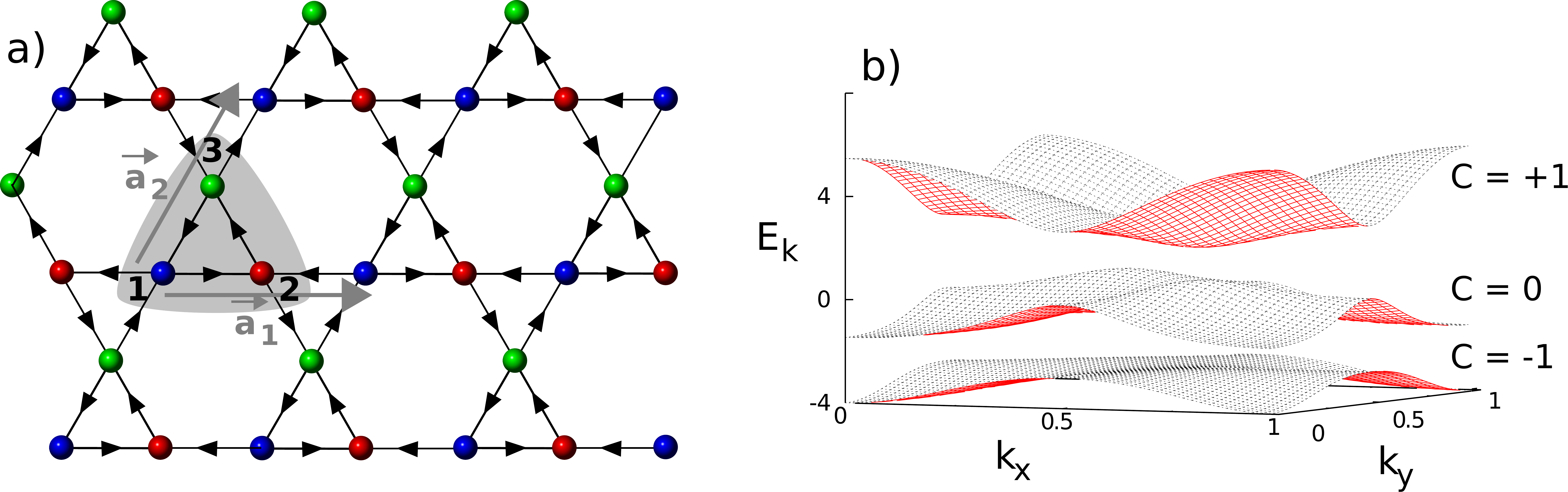}
\caption{a) The kagome lattice model. The three sublattices 1, 2, 3, are colored in blue, red, green respectively. The arrows give the sign convention for the phase of the nearest neighbor hopping term; the hopping amplitude is $\exp(i \varphi)$ in the direction of the arrows. The lattice translation vectors are $\boldsymbol  a_1$ and $\boldsymbol  a_2$. b) Band structure of the kagome lattice model with a nearest neighbor hopping phase $\varphi = \pi/4$. The three bands are separated by an energy gap, and have Chern number $C = -1,  \ 0, \ +1$ respectively.}\label{fig:lattice}
\end{figure}

Our model is based on two copies of the kagome lattice~\cite{tang-PhysRevLett.106.236802} Chern insulator model with only nearest neighbor hopping. The kagome lattice model has three atoms per unit cell and is spanned by $\boldsymbol a_{1}$ and $\boldsymbol a_{2}$(see Fig.~\ref{fig:lattice}a). The Bloch Hamiltonian of the single copy is given by
\begin{align}
\label{eq:OneBodyFCI}
\begin{BMAT}{ccc}{cc}
& \begin{BMAT}(@, 55pt, 10 pt){ccc}{c}
\ket{ 1 } \ \ \ \  & \ \  \ \  \ket{ 2 }  \ \ \ \  &  \ \ \ \  \ket{ 3 }   \ \ \ \  \\
\end{BMAT}
&
\\
h_{\rm CI}(\boldsymbol k) =& \left(\begin{BMAT}(@, 25pt, 20 pt){ccc}{ccc}
0 & e^{i\varphi}(1 + e^{-ik_x}) &  e^{-i\varphi}(1 + e^{-ik_y}) \\
 & 0 & e^{i\varphi}(1 + e^{i(k_x -k_y)})\\
h.c. &  & 0 \\
\end{BMAT}\right)
&
\begin{BMAT}(@, 0pt, 20 pt){c}{ccc}
\ket{1} \\  \ket{2} \\ \ket{3} \\
\end{BMAT}
 \end{BMAT}
\end{align}

Here we take the magnitude of the hopping term to be 1. Except for $\varphi = 0$ and $\varphi = \pm2\pi/3$ (at which the system is gapless), the model has two non-trivial bands with Chern number $C =\pm1$, in addition to a trivial band ($C = 0$) (see Fig.~\ref{fig:lattice}b). For the sake of simplicity, we set $\varphi = \pi / 4$. The model with $\varphi= \pi / 4$ is known to stabilize a Laughlin-like phase when interactions are switched on, as shown numerically in the fermionic case in Ref.~\cite{Wu-2012PhysRevB.85.075116}.
To build a two dimensional time reversal invariant topological insulator, we put together two copies of this model. A pseudospin is used to label the two copies, the $i^{th}$ atoms of both pseudospins being geometrically at the same position as depicted Fig.~\ref{fig:latticeFTIInteraction}. We wish to  study this system in the presence of two coupling terms;  an interpseudospin two-body interaction term, and a one-body term that couples the pseudospins at the band structure level. The single particle Bloch Hamiltonian for this model can be written in the form of a block matrix:
\begin{eqnarray}
\begin{BMAT}{ccc}{cc} 
 & \begin{BMAT}(b){cc}{c}
 \ket{\uparrow} \ \ \ &  \ \ \ \ket{\downarrow}
\end{BMAT}
&
\\
H_R(\boldsymbol k)  =  & \left(\begin{BMAT}(b){cc}{cc}
 h_{\rm CI}(\boldsymbol k) &  R \\
 R^{t} & h^{*}_{\rm CI}(-\boldsymbol k) \\
				\end{BMAT}\right)
&
\begin{BMAT}(b){c}{cc}
\ket{\uparrow} \\  \ket{\downarrow}\\
\end{BMAT}
 \end{BMAT}
\end{eqnarray}
where the top left block acts on the particles with pseudospin up only while the bottom right block acts on the particles with pseudospin down. In addition to these diagonal blocks, there is a coupling term which, in realistic models, is usually a Rashba term that depends on the momentum. We note $\cal T$ the time reversal operator (${\cal T}^2 = -1$). TRI requires that ${\cal T}H_R(\boldsymbol k){\cal T}^{-1} = H_R(-\boldsymbol k)$. This condition forces the coupling term to be antisymmetric ($R=-R^t$). When dealing with bosons, we have to keep in mind that $\cal T$ is just an anti-unitary discrete symmetry that we impose on our system and is not the true time reversal symmetry (which satisfies ${\cal T}^2 = 1$ for bosons). While this is a misnomer,  we will still refer to this symmetry as the time reversal symmetry. 

The lattice has $N_s$ unit cells with periodic boundary conditions. The periodic boundary conditions result in the quantization of the momentum $\boldsymbol k = (k_x, k_y)$, that can thus be labeled by two integers: $k_x = 0, \ ..., \ (N_x  - 1)$, and $k_y = 0, \ ..., \ (N_y - 1)$ with $N_s = N_x \times N_y$. When interactions are switched on, the geometric aspect ratio of the system has a critical influence on the magnitude of the manybody gap, as shown in the case of FCI in Refs.~\cite{regnault-PhysRevX.1.021014, Lauchli-PhysRevLett.111.126802}. In order to minimize this effect while studying the evolution of the gap with the system size, we follow the approach introduced in Ref.~\cite{Lauchli-PhysRevLett.111.126802} and use tilted boundary conditions. This technique allows the system to have a geometric aspect ratio very close to one for any number of unit cells $N_s$. The details of the periodic boundary conditions used in this paper are described in Appendix~\ref{app:TiltedBoundaryConditions}.

\subsection{The R term}
The coupling term $R$ is a $3\times 3$ antisymmetric matrix. In realistic systems, the Rashba effect emerges from spin-orbit coupling, and is momentum dependent. In this article, we want to probe the effect of a term that both couples the two layers at the single particle level, and breaks inversion symmetry.
% Here, we are discussing the perturbative effect of the {\color{red}Rashba} term. 
Since we are not tight to a physically realistic model, we will replace the Rashba term with a real, $\boldsymbol k$-independent matrix.
Using this approximation, and the fact that $R^t = -R$, the inversion symmetry breaking term is controlled by three parameters. We will study their influence independently. Examining how the inversion breaking term transforms under spatial symmetry transformations shows us that there is a natural choice for one of the axes. The following coupling matrix 
\begin{equation}
R_1  =  \frac{1}{\sqrt 6}\left[\begin{array}{c c c}
0 & 1 & -1 \\
-1 & 0 & 1 \\
1 & -1 & 0
\end{array}\right]
\label{eq:M1}
\end{equation}
preserves the $C_3$ rotational invariance present in the kagome lattice model. For that reason, we choose this matrix to be one of the axes of the phase diagram. We define the two other directions $R_2$ and $R_3$ so that $(R_1, R_2, R_3)$ is an orthonormal basis under the scalar product $(M,M') \rightarrow Tr(M M'^t)$.
\begin{eqnarray}
R_2 & = & \frac{1}{\sqrt {12}} \left[\begin{array}{c c c}
0 & 1 & 2 \\
-1 & 0 & 1 \\
-2 & -1 & 0
\end{array}\right]\\
R_3 & = & \frac{1}{2} \left[\begin{array}{c c c}
0 & 1 & 0 \\
-1 & 0 & -1 \\
0 & 1 & 0
\end{array}\right]
\end{eqnarray}
The total coupling term as a function of the matrices $R_i$ is given below
\begin{equation}
R = \alpha_1 R_1 + \alpha_2 R_2 + \alpha_3 R_3
\end{equation}
where $\alpha_1$, $\alpha_2$, $\alpha_3$ are real parameters. The one-body Hamiltonian has two symmetries with respect to $\alpha_1$, $\alpha_2$, $\alpha_3$ that we make explicit in the following. Note that for a generic case, $R$ breaks the mirror symmetry, even for $\a_2 = \a_3 = 0$.

\subsubsection{Symmetry under the transformation $R\rightarrow -R$} 

The Hamiltonians with opposite coupling
terms $H_{R}$ and $H_{-R}$ are related by the following unitary transformation
\ba
H_{R} = U^t H_{-R} U 
\ea
with\ba
U = \left[\begin{array}{c c }
I_3 & 0  \\
0 & -I_3 
\end{array}\right]
\ea
where $I_3$ is the $3 \times 3$ identity matrix. Therefore the transformation $R\rightarrow -R$ is a symmetry of the Hamiltonian.

\subsubsection{Symmetry under the transformation $\alpha_3 \rightarrow -\alpha_3$} 

Rotating the lattice by an angle $2\pi/3$ and interchanging the sublattices $1$ and $3$ leaves the system invariant. In the Hamiltonian, this transformation changes the sign of $\alpha_3$. Consequently, the transformation $\alpha_3 \rightarrow -\alpha_3$ is a symmetry of the Hamiltonian.

Using both symmetries, we can reduce the parameter space to span by fixing the sign of two parameters. We choose $\alpha_1, \alpha_3  >0$.

\begin{figure}
\includegraphics[width = 0.24\linewidth]{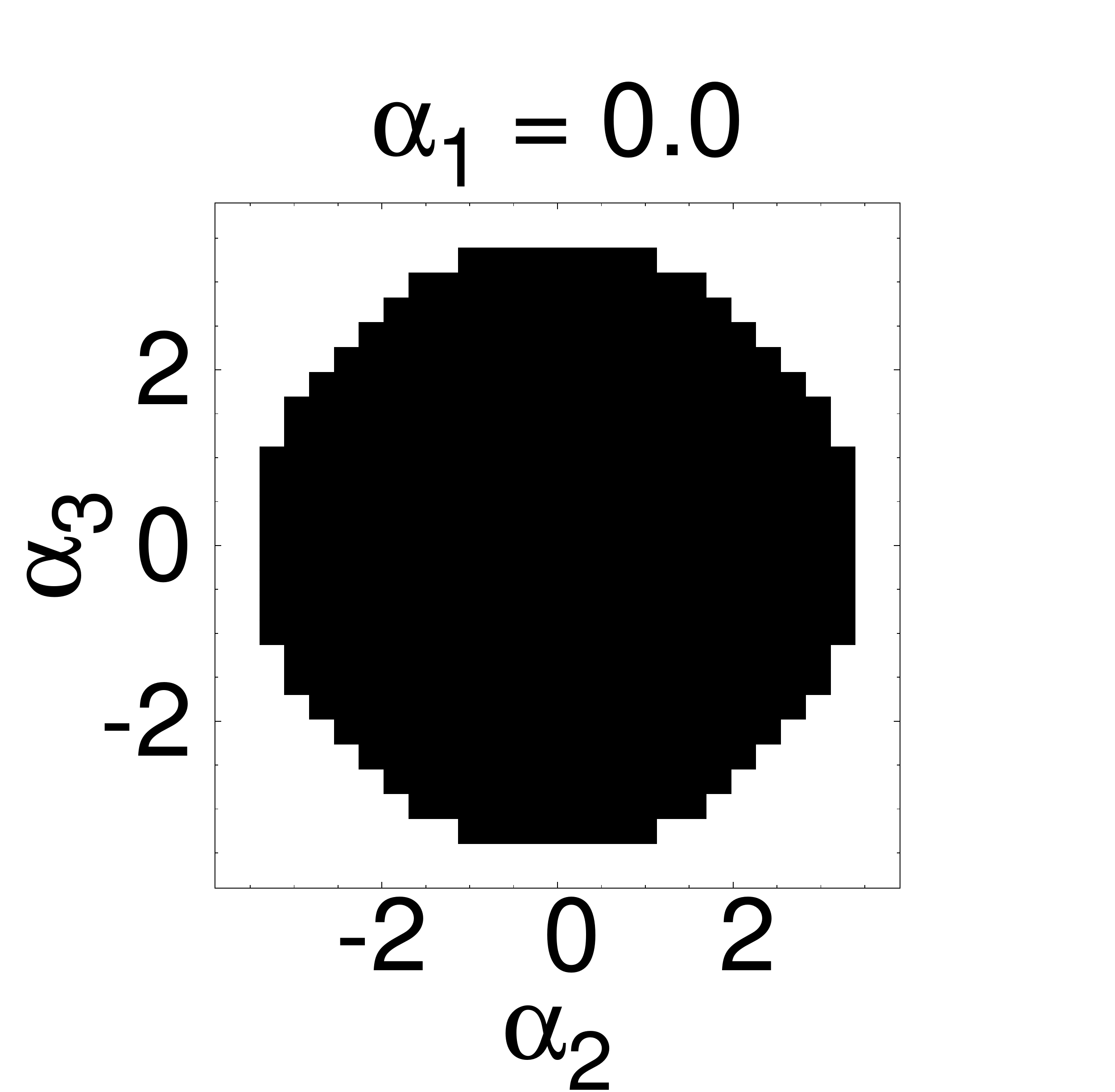}
\includegraphics[width = 0.24\linewidth]{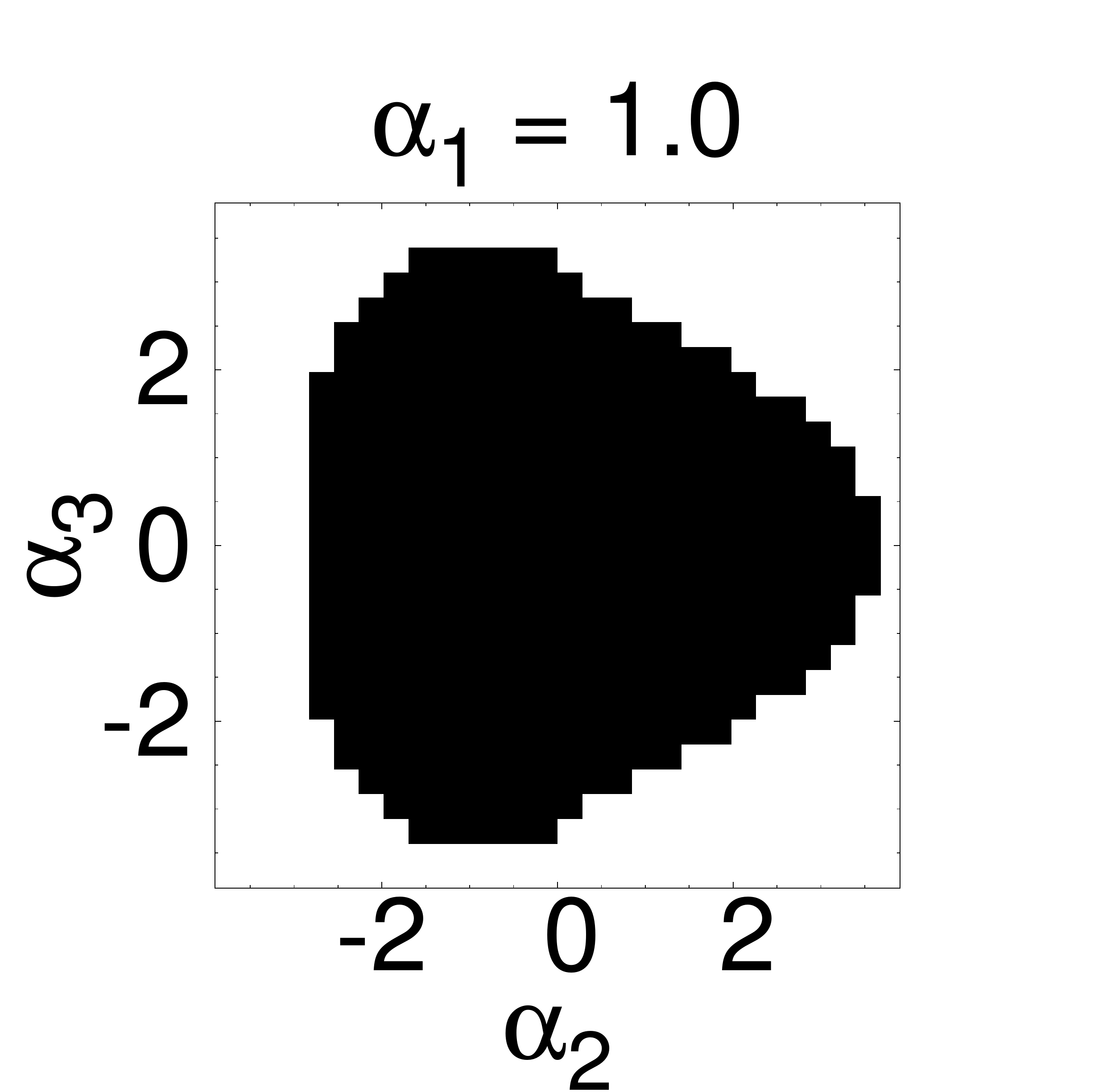}
\includegraphics[width = 0.24\linewidth]{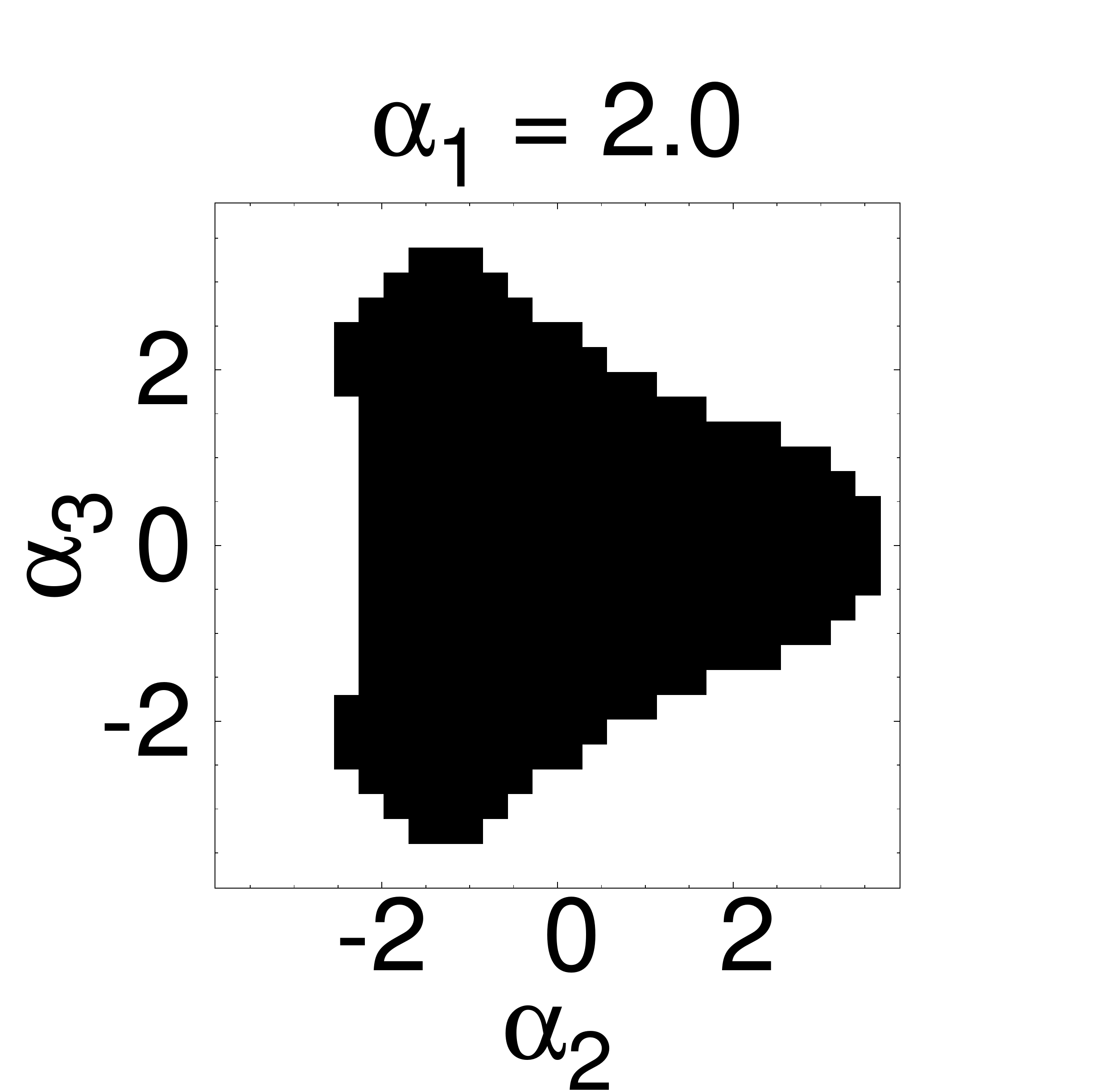}
\includegraphics[width = 0.24\linewidth]{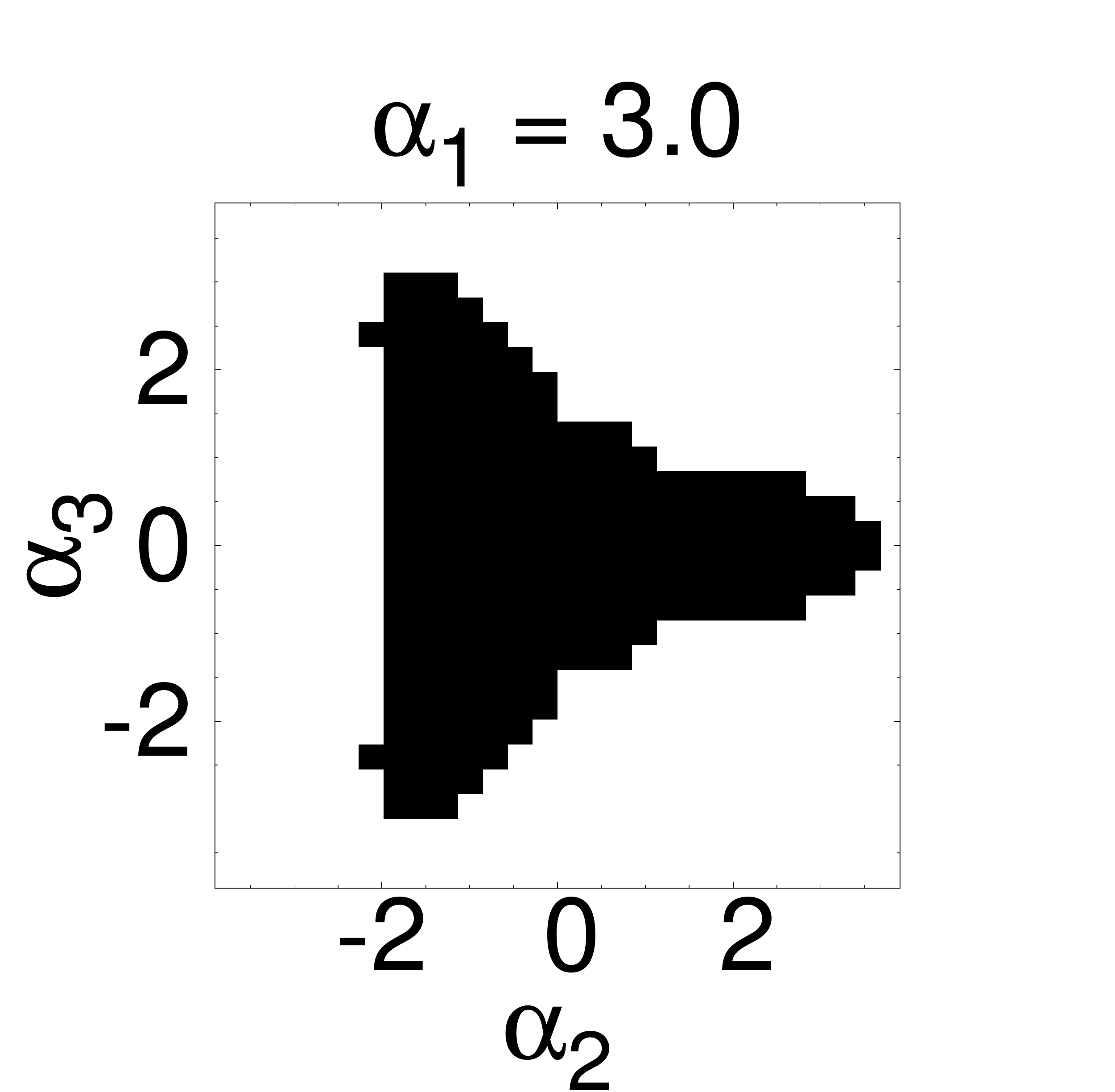}
\includegraphics[width = 0.24\linewidth]{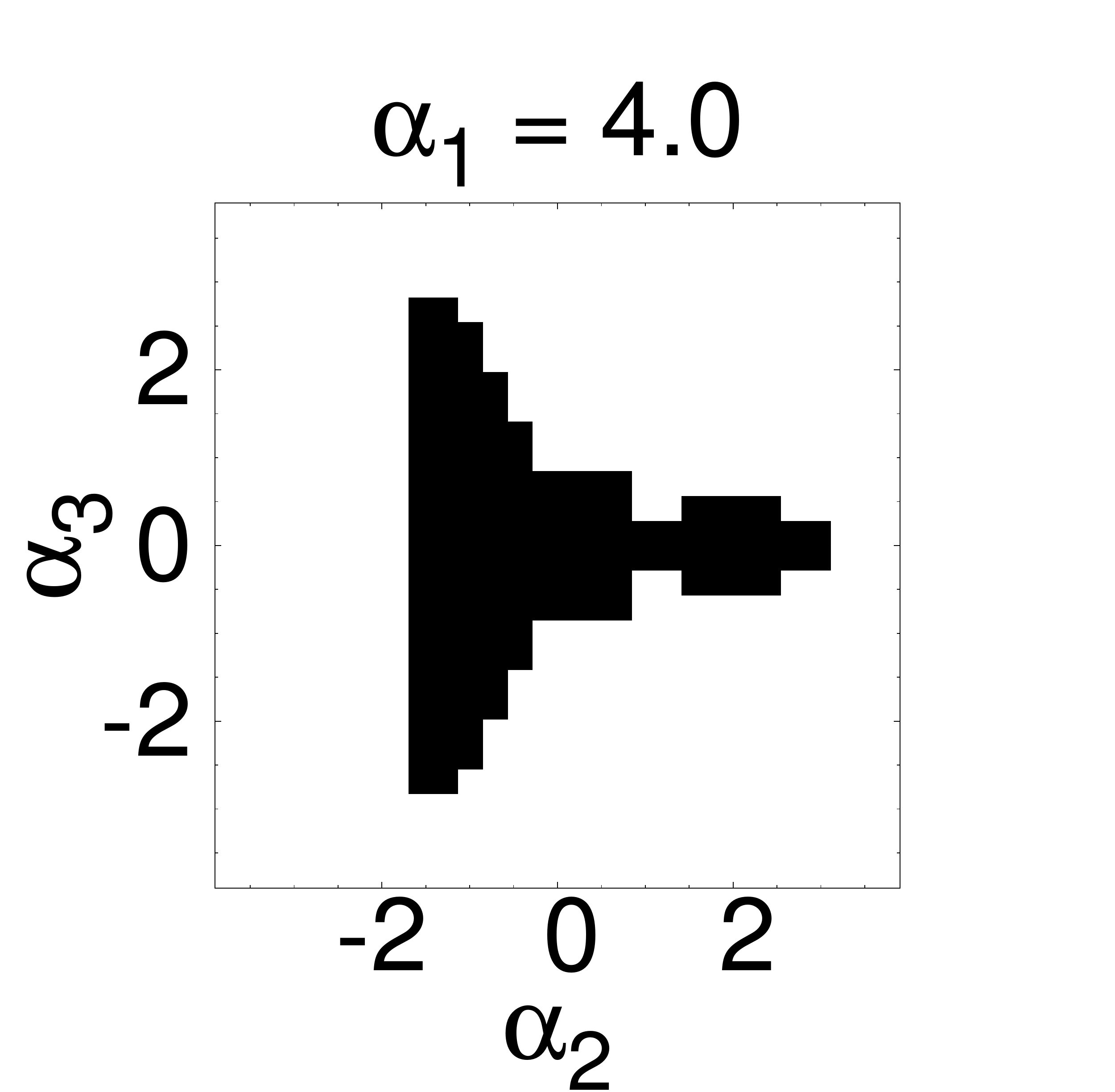}
\includegraphics[width = 0.24\linewidth]{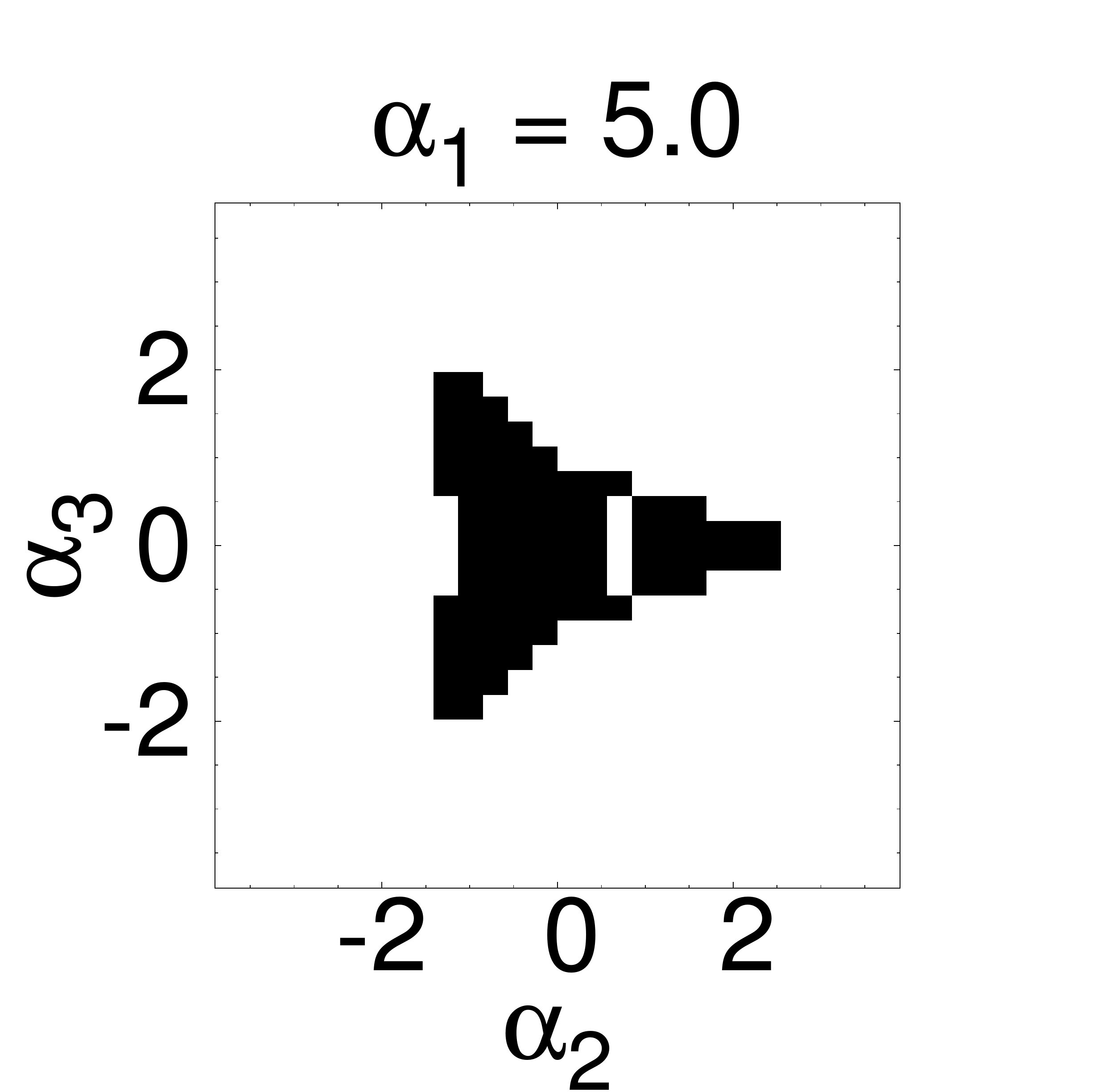}
\includegraphics[width = 0.24\linewidth]{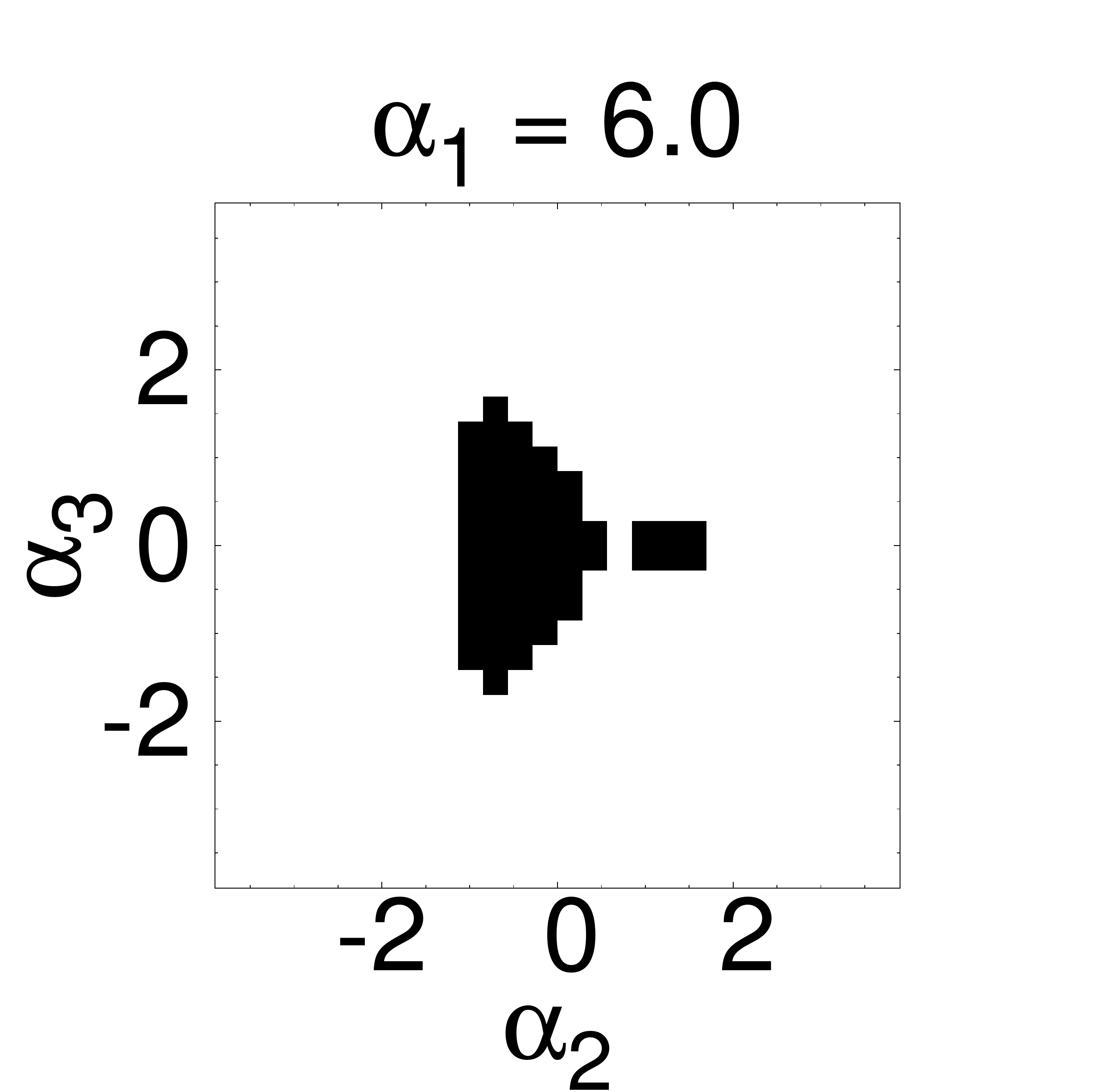}
\includegraphics[width = 0.24\linewidth]{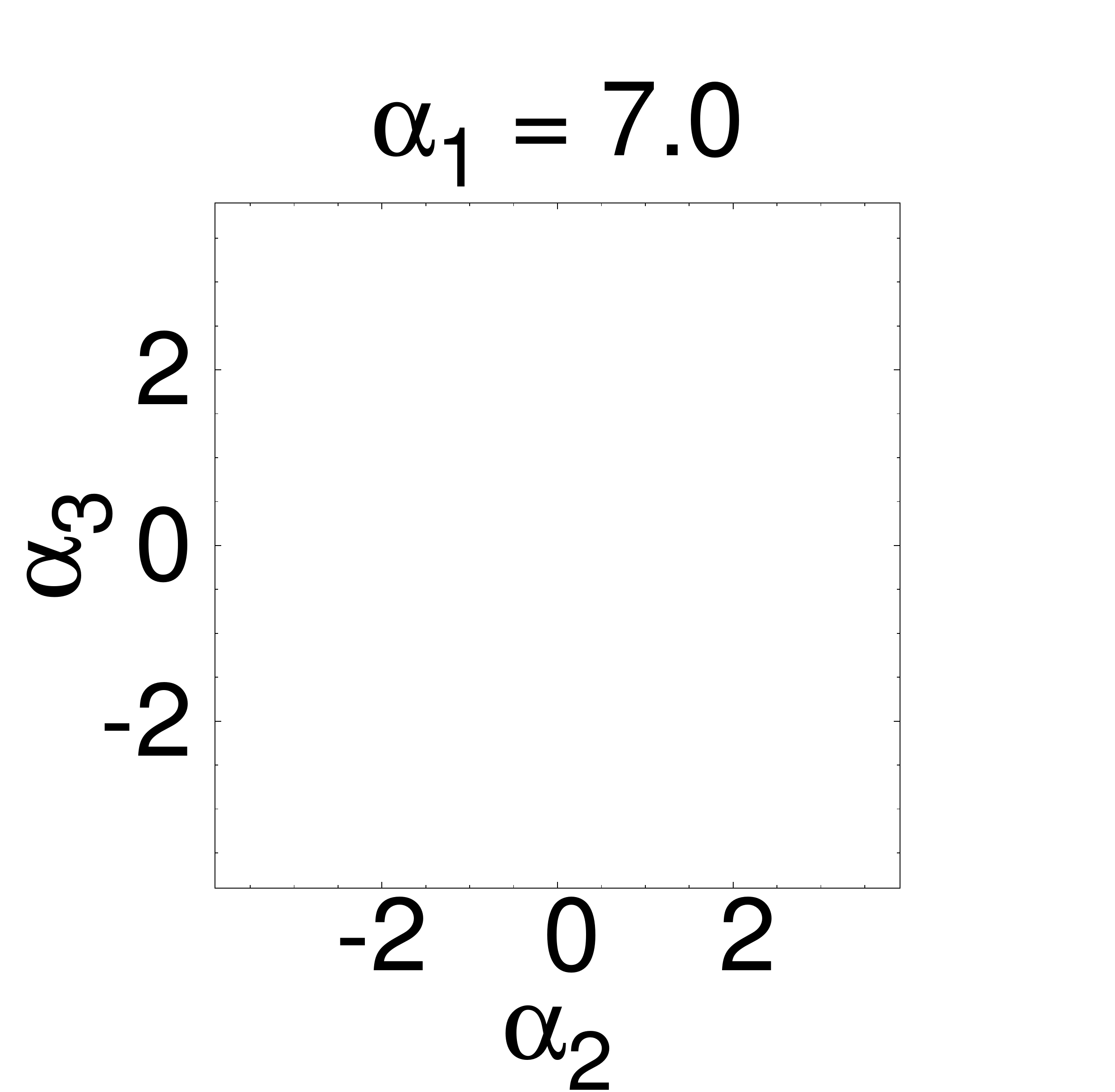}
\caption{Different cuts of the single particle phase diagram along the $C_3$ invariant axis. The one body model is non-trivial in the black region of the phase diagram. $\alpha_1$ is the amplitude of the $C_3$ invariant term and ranges from $-7$ to $7$, limits for which the kagome lattice model insulator is trivial for any value of $\alpha_2$ and $\alpha_3$. }
\label{fig:1bodyPhaseDiagram}
\end{figure}

In principle, we do not expect that adding interactions to a topologically trivial insulator will easily create a fractional phase. Therefore, we need to address the stability of the topological phase at the single particle level before attacking the problem of the stability of a fractional topological phase. When $R=0$, the single particle model is topologically non-trivial because one copy has a Chern number $+1$, the other $-1$~\cite{PhysRevLett.95.146802,kane-PhysRevLett.95.226801}. 
Increasing the strength of the coupling term $R$ to infinity will drive the system into a trivial phase that corresponds to the atomic limit. In the generic case ($R \neq 0$), the $\mathbb Z_2$ invariant can be effectively computed using the method developed in Ref.~\cite{PhysRevB.84.075119}. Fig.~\ref{fig:1bodyPhaseDiagram} shows some slices of the three dimensional phase diagram along the $\alpha_1$ axis. Note that for $|\alpha_1| > 7$, the kagome lattice model insulator is trivial for any value of $\alpha_2$ and $\alpha_3$.

\subsection{Interaction}
\begin{figure}
\includegraphics[width=0.8\linewidth]{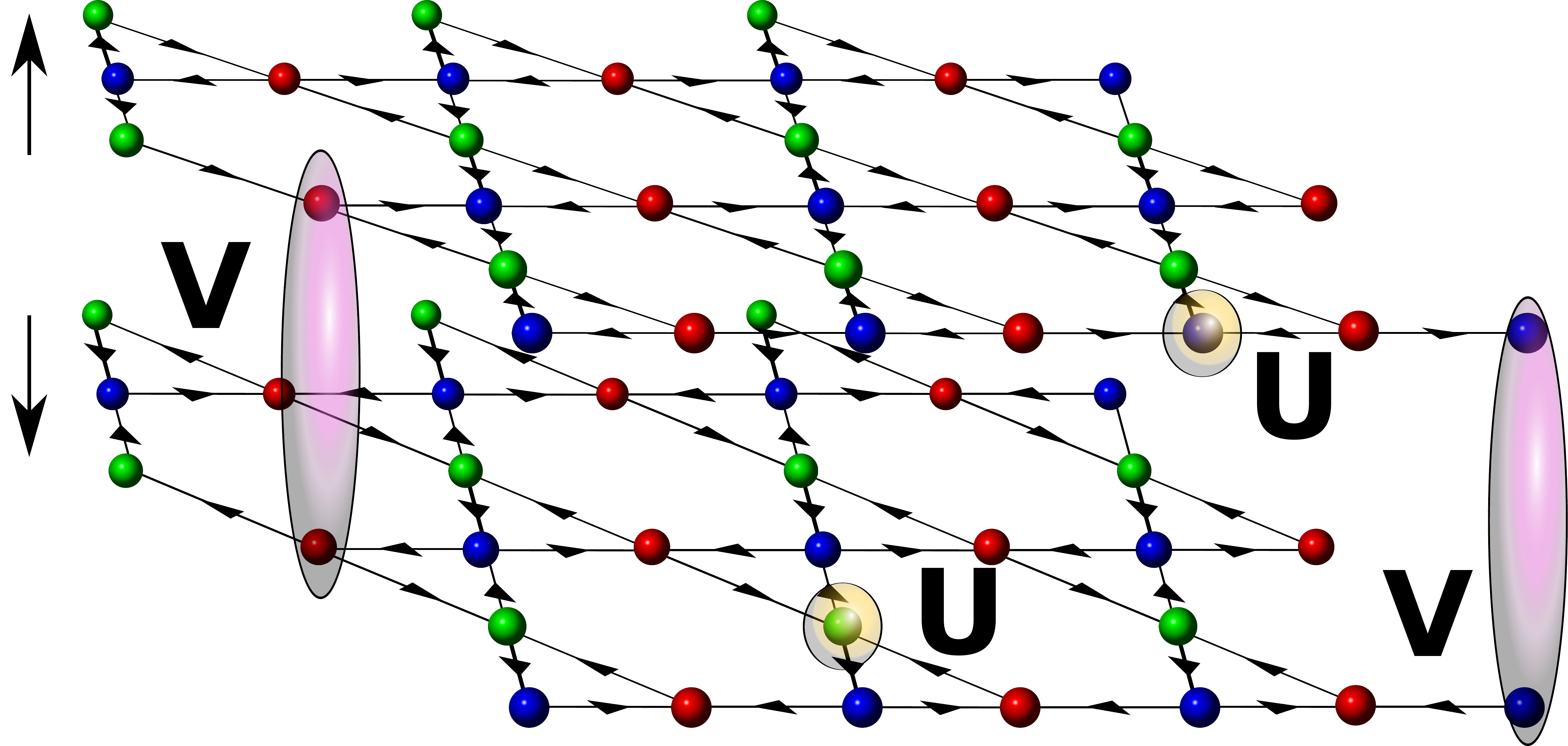}
\caption{Schematic representation of the interaction in our FTI kagome lattice system. The interaction strength is $U$ between bosons within the same layer (identical pseudospin) and $V$ between bosons in different layers (opposite pseudospin).}
\label{fig:latticeFTIInteraction}
\end{figure}

We consider $N$ bosons interacting through an on-site interaction. Fig.~\ref{fig:latticeFTIInteraction} gives a schematic representation of the interaction: the interaction has a strength $U$ for bosons within the same layer (same pseudospin), while it has a strength $V$ for bosons in different layers (opposite pseudospins). The interaction Hamiltonian is then
\begin{eqnarray}
\label{eq:interaction}
H_{\text{int}}&=&U\sum_{i,\sigma}\colon n_{i\sigma} n_{i\sigma}\colon+ 2V \label{eq:spininteraction}\sum_{i}\colon n_{i\uparrow}n_{i\downarrow}\colon
\end{eqnarray}
where $::$ denotes normal ordering, and $\sigma$ represents the pseudospin index (up or down). The filling fraction $\nu = N/(2N_s)$ is defined with respect to the partially filled two lowest bands, which together carry a $\mathbb Z_2$ invariant 1. In conventional FQH systems, the filling fraction is usually defined with respect to the fully polarized Landau level $\nu_\text{FQH} = N/N_s = 2 \nu$. We draw the reader's attention to this unusual convention, to avoid any confusion.

We use the so-called flat-band limit~\cite{regnault-PhysRevX.1.021014, neupert-PhysRevLett.106.236804} to remove the effect of band dispersion and band mixing. The kagome lattice model with pseudospin $1/2$ has six bands. The original Bloch Hamiltonian is $H(\mathbf{k})=\sum_n E_n(\mathbf{k})P_n(\mathbf{k})$ where $E_n(\mathbf{k})$ and $P_n(\mathbf{k})$ are the dispersion  and the projector onto the $n$-th band, respectively\footnote{Notice that the two bands with indices $2n$ and $2n + 1$ are not necessarily degenerate.}. We remove the effect of dispersion by considering the flat-band Bloch Hamiltonian restricted to the  two lowest bands $H(\mathbf{k})=1 - (P_0(\mathbf{k}) + P_1(\mathbf{k}))$. We thus consider the effective Hamiltonian:
\begin{equation}
H_{\rm eff}(\mathbf{k})=  {\cal P}(\mathbf{k}) H_\text{int}(\mathbf{k}) {\cal P}(\mathbf{k})
\end{equation}
where ${\cal P}(\mathbf{k})$ is the projector onto the two lowest bands.

\section{Stability of the fractional phase at half filling with pseudospin conservation}
\label{sec:HalfFilling}
We first consider the case of a single FCI copy on the kagome lattice model. Our model has a Bloch Hamiltonian $h_{\rm CI}(\boldsymbol k)$ and spinless bosons with an on-site interaction. At half filling (with respect to the lowest band), the fractional Chern insulator ground state is a two-fold almost degenerate ground state akin to a bosonic $1/2$ Laughlin phase~\cite{sheng-natcommun.2.389}. The ground state manifold has an energy splitting $\delta$ and is separated from higher energy states by a many-body gap $\Delta$. A well defined ground state (ie $\Delta > 0$ and $\delta \ll \Delta$) is an indication of the stability of the topological phase. Exact diagonalization of our model at half filling reveals that it has a well defined ground state, with little to no finite size effect. For instance, the system with $N = 6$ bosons and $N_s = 12$ unit cells with an aspect ratio $\k = 0.65$ is characterized by $\Delta = 0.17$ and $\delta/\Delta = 6.5 \ 10^{-3}$. The system with $N = 10$ bosons, $N_s = 20$ unit 
cells with an aspect ratio $\k = 0.93$ is characterized by $\Delta = 0.16$ and $\delta/\Delta = 1.0 \ 10^{-2}$. The ground state of the decoupled FTI system at $V = 0$ is the tensor product of two such FCI phases with opposite chiralities. The properties of the FCI ground state thus make our model an excellent candidate for the realization of FTI.

\subsection{Energy spectra}
\label{sec:ExactDiagHalfFilling}
\begin{figure}
\includegraphics[width = 0.49\linewidth]{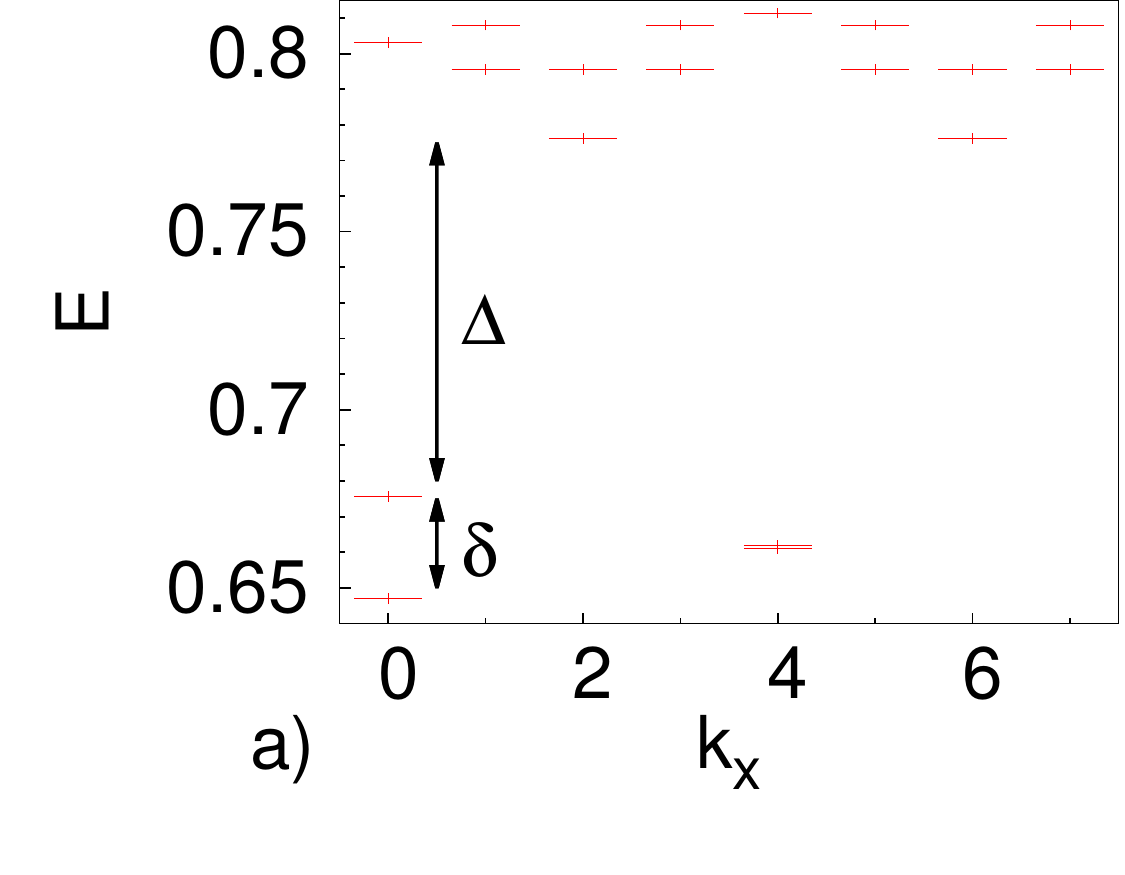}
\includegraphics[width = 0.49\linewidth]{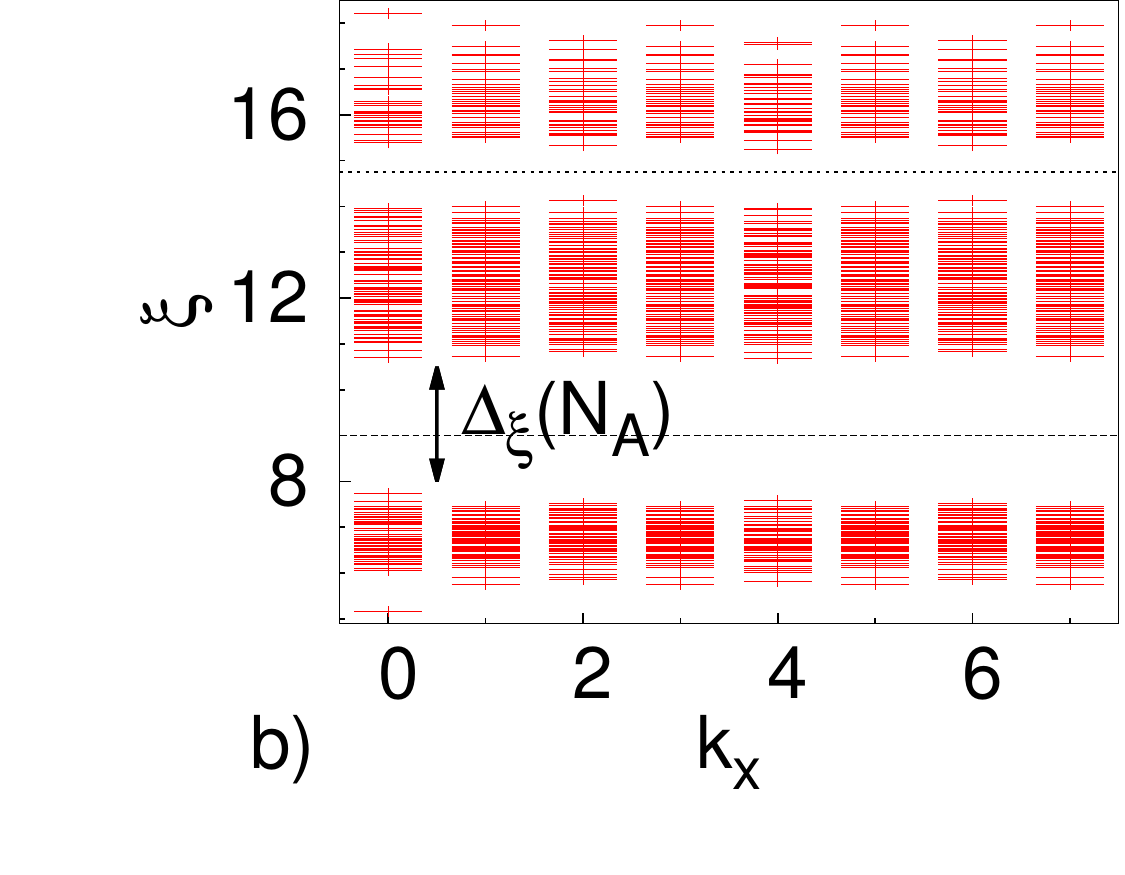}
\caption{Half filling FTI phase at $N = N_s = N_x = 8$ ($N_y = 1$), in the case where there is no band structure coupling, and with an interlayer interaction of strength $V/U = 0.5$. \emph{Left panel} Low energy spectrum. We observe an almost four-fold degenerate ground state lying in momentum sectors $(k_x, k_y) = (0,0)$ (two states) and $(k_x, k_y) = (4,0)$ (two states), with a gap $\Delta$ to higher excitations. The highest and lowest energy states of the ground state manifold have an energy splitting $\delta$. \emph{Right panel} PES of the ground state manifold, for a partition with $N_A = 4$ particles, in the sector $S_{zA} = 0$. The number of states below the lowest gap is $400$.  The number of states between the two dotted lines is $640$. Both countings are simply related to the number of states in the PES of the FCI system with $N_{\uparrow} = 4$ bosons, $N_s = 8$, and $N_{A}^{\uparrow} = 2$ bosons in the $A$ partition.}
\label{fig:EnergySpectrumDecoupled}
\end{figure}
\begin{figure}
\includegraphics[width = 0.49\linewidth]{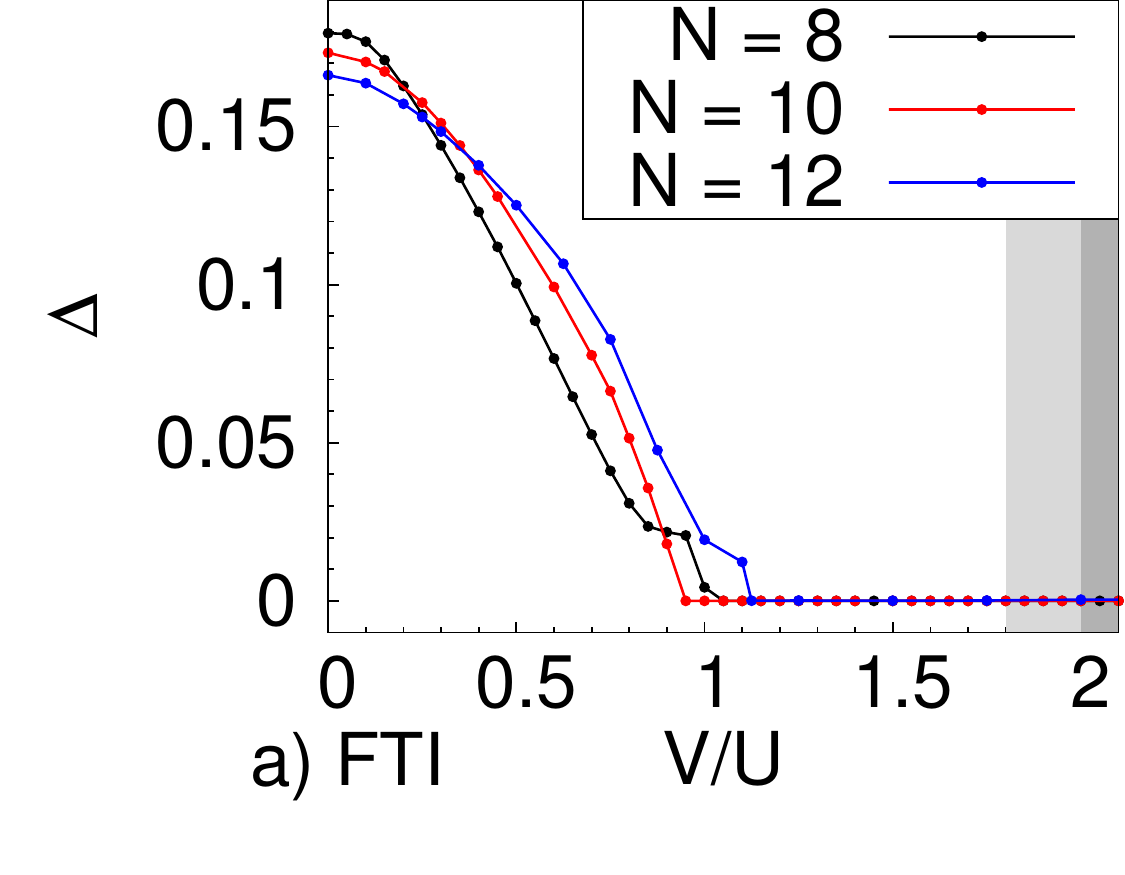}
\includegraphics[width = 0.49\linewidth]{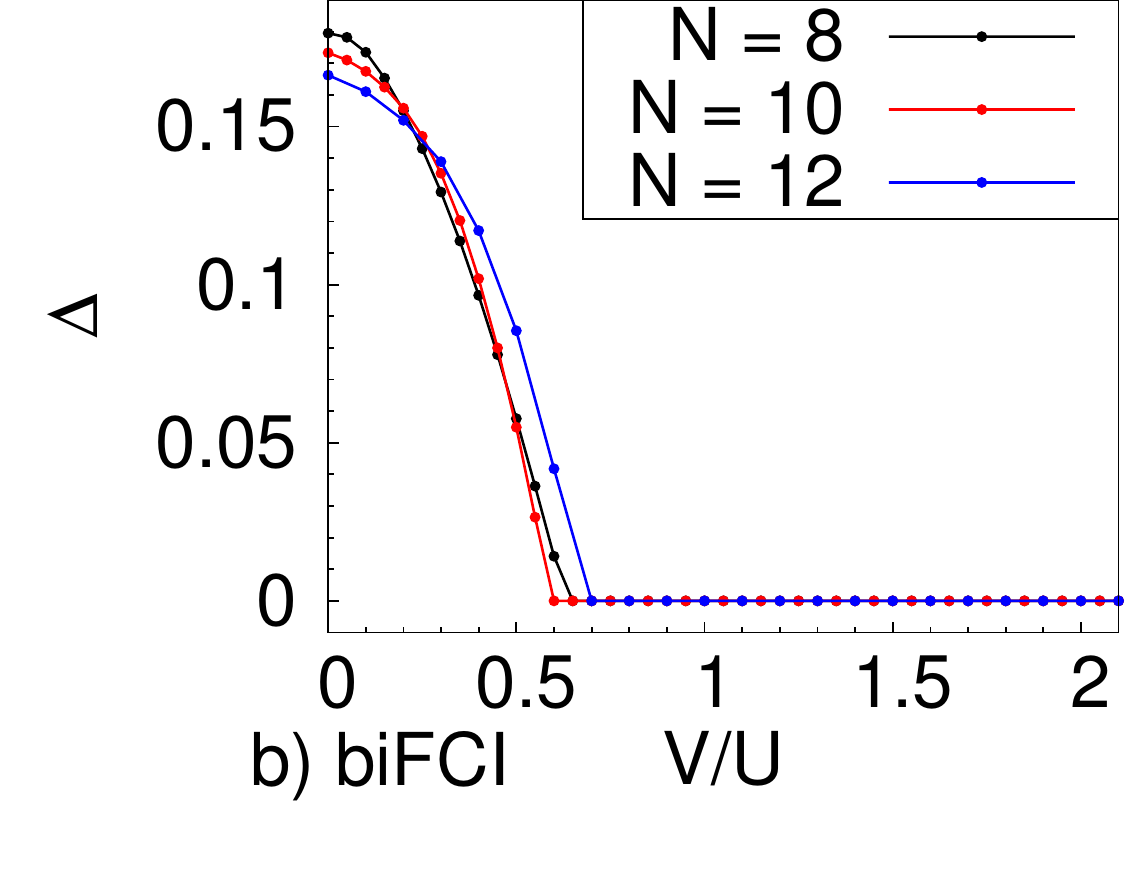}
\caption{Evolution of the many-body gap of the systems with $N = 8, \ 10, \ 12$ bosons and $N_s = N$ unit cells with respect to the magnitude of the interaction between bosons of opposite pseudospin ($V/U$), for the TRI FTI (a) and for the bilayer FCI (b). The shaded area in (a) corresponds to a full polarization of the FTI system, for $N = 10, \ 12$ (light grey) and $N = 8$ (dark grey). The bilayer FCI system only becomes fully polarized for values of $V/U$ that are beyond the scope of this graph (the transition happens for $20 < V/U < 30$ depending on the system size).}
\label{fig:DecoupledPhaseDiagramEnergy}
\end{figure}
We first focus on systems where the two Chern insulators copies are only coupled by the interaction, and not at the band structure level ($R = 0$). We define $S_z$, the difference of population between the two layers:
\begin{equation}
S_z = \frac{N_{\uparrow} - N_{\downarrow}}{2}
\end{equation}
where $N_{\uparrow}$ (respectively $N_{\downarrow}$) is the number of particles with pseudospin up (respectively down). 
This is thus a good quantum number in the case where no interlayer hopping is allowed.
We apply the interaction defined in Eq.~\pref{eq:interaction}, and study the stability of the fractional phase with increasing $V$, the amplitude of the interaction between bosons in different layers. When $V = 0$, the energy spectrum can be trivially induced from the spectrum of the FCI with the same number of unit cells. The energy spectrum of the decoupled system is an almost degenerate fourfold ground state (each bosonic $1/2$ Laughlin state being twofold degenerate), with a gap $\Delta$ to higher energy excitations. When we add some on-site interaction between bosons of opposite pseudospins, the gap is still clearly traceable (see Fig.~\ref{fig:EnergySpectrumDecoupled}a). We plot $\Delta$ for different values of $V/U$ (see Fig.~\ref{fig:DecoupledPhaseDiagramEnergy}a). We can also notice that the energy spread $\delta$ between the highest and lowest energy states of the ground state manifold increases.  

We keep track of the momentum sectors where the ground state manifold lies. These quantum numbers can be deduced from the ground state momentum sectors of the FCI system. We set the gap to zero whenever at least one of the four lowest lying states in the spectrum does not lie in the expected momentum sector. 
 There is a limit where $V/U$ is large enough that the system becomes fully polarized (all bosons occupy the same layer), as seen in Fig.~\ref{fig:DecoupledPhaseDiagramEnergy}a at $V/U = 1.8$ for $N=8$ and at $V/U = 2.0$ for $N_s = 10, \ 12$. Our results show that the gap amplitude decreases slowly, and does not vanish until $V/U \simeq 1.0$. This suggests that the FTI phase is remarkably robust to the introduction of a coupling interaction. These results barely exhibit any finite size effect, as the range of stability does not show any significant variation from $8$ to $10$ to $12$ particles. 
 
 \begin{figure}
\includegraphics[width = 0.49\linewidth]{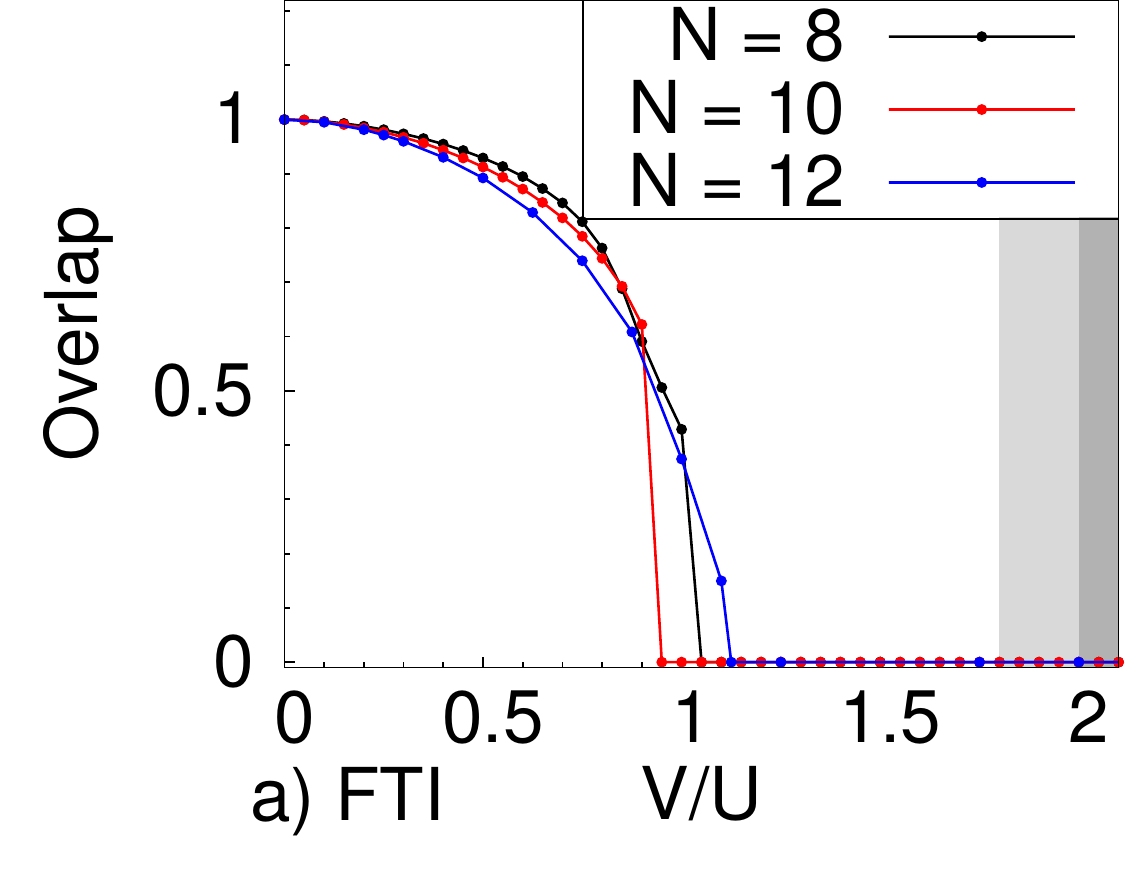}
\includegraphics[width = 0.49\linewidth]{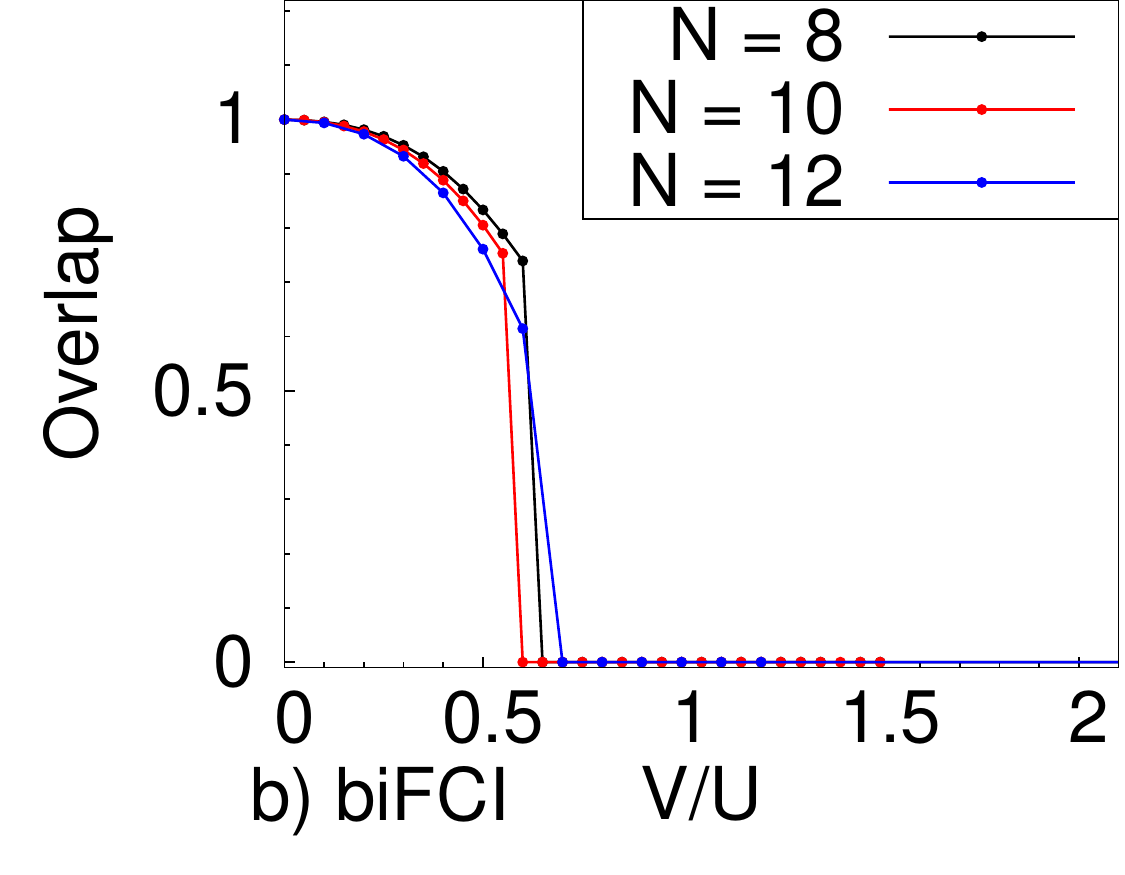}
\caption{Overlap of the fourfold ground state wavefunction with the wavefunction of the decoupled system. Evolution of the overlap of the systems with $N = 8, \ 10, \ 12$ bosons and $N_s = N$ unit cells with respect to the magnitude of the interaction between bosons of opposite pseudospin ($V/U$), for the TRI FTI (a) and for the bilayer FCI (b). The shaded areas correspond to the fully polarized phase as discussed in Fig.~\ref{fig:DecoupledPhaseDiagramEnergy}. We set the overlap to zero as soon as one of the four lowest lying states falls out of the expected momentum sectors.}
\label{fig:DecoupledPhaseDiagramOverlap}
\end{figure}
 
We also computed the overlap of the ground state eigenvectors with the eigenvectors of the decoupled system (see Fig.~\ref{fig:DecoupledPhaseDiagramOverlap}a). This shows that as long as there is a well defined ground state, the overlap remains close to 1, but drops to a value close to zero when the gap becomes smaller, thus confirming the results given in Fig.~\ref{fig:DecoupledPhaseDiagramEnergy}. Again, the results do no show any finite size effect. In the following paragraph, we will show that time reversal invariance is in fact crucial to such stability.

\subsection{Comparison with the system of two FCI copies with no time reversal invariance}
\label{sec:FCIBilayer}
 The study we carried out can be repeated for a system of two Chern insulators with identical chiralities for pseudospin up and down. The physics of this system is the same as the physics of the bilayer FQH. The single particle Hamiltonian reads 
\begin{eqnarray}
\begin{BMAT}{ccc}{cc} 
& \begin{BMAT}(b){cc}{c}
 \ket{\uparrow} \ \ \ &  \ \ \ \ket{\downarrow}
\end{BMAT}
&
\\
H_\text{Bilayer}(\boldsymbol k)  = & \left(\begin{BMAT}(b){cc}{cc}
 h_{\rm CI}(\boldsymbol k) &  0 \\
 0 & h_{\rm CI}(\boldsymbol k) \\
				\end{BMAT}\right)
&
\begin{BMAT}(b){c}{cc}
\ket{\uparrow} \\  \ket{\downarrow}\\
\end{BMAT}
 \end{BMAT}
\end{eqnarray}
where $h_{\rm CI}(\boldsymbol k)$ is defined in Eq.~\pref{eq:OneBodyFCI}. This system breaks time reversal symmetry. We remind the reader that we defined the FTI filling fraction with respect to the two lowest bands ($\nu = \nu_\text{FQH}/2$).

The interaction Hamiltonian is unchanged and given by Eq.~\pref{eq:interaction}. The physics of a bilayer FCI is similar to the one of a bilayer FQH effect. In particular, at $\nu = 1/2$ and $V = 0$, the ground state is similar to the Halperin $(2,2,0)$ state~\cite{Halperin83}. When the two layers are completely decoupled, the energy spectrum of the bilayer FCI and the energy spectrum of the FTI are identical, due to the inversion symmetry of the kagome lattice model. Therefore, the starting point of the stability analysis is exactly the same in both cases.

We perform exact diagonalizations on the bilayer FCI system. Again, we expect a fourfold quasidegeneracy of the ground state. We look at the many-body gap $\Delta$, and at the ground state manifold energy splitting $\delta$ to see how well defined the ground state is. The evolution of the gap for systems with $N = 8, \ 10$ and $12$ particles is shown Fig.~\ref{fig:DecoupledPhaseDiagramEnergy}b, and also exhibits negligible finite size effects. At $V/U \simeq 0.5$, it is no longer possible to distinguish a clear low energy manifold from higher energy excitations. This transition occurs near $V/U \simeq 1.0$ in the FTI system. Note that for $V/U=1.0$, the bilayer system recovers a full $SU(2)$ symmetry. Such a symmetry is absent in the FTI case.
This tends to indicate that the stability of the FTI is not 
trivially supported by the stability of the FCI phase, but rather that the time reversal invariance plays a crucial role. 
We also looked at the overlap of the low energy manifold with the ground state of the decoupled system (see Fig.~\ref{fig:DecoupledPhaseDiagramOverlap}b). As expected, the overlap decreases faster in the case of a bilayer FCI than in the case of a FTI.

There is another striking difference between the bilayer FCI and the FTI phase diagrams at large interlayer interaction. As discussed in Sec.~\ref{sec:ExactDiagHalfFilling}, at large $V/U$, we expect the system to be fully polarized. In the FTI case, such a transition occurs around $V/U \simeq 2$. For the bilayer FCI, the transition happens at a value that is one order of magnitude greater ($V/U = 26.4$, $21.8$ and $19.6$ for respectively $N = 8, \ 10, \ 12$).
This is another fundamental characteristic that differentiates the phase diagrams of these two systems.

\subsection{Particle entanglement spectrum}

\begin{figure}
\includegraphics[width = 0.48\linewidth]{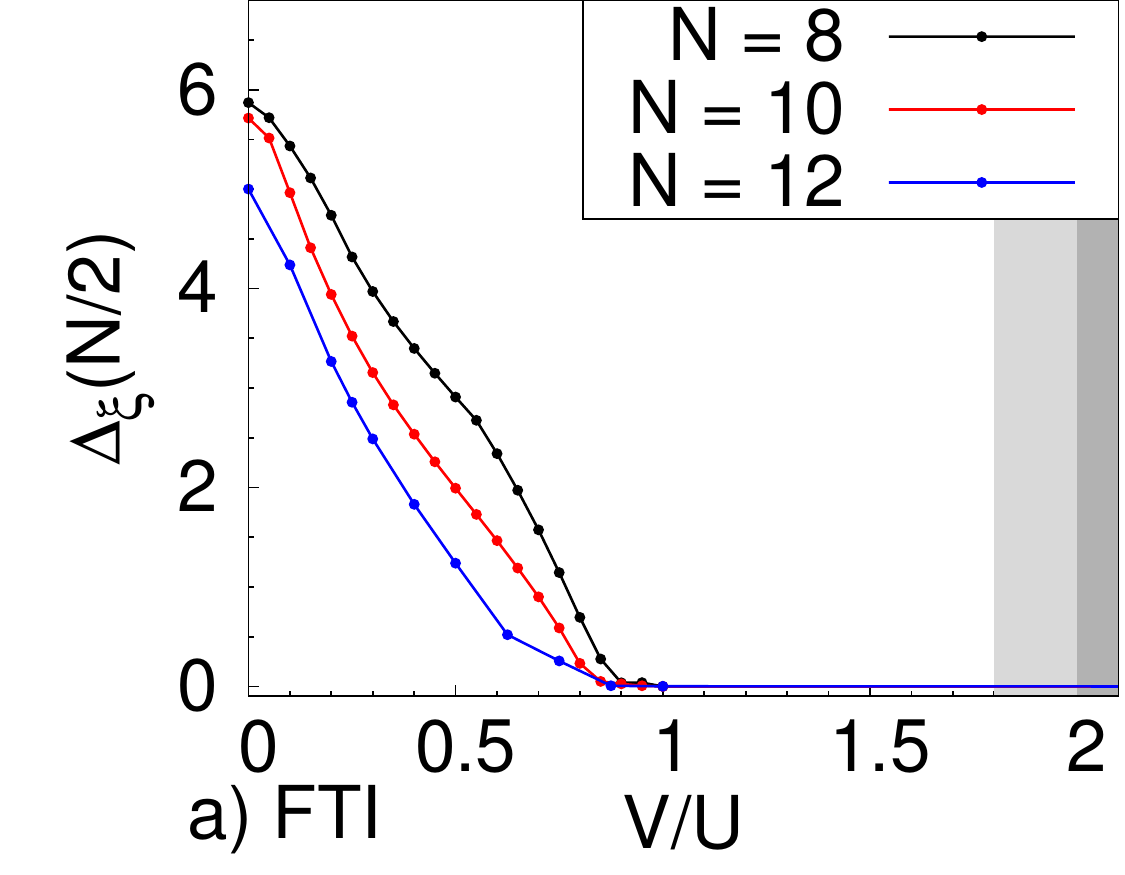}
\includegraphics[width = 0.48\linewidth]{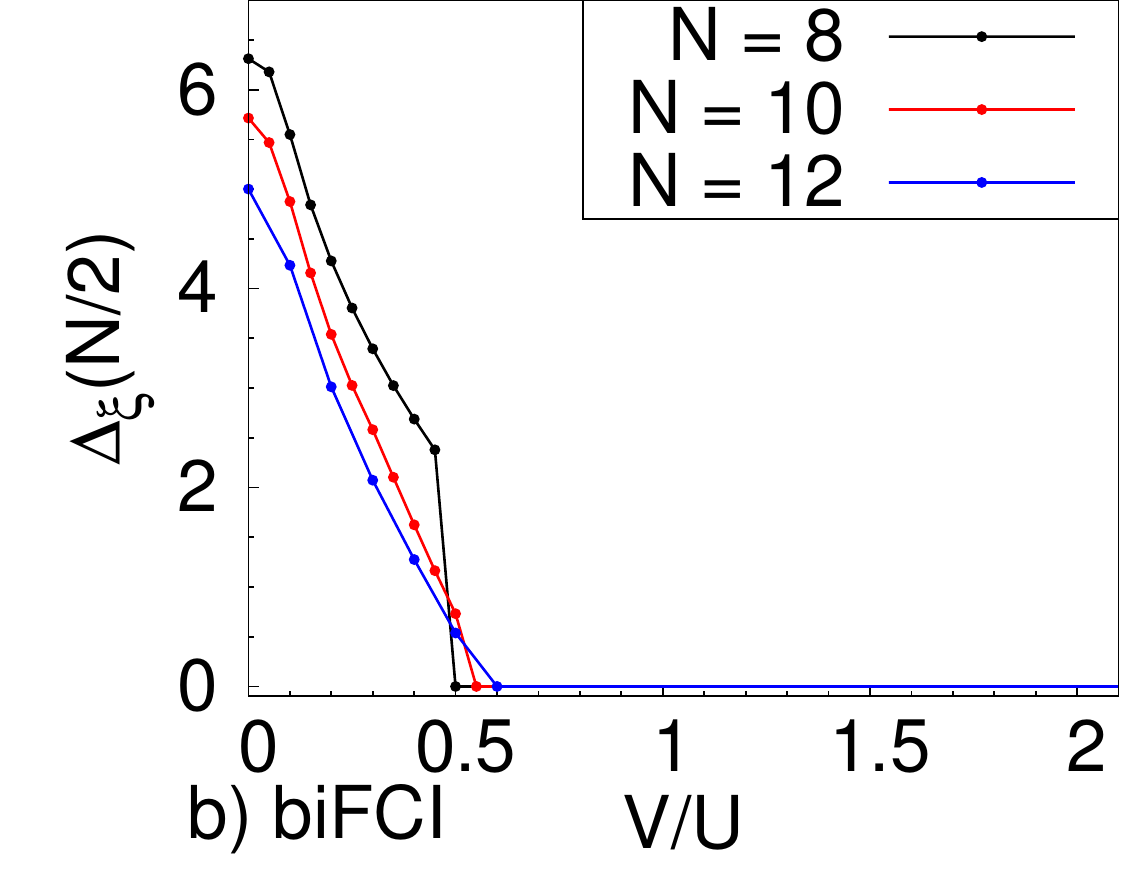}
\caption{Evolution of the entanglement gap of the systems with $N = 8, \ 10, \ 12$ bosons and $N_s = N$ unit cells for a partition $N_A = N/2$, for the TRI FTI (a) and for the bilayer FCI (b). $V/U$ is the amplitude of the interlayer interaction.  The shaded areas correspond to the fully polarized phase as discussed in Fig.~\ref{fig:DecoupledPhaseDiagramEnergy}.}
\label{fig:PESGap}
\end{figure}

The signatures of a Laughlin-like phase in a FCI appear in different manners, the most simple one being the ground state quasidegeneracy. Unfortunately, the energy spectrum of a charge density wave (CDW) could present a similar property. The PES~\cite{li-08prl010504, sterdyniak-PhysRevLett.106.100405} allows one to distinguish FQH-like phases from CDW and thus offers better signatures of the FCI phases. 
For a $d$-fold degenerate state $\{\ket{\psi_i}\}$, we consider the density matrix $\rho=\frac{1}{d}\sum_{i=1}^{d}\ket{\psi_i}\bra{\psi_i}$. We divide the $N$ particles into two groups $A$ and $B$ with respectively $N_A$ and $N_B$ particles. Tracing out on the particles that belong to $B$, we compute the reduced density matrix $\rho_A={\rm Tr}_B \rho$. This operation preserves the geometrical symmetries of the original state, so we can label the eigenvalues $\exp(-\xi)$ of $\rho_A$ by their corresponding momenta $k_{xA}, \ k_{yA}$. $S_{zA} = \frac{N_A^{\uparrow} - N_A^{\downarrow}}{2}$ is also a good quantum number as long as $S_z$ is one. The PES of the ground state of the system at $V/U = 0.5$ with $N = 8$ bosons is shown Fig.~\ref{fig:EnergySpectrumDecoupled}b for a particle partition with $N_A = 4$. 

When the entanglement spectrum is gapped, the number of states below the gap is a signature of a given topological phase. 
It is related to the number of quasihole excitations, a hallmark of the fractional phase. 
Refs.~\cite{Bernevig-2012PhysRevB.85.075128, Wu-2012arXiv1206.5773W, regnault-PhysRevX.1.021014} have argued that the low lying part of a FCI PES captures the physics of the corresponding model wavefunction. In constrast, the high entanglement energy structure above the gap is non-universal. We assume that this property holds true for FTI, and focus on the states that are separated by an entanglement gap $\Delta_{\xi}(N_A)$ from the rest of the PES (see Fig.~\ref{fig:EnergySpectrumDecoupled}b). Their number can be derived from the number of states below the entanglement gap of the FCI system, plus the condition that the original ground state is characterized by $S_z = 0$. This results in additional constraints that affect the counting (see Ref.~\cite{sterdyniak-PhysRevB.87.205137}). We give the counting that results from these constraints and some additional properties of the FTI PES in appendix \ref{app:FTIPESProp}.

We check that in the fully decoupled case, the PES has the expected counting. We note $\Delta_{\xi}(N_A)$ the smallest of the entanglement gaps in all sectors $(S_{zA}, \mbf k_A)$ 
\ba
\Delta_{\xi}(N_A) = \text{min}\{ \Delta_{\xi}(N_A, S_{zA}, \mbf k_A)\}
\ea
We follow the evolution of $\Delta_{\xi}(N_A)$ upon increasing interlayer interaction $V/U$. In addition to the persistence of the entanglement gap, the consistent counting below this gap with increasing $V/U$ allows us to conclude to the stability of the FTI phase in a significant interval of $V/U$. In Fig.~\ref{fig:PESGap}a we show the evolution of the entanglement gap for three system sizes ($N = 8, \ 10, \ 12$). We show the evolution of $\Delta_\xi (N_A = N/2)$, as this gives the most conservative estimation of the stability of the phase ($N_A = N/2$ gives the smallest entanglement gap). Additional results are presented in appendix \ref{app:HalfFillingSuppMat}. These results confirm the topological nature of the phase for values of the interlayer interaction up to $V/U \simeq 0.8$, with negligible finite size effect.

We also compute the PES of the ground state of the FCI bilayer system discussed in section \ref{sec:FCIBilayer}. The number of states below the gap is the same as for the TRI FTI, due to inversion symmetry of the kagome lattice model. As we can see in Fig.~\ref{fig:PESGap}b, $\Delta_{\xi}(N_A = N/2)$ decreases much faster than it does in the TRI case for all the system sizes that we have looked at. This confirms the results from the energy spectrum study, in so far as it gives additional evidence that the time reversal invariance is crucial to the stability of the FTI phase.

\section{Stability of the fractional phase at half filling without pseudospin conservation}
\label{sec:HalfFillingRashba}
\begin{figure}
\includegraphics[width = 0.49\linewidth]{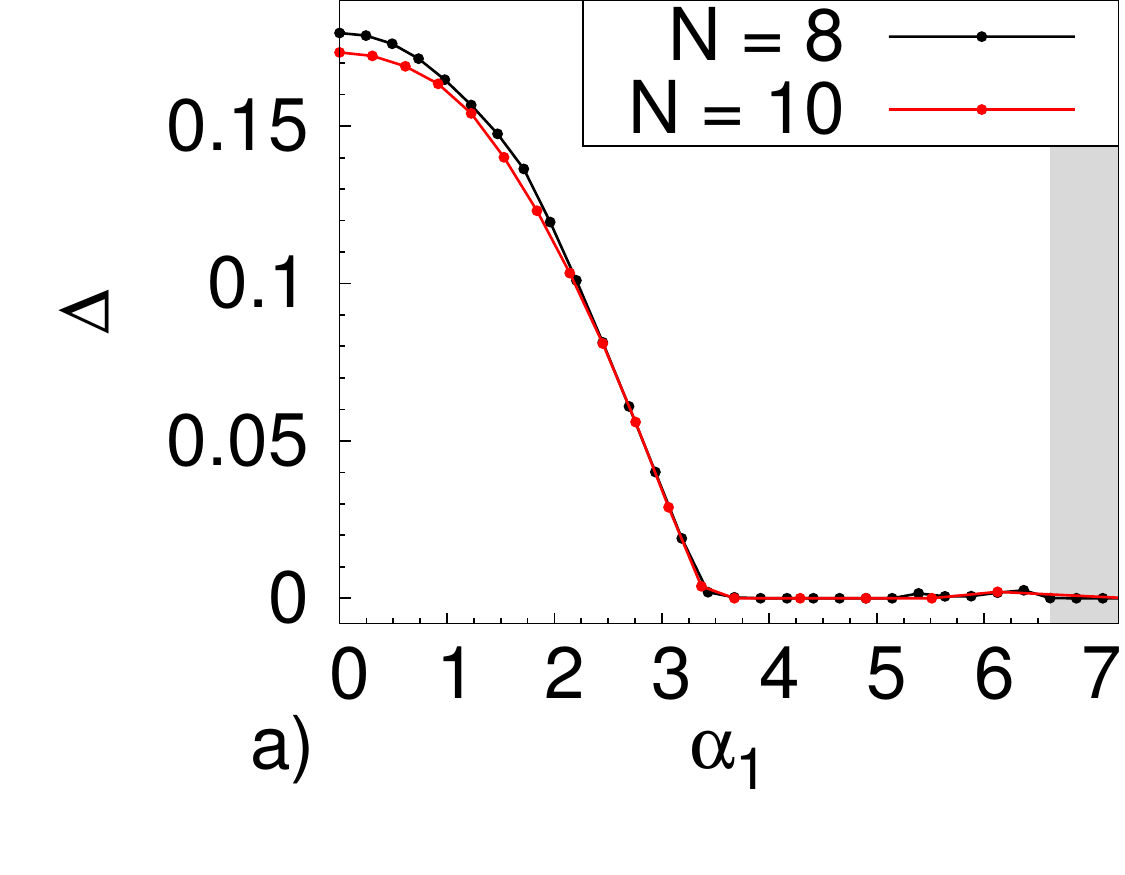}
\includegraphics[width = 0.49\linewidth]{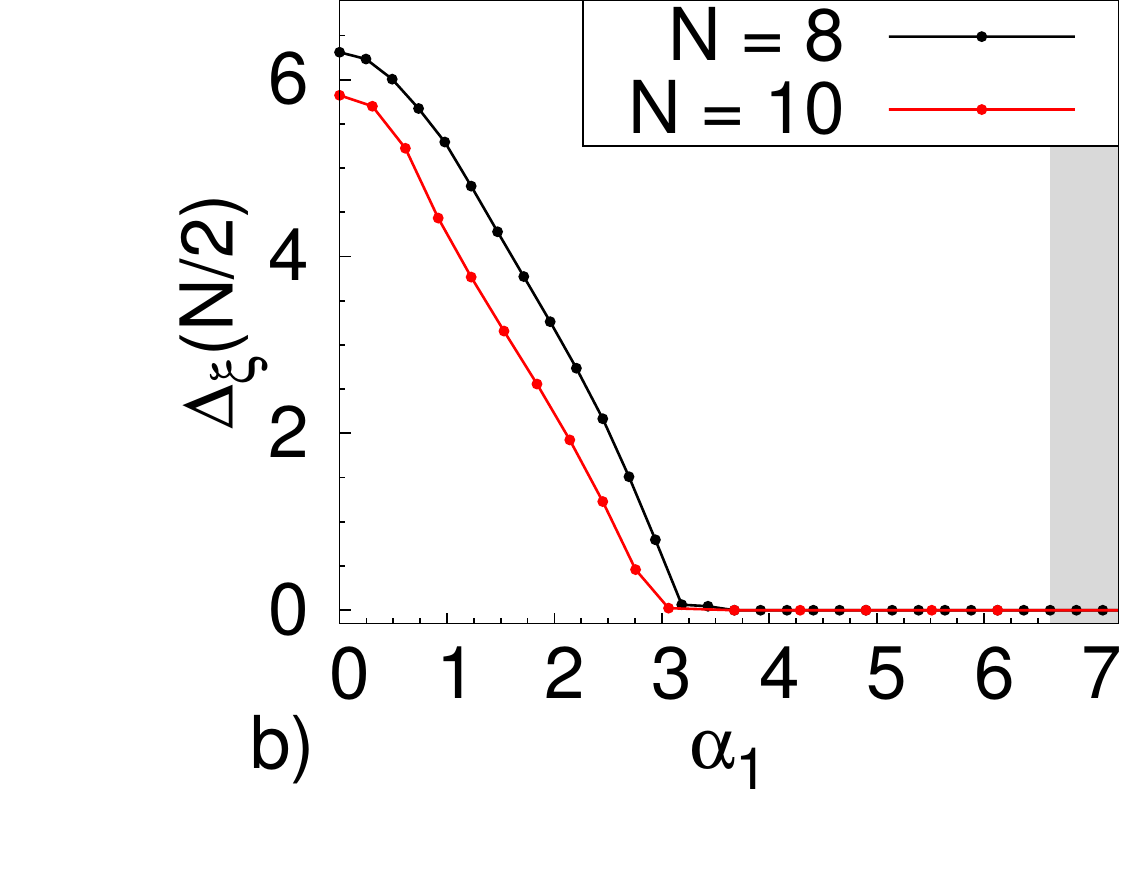}
\includegraphics[width = 0.49\linewidth]{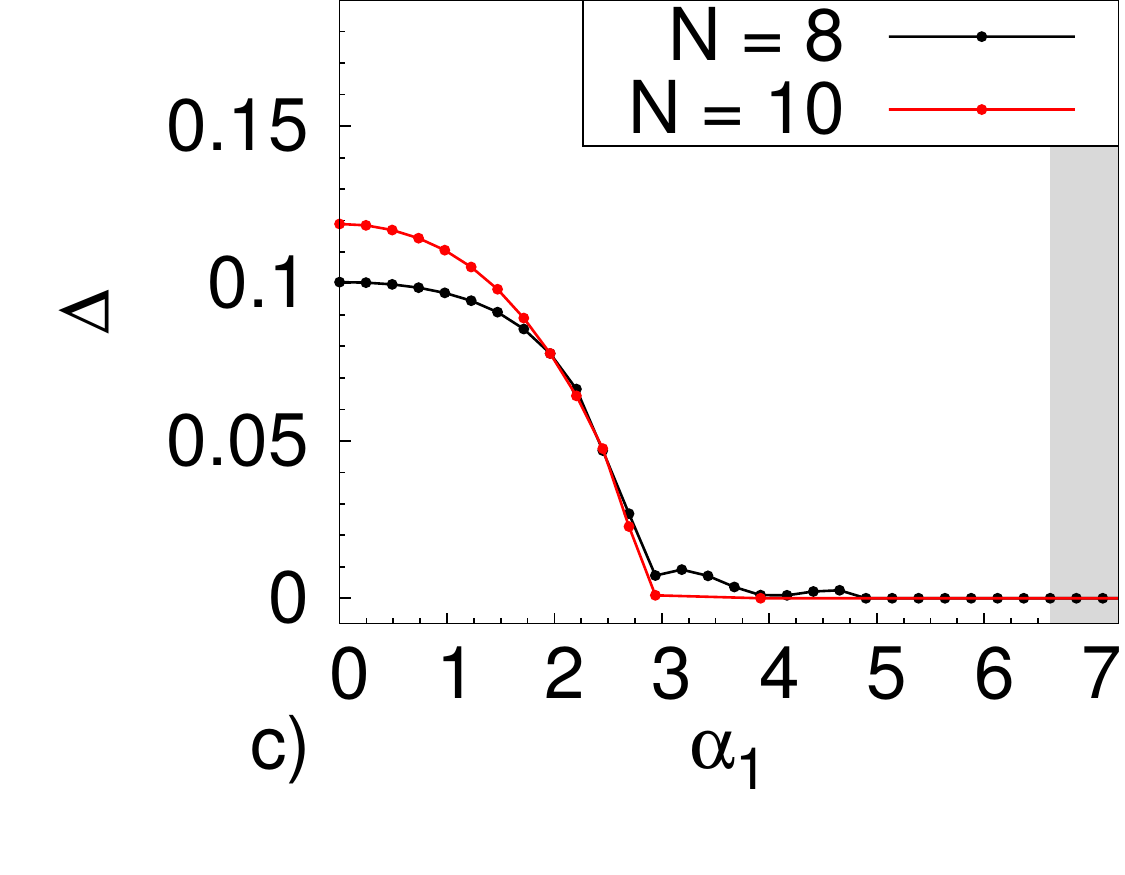}
\includegraphics[width = 0.49\linewidth]{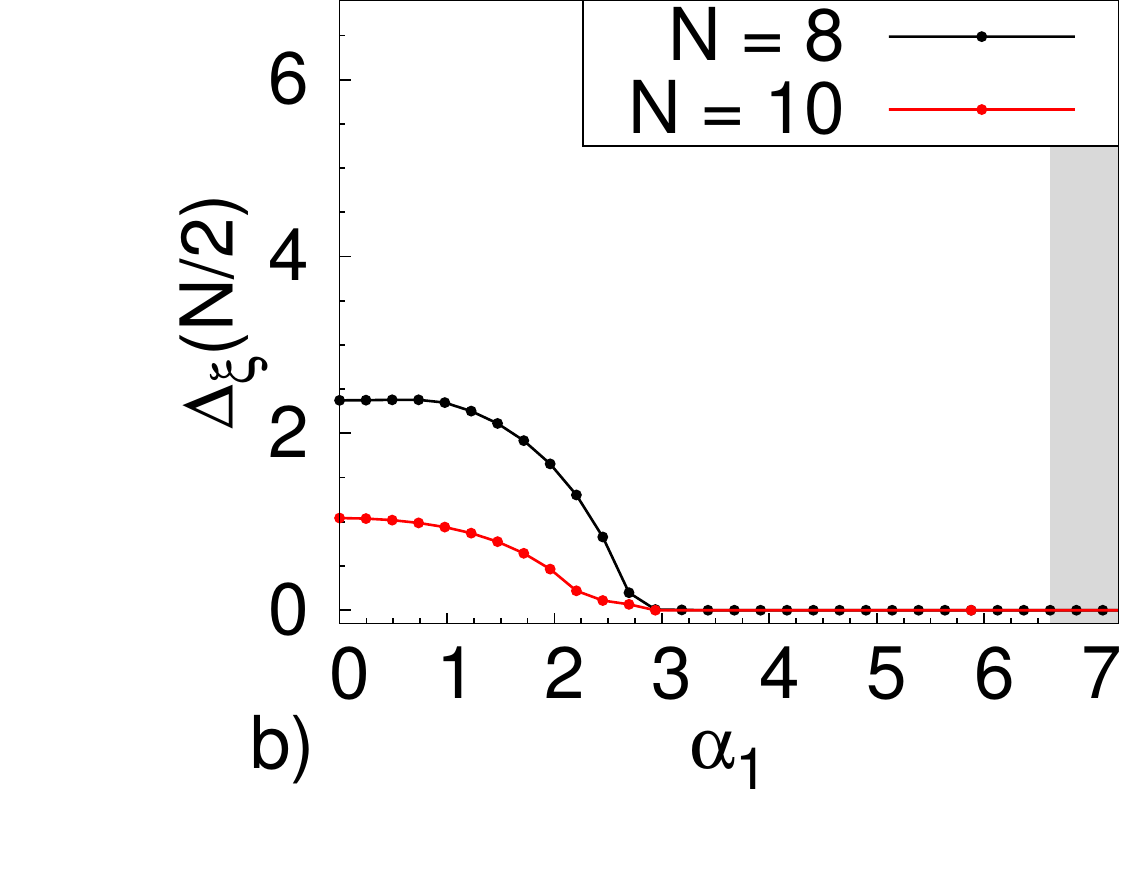}
\caption{Stability of the FTI phase at half filling upon addition of a $C_3$ symmetry preserving coupling term of amplitude $\alpha_1$, in the absence of interlayer interaction (a and b), and with $V/U = 0.5$ (c and d). The systems have $N = 8$ and $N=10$ particles, $N_s = N$ unit cells. The shaded area corresponds to the region where the one-body model is a trivial insulator. a and c: Evolution of the manybody gap with $\alpha_1$. b and d: Evolution of the PES gap with $\alpha_1$, for a particle partition $N_A = N/2$.}
\label{fig:PhaseDiagramMixingC3}
\end{figure}
We now investigate the system where the two kagome lattice layers are coupled via both the interaction and the inversion symmetry breaking term $R$. We study here the influence of the magnitude of that coupling term on the stability of the fractional phase at half filling. $S_z$ is no longer a good quantum number, which drastically increases the computational effort compared to the situation in Sec.~\ref{sec:HalfFilling}. For instance at $N = 12$, the dimension of the largest subspace was $1.3 \times 10^7$ previously, it is now $7.0 \times 10^7$. For $N = 10$, the dimension of the largest subspace raises from $4 \times 10^5$ to $2 \times 10^6$. Meanwhile, we have a larger parameter space to explore, with $\alpha_1, \ \alpha_2$, $\alpha_3$, and $V/U$ that can all vary independently. We will limit ourselves to systems 
with $N = 8$ and $10$ particles.

We first study the stability of the topological phase in the direction $\alpha_2 = \alpha_3 = 0$, for different values of $\alpha_1$. In that case, the inversion symmetry is broken, but the $C_3$ rotational invariance is preserved. The evolution of the gap $\Delta$ is represented Fig.~\ref{fig:PhaseDiagramMixingC3}a, c for the systems with respectively $N = 8$ and $N = 10$ particles. 
We also compute the PES of these systems and find that an entanglement gap exists for values of $\alpha_1$ up to $\alpha_1 \simeq 3.0$.
The evolution of the entanglement gap is represented Fig.~\ref{fig:PhaseDiagramMixingC3}b, d for partitions with $N_A = N/2$ particles. We use two different values of the interlayer interaction: $V/U = 0$ (a and b) and $V/U = 0.5$ (c and d). The stability region of the FTI lies well within the topological region of the one-body model ($\a_1 \in [0,6.5]$). Interestingly, it covers a significant part (more than a third) of the non-interacting TI stability 
zone. Note that the gap vanishes for similar values of $\alpha_1$ for both interactions, although it is very different at $V/U = 0$ and $V/U = 0.5$ when $R = 0$.

\begin{figure}
\includegraphics[width = 0.49\linewidth]{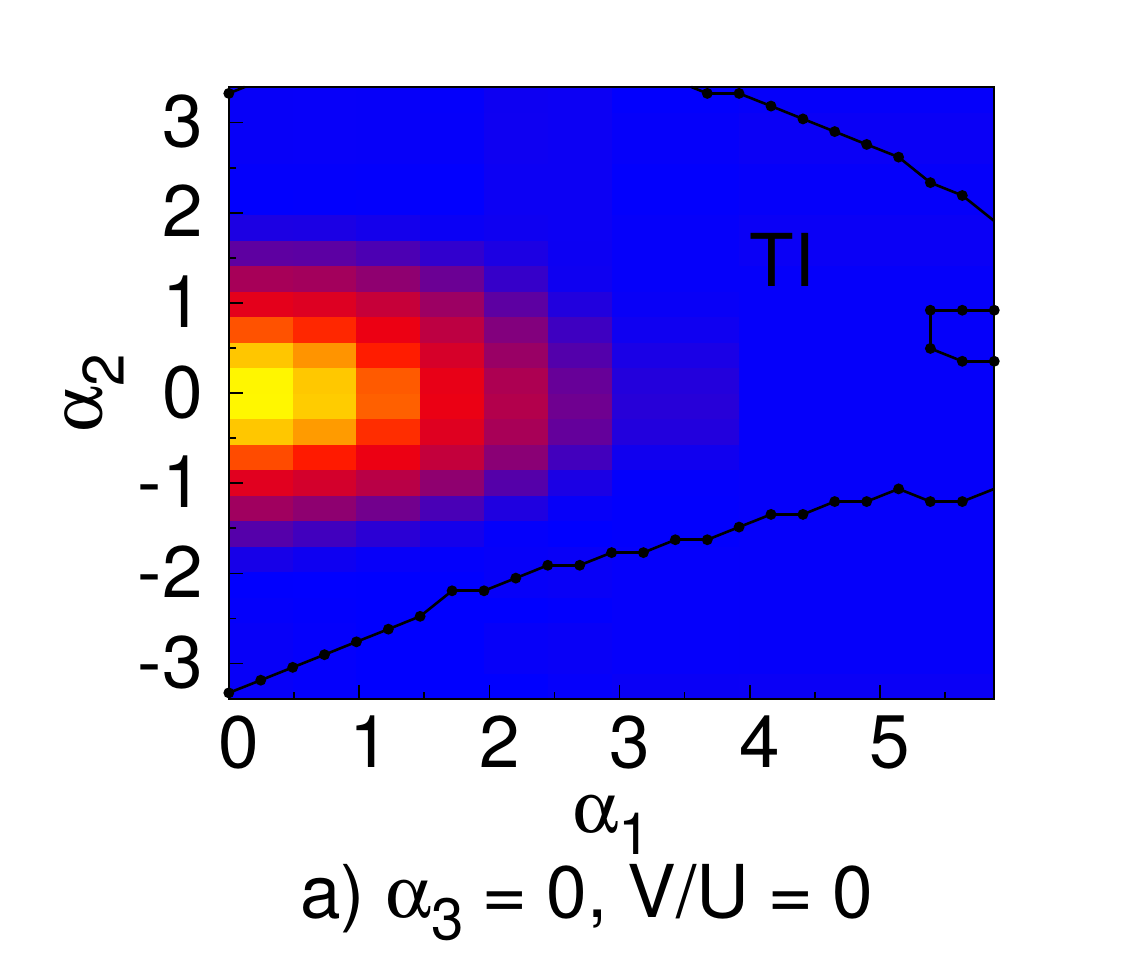}
\includegraphics[width = 0.49\linewidth]{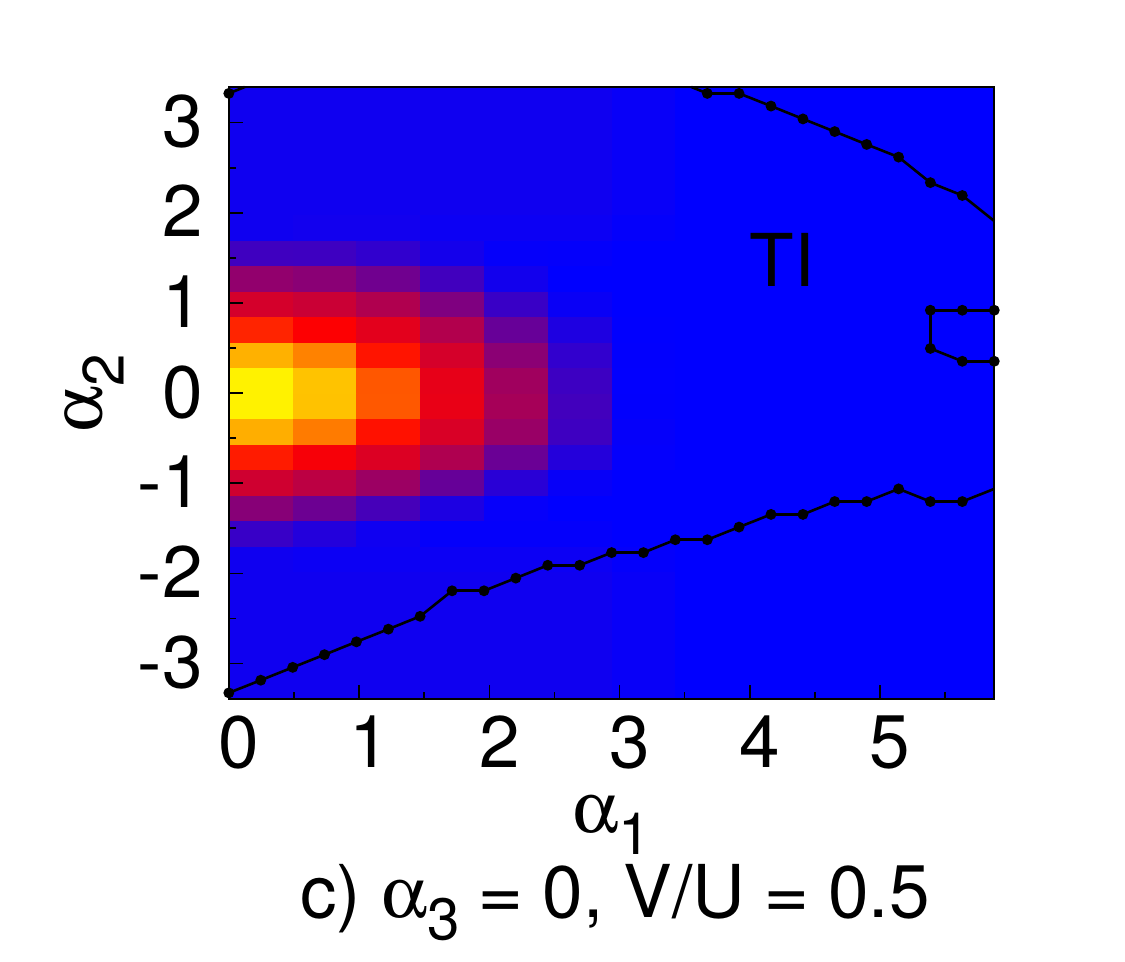}
\includegraphics[width = 0.49\linewidth]{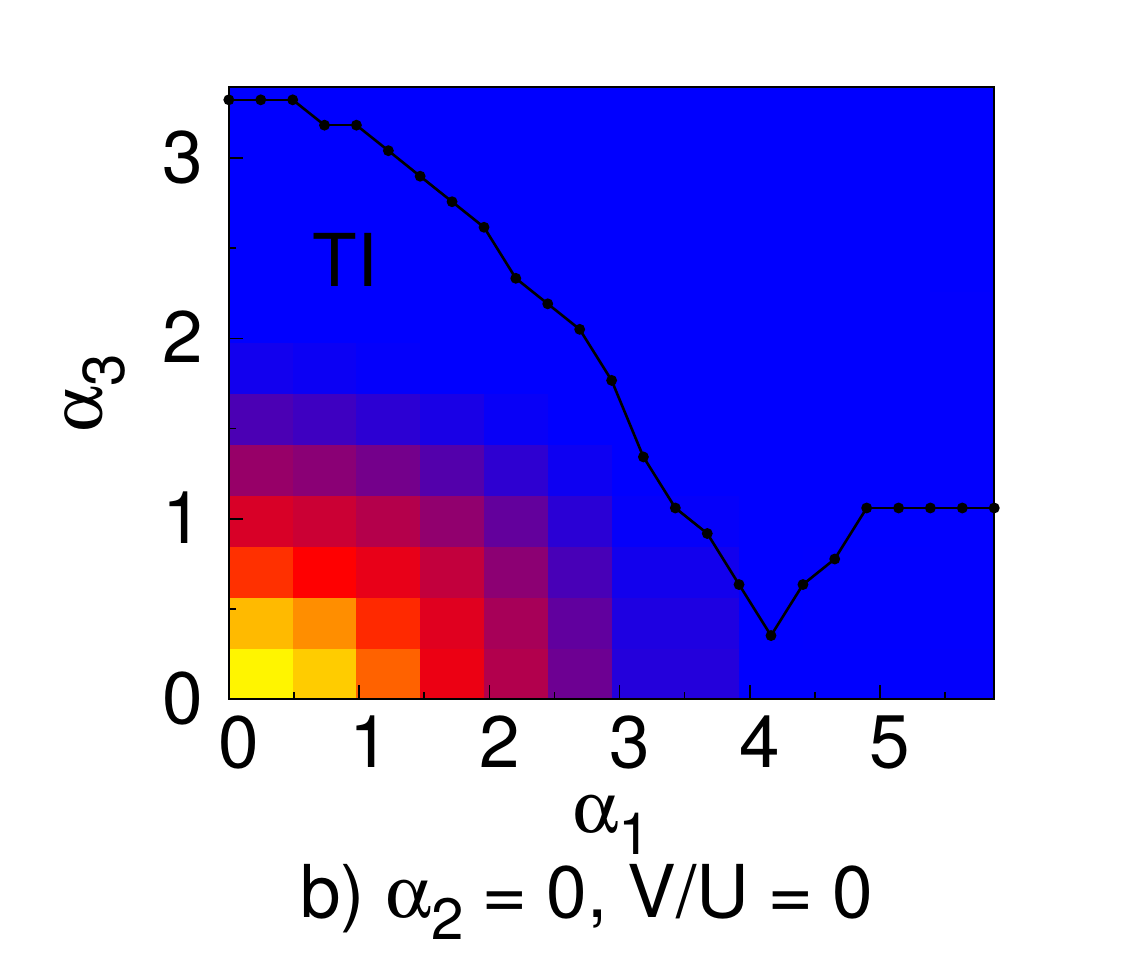}
\includegraphics[width = 0.49\linewidth]{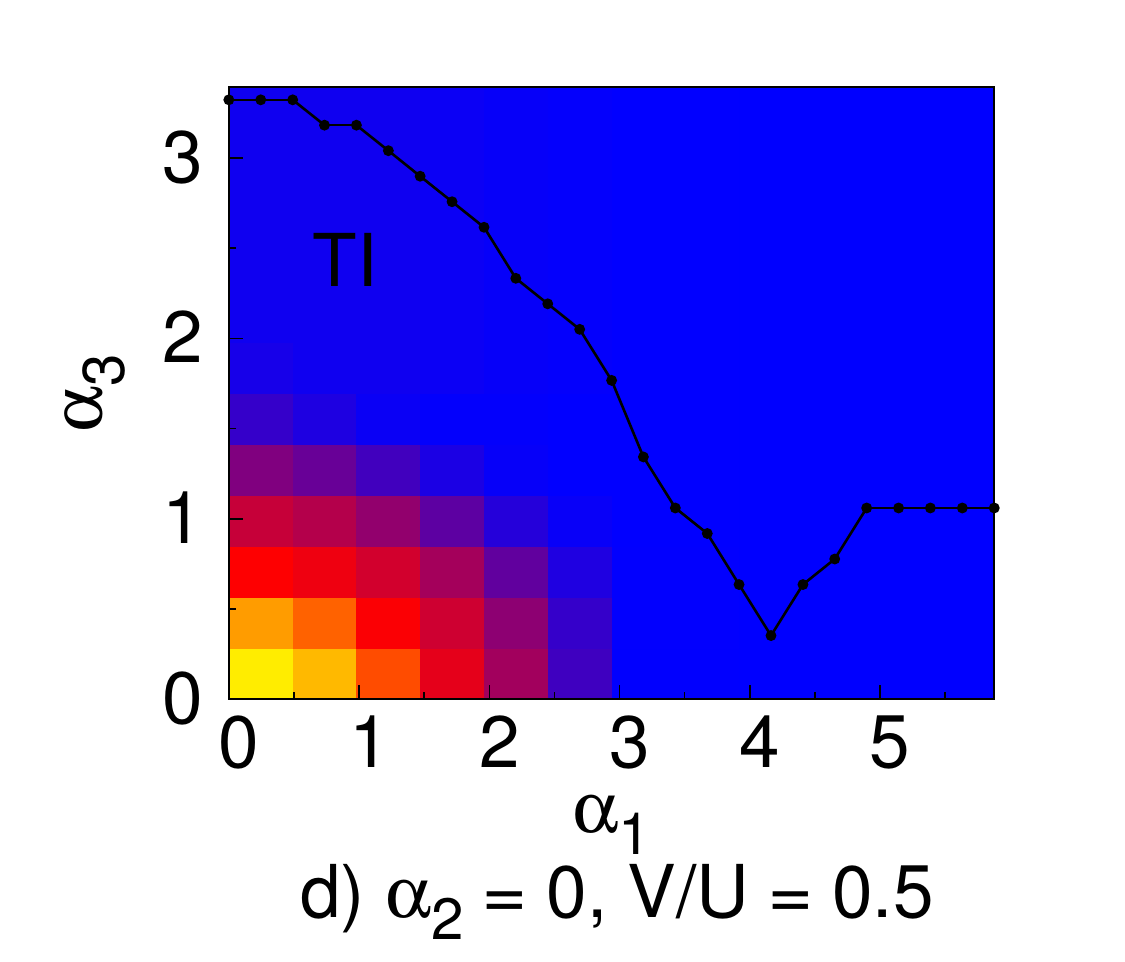}
\includegraphics[width = 0.49\linewidth]{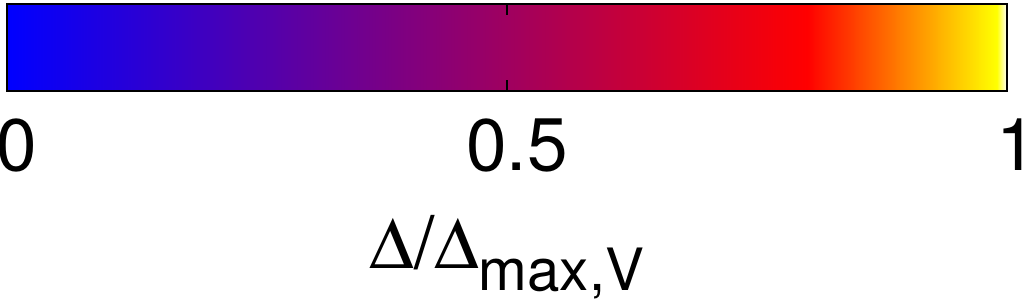}
\caption{Gap of the kagome lattice model topological insulator at half filling with $N = 10$ bosons, in the plane $\alpha_3 = 0$ (a and b) and in the plane $\alpha_2 = 0$ (c and d), for $V/U = 0$ (a and c) and $V/U = 0.5$ (b and d).$\Delta_\text{max,V}$ is the amplitude of the gap in the case where $R = 0$, with an interlayer interaction is $V$. We use the symmetries of the system with respect to the coupling elements $\alpha_i$ and show only the zones $\alpha_1 > 0$ and $\alpha_3 > 0$. The dotted line indicates the boundary between the trivial and topological insulator phases in the non-interacting model.}
\label{fig:PhaseDiagramMixingC3Delta1Delta3}
\end{figure}

\begin{figure}
\includegraphics[width = 0.49\linewidth]{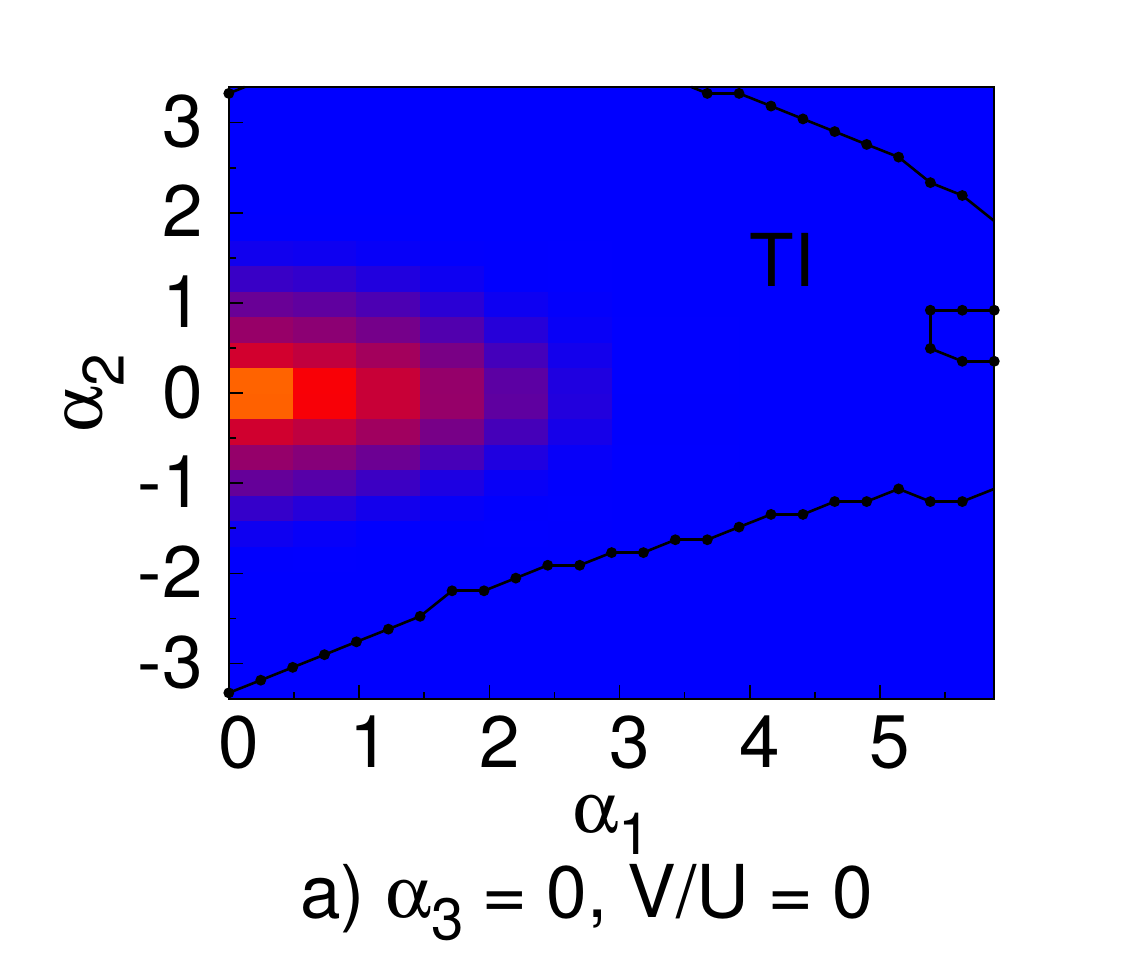}
\includegraphics[width = 0.49\linewidth]{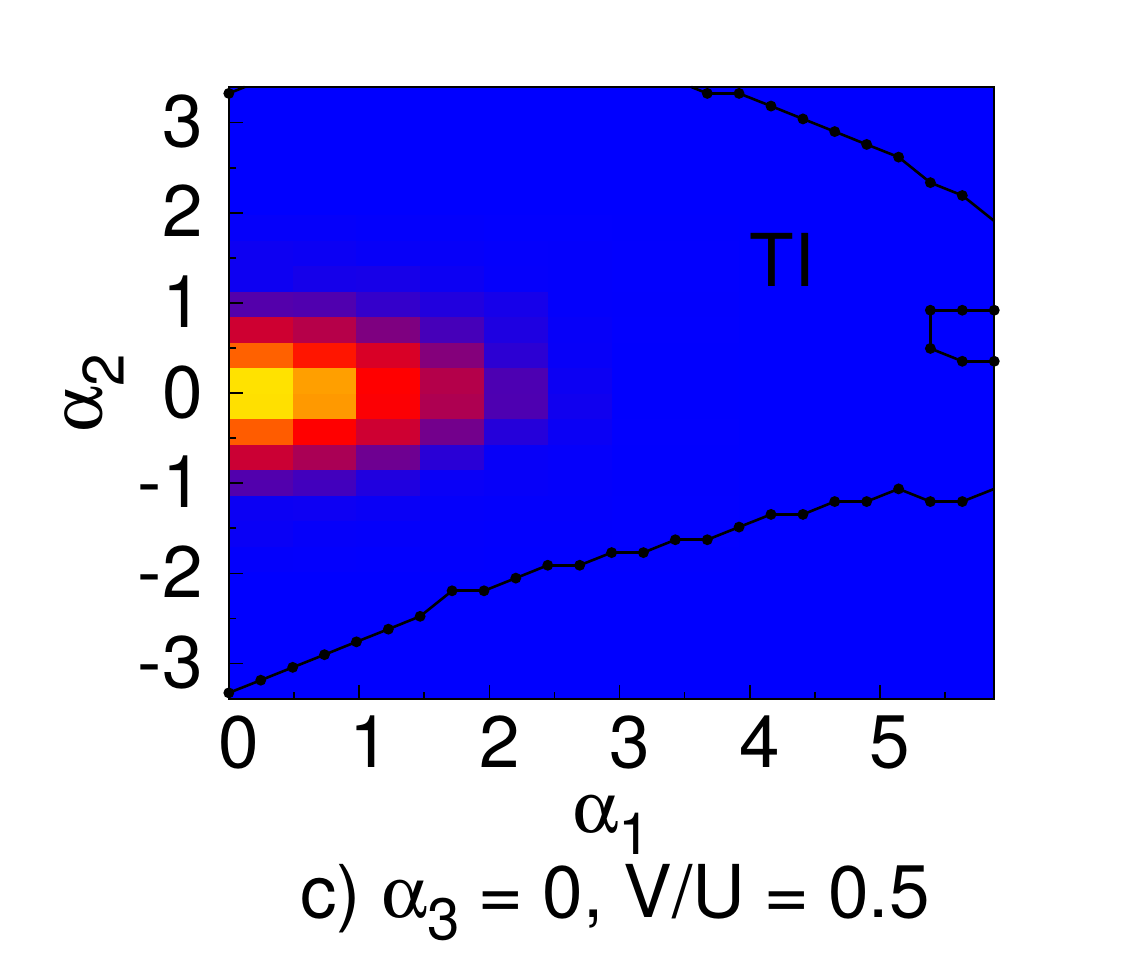}
\includegraphics[width = 0.49\linewidth]{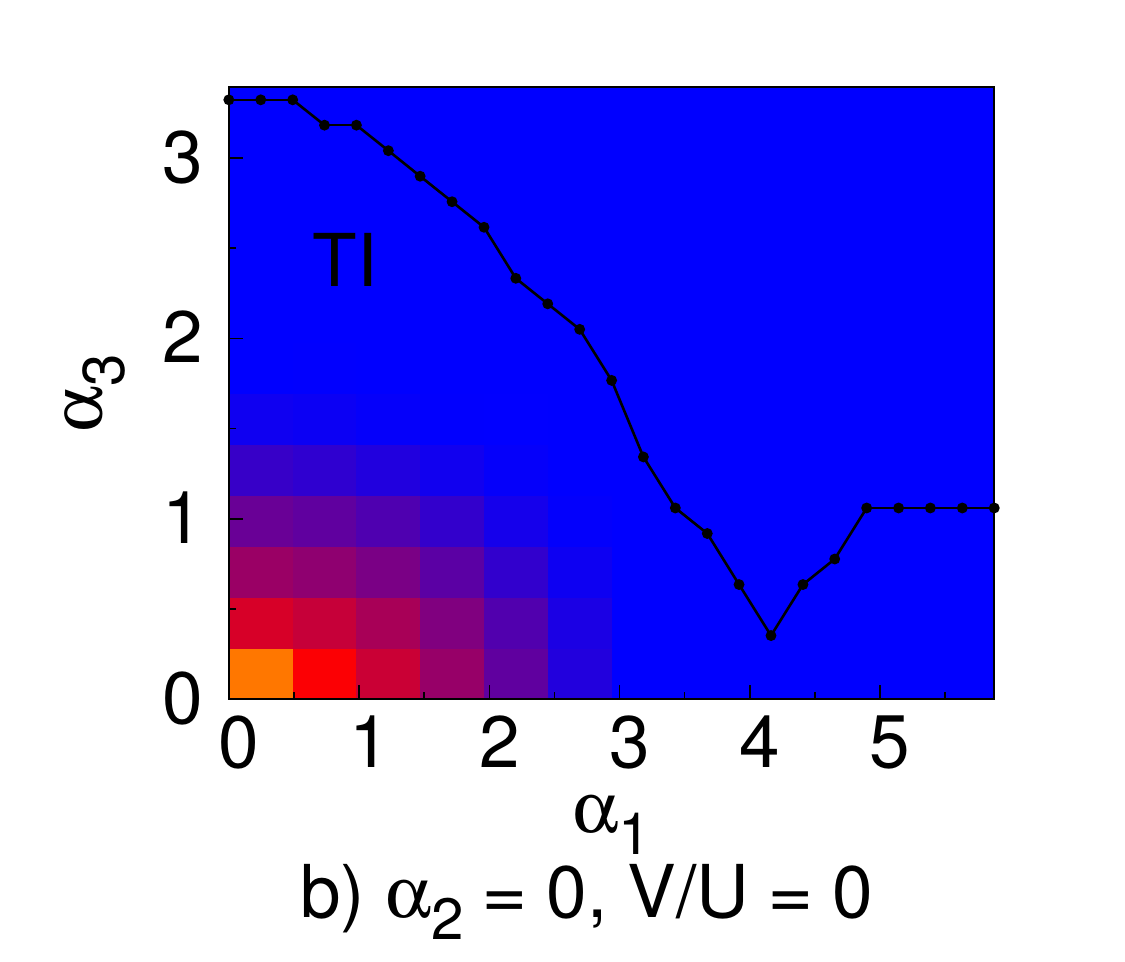}
\includegraphics[width = 0.49\linewidth]{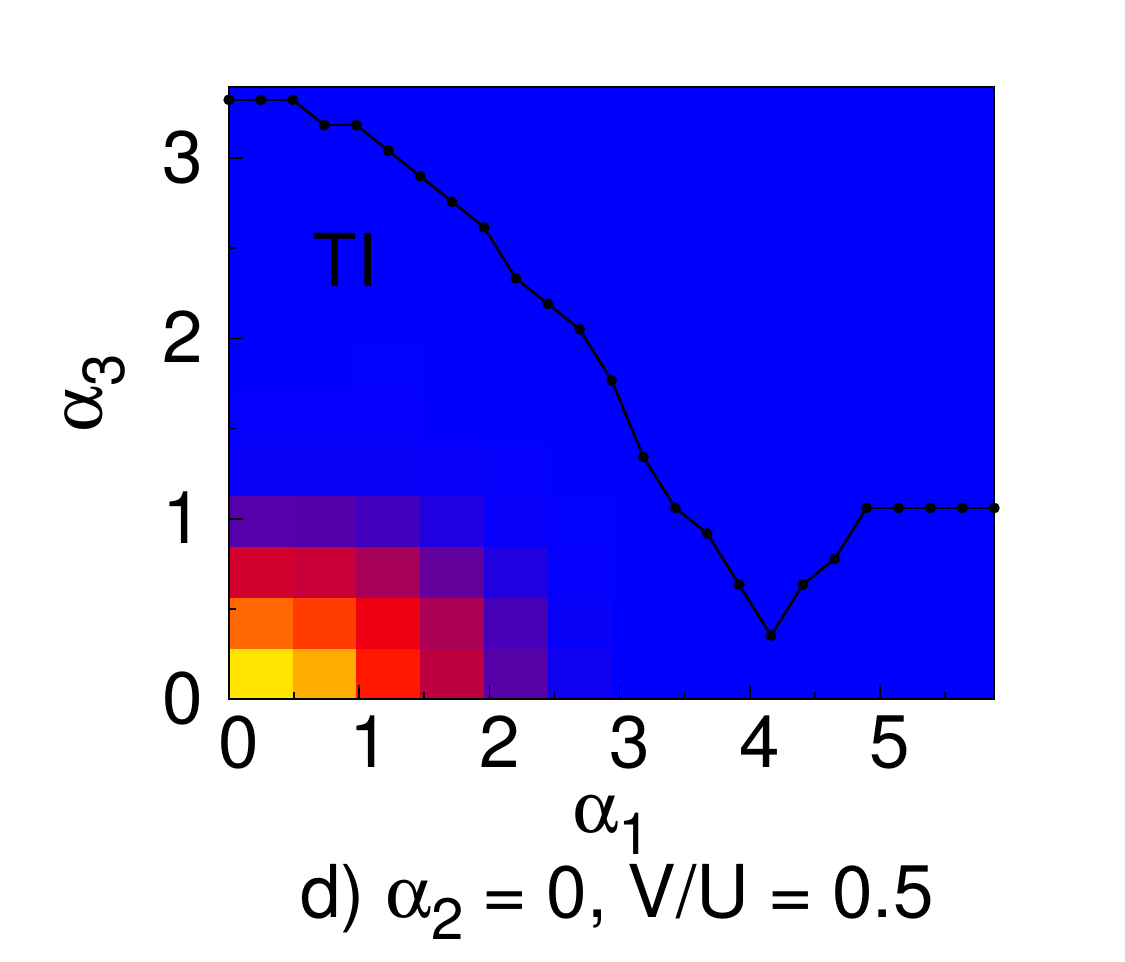}
\includegraphics[width = 0.49\linewidth]{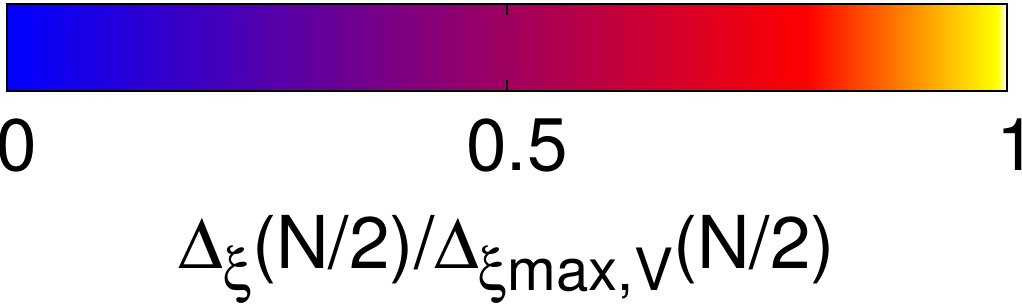}
\caption{PES gap of the kagome lattice model topological insulator at half filling with $N = 10$ bosons, in the plane $\alpha_3 = 0$ (a and b) and in the plane $\alpha_2 = 0$ (c and d), for $V/U = 0$ (a and c) and $V/U = 0.5$ (b and d). $\Delta_{{\xi}\text{max,V}}(N/2)$ is the amplitude of the PES gap in the case where $R = 0$, with an interlayer interaction is $V$). The number of particles in the partition is $N_A = 5$. We use the symmetries of the system with respect to the coupling elements $\alpha_i$ and show only the zones $\alpha_1 > 0$ and $\alpha_3 > 0$. The dotted line indicates the boundary between the trivial and topological insulator phases in the non-interacting model.}
\label{fig:PhaseDiagramMixingC3Delta1Delta2PES}
\end{figure}
 
We also look at the stability of the topological phase at half filling along two different planes: $\alpha_2 = 0$ and $\alpha_3 = 0$. 
We plot the gap $\Delta$ (see Fig.~\ref{fig:PhaseDiagramMixingC3Delta1Delta3}) and the PES gap (see Fig.~\ref{fig:PhaseDiagramMixingC3Delta1Delta2PES}) of the $N = 10$ system. Similar plots are displayed in appendix \ref{app:HalfFillingSuppMat} for the $N = 8$ system. The boundary between the trivial and topological regions in the one-body model is shown as a dotted line on the same graph. Again, the FTI stability regions covers a large part of the one-body TI stability zone.
Phase diagrams are presented for both $V/U = 0$ and $V/U = 0.5$. We can then see how the FTI phase survives the introduction of both types of interlayer coupling (i.e. interaction and inversion symmetry breaking term) for a significant amplitude of both terms. Interestingly, the gap vanishes for similar values of $\alpha_1$, $\alpha_2$ and $\alpha_3$ for both values of the interaction, although for $R=0$, the gaps at $V/U = 0$ and at $V/U = 0.5$ have very different values. Their ratio is $0.56$ at $N=8$, $0.69$ at $N = 10$. As expected, the FTI stability regions are of similar area and lie in the same zone of the phase diagram for both the $N = 8$ and $N = 10$ systems. This is a good indication that the FTI phase might survive the coupling beyond the finite size case.

\section{Beyond the half filling FTI -- Exploring other fractions}
As discussed in Sec.~\ref{sec:HalfFilling}, the kagome lattice model hosts a very stable Laughlin-like $\nu = 1/2$ phase. Exploring other fractions (in particular the composite fermion series) would be very interesting, especially in the cases where Levin and Stern's theory~\cite{levin-PhysRevLett.103.196803, levin-PhysRevB.86.115131} predicts an unstable FTI phase, like $\nu = 2/3$. Unfortunately, the FQH bosonic composite fermion~\cite{jain89prl199} phase at $\nu = 2/3$ is less stable than the Laughlin phase. Although the kagome lattice model hosts a rather stable FCI bosonic phase at $\nu = 2/3$ (see Ref.~\cite{Liu-2013PhysRevB.87.205136}), its threefold low energy manifold has a large splitting. As a result, the FTI low energy spectrum does not have a ninefold almost degenerate ground state, even in the fully decoupled case. Note that even for an ideal FQH system, the bosonic composite fermion phase is not as stable as its fermionic counterpart, as 
shown in Ref.~\cite{Chang-PhysRevA.72.013611} for the sphere geometry.
Moreover, it would be interesting to look at phases that are not the tensor product of two FQH phases.

\begin{figure}
\includegraphics[width = 0.49\linewidth]{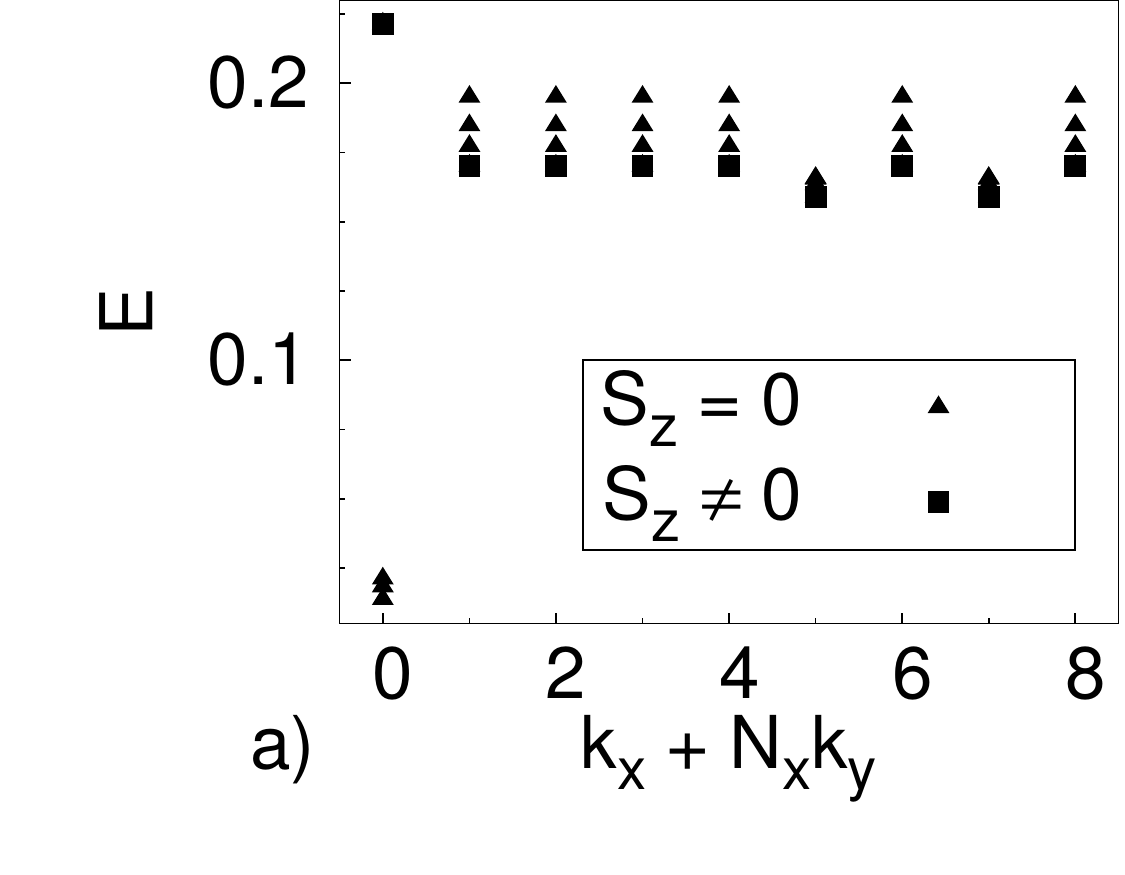}
\includegraphics[width = 0.49\linewidth]{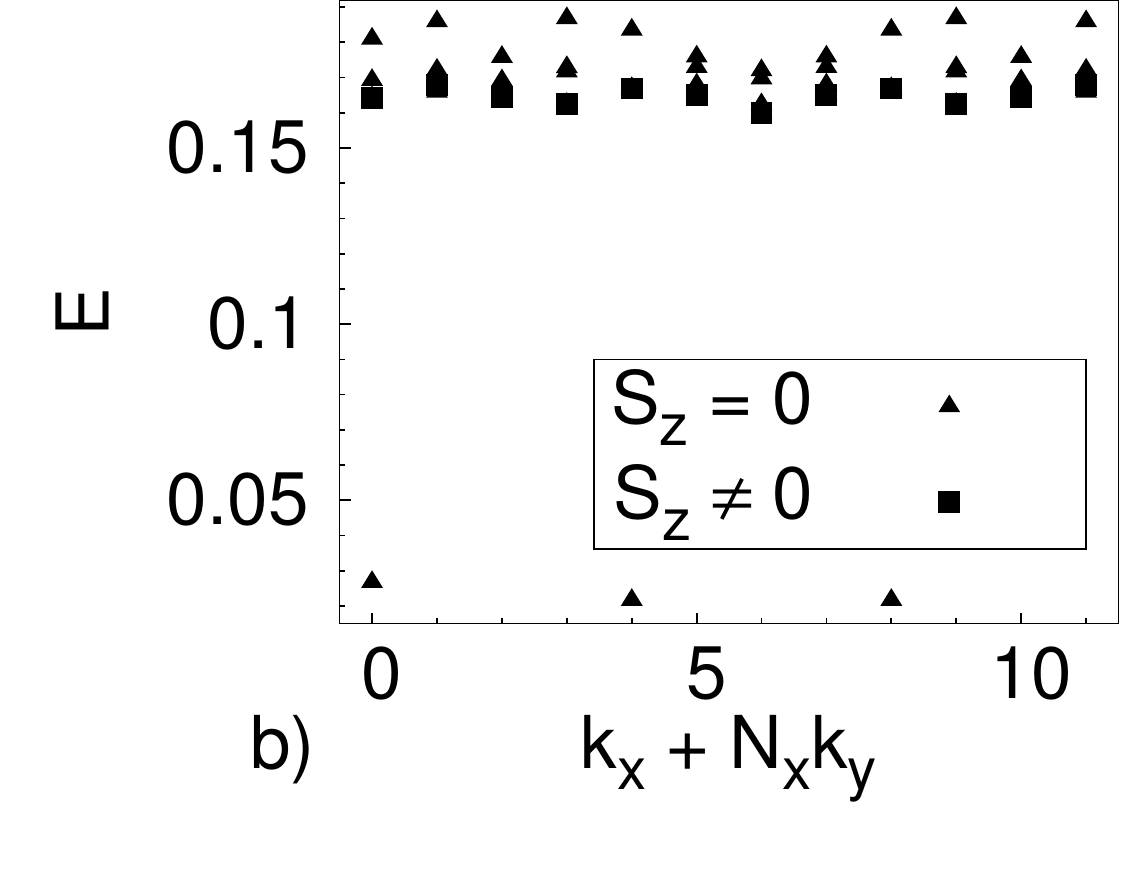}
\caption{Low energy spectrum of the bilayer FCI model at a filling fraction $\nu = 1/3$ (with respect to the two lowest bands) for $N = 6$, $N_s = 9$, $N_x = N_y = 3$ (a) and $N = 8$, $N_s = N_x = 12$, $N_y = 1$ (b) There is a threefold almost degenerate ground state, separated from higher energy excitations by a large gap. }
\label{fig:Halperin221}
\end{figure}

The physics of two FCI copies coupled by an interaction term $V$ as described in Eq.~\pref{eq:interaction} is similar to that of a FQH bilayer. We focus on the filling fraction $\nu = 1/3$ (with respect to the two lowest bands). The physics of the bilayer FQH system at $V = 0$ can be deduced from the simple FQH case, which is a Fermi sea of composite fermions~\cite{Chung-PhysRevA.77.043608}. At $V/U = 1.0$, the bilayer FQH system has an additional symmetry, the $SU(2)$ symmetry. It has a threefold degenerate ground state described by the Halperin $(2,2,1)$ wavefunction~\cite{Halperin83}. The FCI bilayer with an interaction term $V/U = 1.0$ is analogous to this system.  Fig.~\ref{fig:Halperin221} shows that it has the expected threefold almost degenerate ground state, with a large gap to higher energy excitations, for $N = 6$ and $N = 8$.

\begin{figure}
\includegraphics[width = 0.49\linewidth]{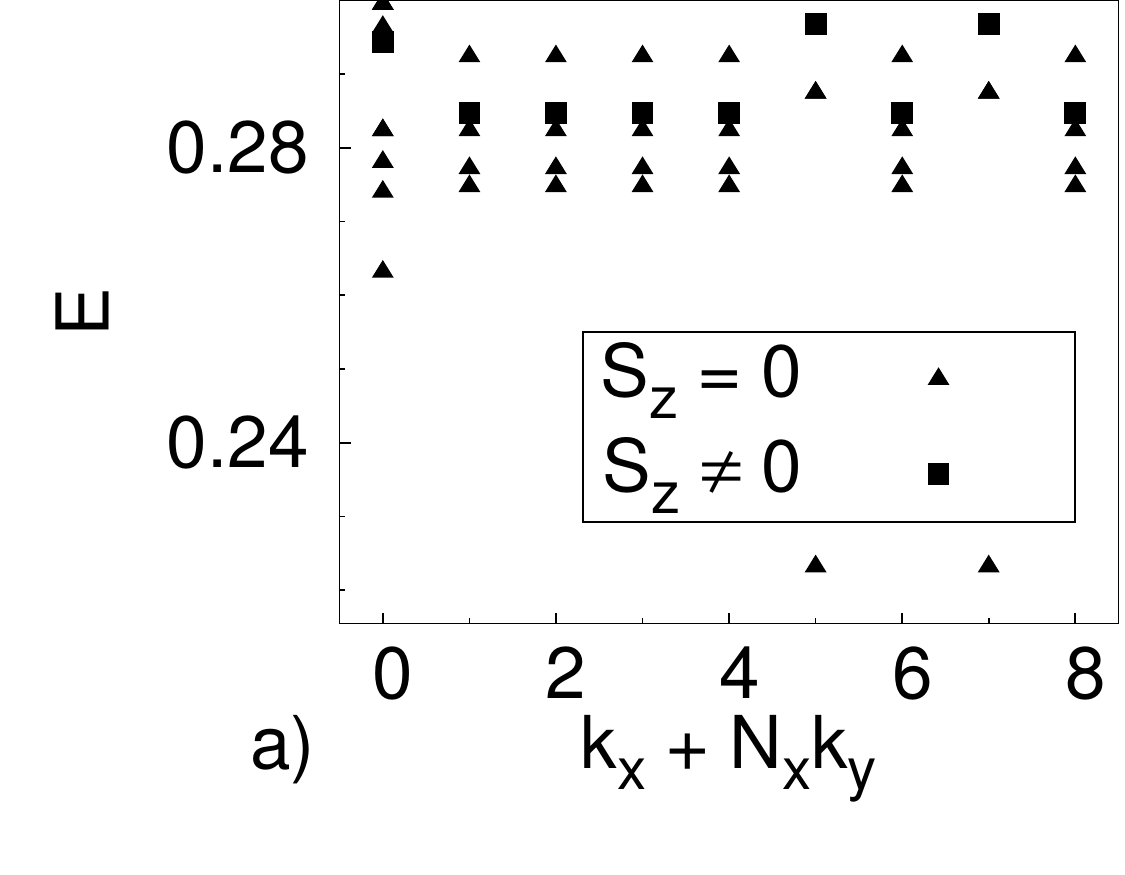}
\includegraphics[width = 0.49\linewidth]{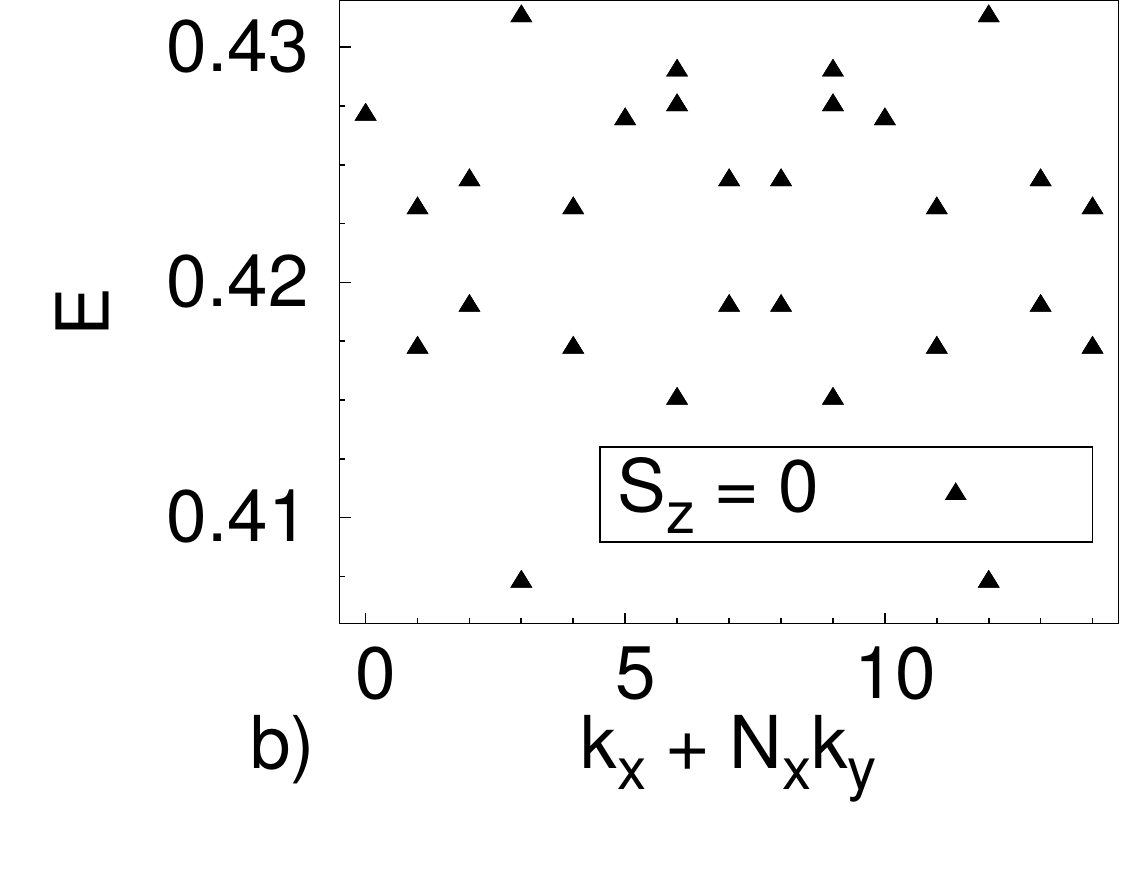}
\caption{Low energy spectrum of the FTI model at $\nu = 1/3$, $V/U = 1.0$. All sectors of the pseudospin projection $S_z$ are represented. \emph{Left panel} $N = 6$, $N_s = 9$, $N_x = N_y = 3$. \emph{Right panel} $N = 10$, $N_s = N_x = 15$, $N_y$. In this latter case, the states with $S_z>0$ are all higher in energy than the two first excited states in each sector, and are therefore not represented here. Both systems have a ground state with a twofold exact degeneracy due to inversion symmetry. Nevertheless, in the $N = 10$ system, the ground state mixes with higher energy state upon flux insertion.}
\label{fig:quantumSpinHall1_3}
\end{figure}

At $V/U = 1.0$, the interaction term of the FTI model is $SU(2)$ invariant. The one-body model, however, breaks the $SU(2)$ symmetry, and so does the full model. Consequently, there is no guarantee that this system can host any topological phase. Still, this is an interesting starting point, as its time reversal symmetry breaking counterpart is so peculiar. We perform exact diagonalizations to compute the low energy spectrum of the systems with $N = 6, \ 8$ and $N = 10$ particles at $\n = 1/3$. 
For $N = 6$ and $N=10$ (ie when $N_s$ is odd), we observe a twofold degenerate ground state, (with the exact degeneracy explained by inversion symmetry) and a gap $\Delta$ to higher energy excitations (see Fig.~\ref{fig:quantumSpinHall1_3}). Unfortunately, for $N = 10$, the ground state mixes with higher energy states upon flux insertion in the $x$ direction. This excludes the possibility that the ground state is of topological nature. For $N = 8$, the low energy spectrum also presents some energy separation, but the energy spread $\delta$ 
between the two lowest energy states (in momentum sectors $(k_x, k_y) = (0,0)$ and $(6, 0)$) is of the 
same 
order of magnitude as the gap $\Delta$ between these states and the state with the closest energy ($\delta / \Delta = 0.82$).

Unlike the case of the bilayer FCI system, choosing $V/U = 1.0$ does not give the FTI Hamiltonian an additional symmetry that would justify to restrict our study to $V/U = 1.0$. We looked at the $\nu = 1/3$ system with $N = 6$, $8$ and $10$, with $0 < V/U < 1.5$ (the decoupled case is a Fermi sea of composite fermions~\cite{Chang-PhysRevA.72.013611}). Even when the low energy spectrum is gapped, the gap does not survive the insertion of flux, nor does the PES show any particular feature. Moreover, the ground state of the $N = 10$ system has a twofold degeneracy for various values of $V/U$, but a variation of $V/U$ of less than $10\%$ will change the sectors it falls in. We also studied the influence of an inversion breaking term $R$ on the $\nu = 1/3$ system for various $V/U$. We observe a gapped almost degenerate ground state for some values of $R$, but the gap never survives the insertion of flux. 
In the absence of a robust phase, we scanned all systems with less than $16$ unit cells and involving a Hilbert space dimension below $11000$. Besides the case of $N = 6$, $N_s = 9$, which corresponds to $\nu = 1/3$, we did not find any indication of a gapped ground state.

To summarize, we did not find any evidence of a robust FTI phase in our model beyond the half filling case.
Note that we expect the stability of an FTI phase to strongly depend on the microscopic model, since FCI systems exhibit strong model dependence~\cite{Wu-2012PhysRevB.85.075116}. Therefore, we cannot rule out the the existence of topological phases beyond the $\nu = 1/2$ case.

\section{Conclusion}
In this article, we have proposed and numerically studied the half filling phase of a stable microscopic FTI model based on the kagome lattice. We proved that the system of two FCI copies with opposite chiralities survives the introduction of coupling terms of different natures. At the single particle level, one can add an inversion symmetry breaking term of a significant amplitude without destroying the FTI phase. The FTI phase also survives the addition of an interlayer interaction term, as long as the amplitude of this term does not exceed the amplitude of the intralayer interaction term. Surprisingly, the FTI model is more robust than the equivalent bilayer FCI model in that regard. Further works will investigate the reasons for this difference.
The comparison of different system sizes allows us to be confident that our results are not simple artefacts of the system's finite size. Moreover, we have used a similar geometric aspect ratio for each system size, thus eliminating the aspect ratio as a 
factor of variation of the energetic quantities. 
We have also looked at other fractions, such as $\nu = 1/3$. We did not find convincing evidence of a robust FTI phase for 
these systems, as there is no consistent pattern in all the studied system sizes. Nevertheless, we do not exclude the possibility of a new FTI phase in other microscopic models. Refs.~\cite{levin-PhysRevLett.103.196803, levin-PhysRevB.86.115131, Cappelli-2013JHEP...12..101C, Koch-2013arXiv1311.6507K} gave some criteria for the stability of the edge states in a TRI system with strong interactions. In contrast, our stability discussion was only based on the system's bulk properties. Probing the stability of the edge states would be an exciting direction for further research.

\begin{center}ACKNOWLEDGEMENTS\end{center}
We thank T. Neupert, A. Sterdyniak and Y.-L. Wu for discussions, and especially T.L. Hughes for earlier collaboration on this work. BAB and NR were supported by NSF CAREER DMR-095242, ONR-N00014-11-1-0635, MURI-130- 6082, Packard Foundation, and Keck grant. NR and CR were supported by ANR-12-BS04-0002-02. CR was supported by SPAWARSCYCEN Pacific.

\appendix
\section{Tilted boundary conditions}
\label{app:TiltedBoundaryConditions}
We apply periodic boundary conditions defined by two vectors $\mbf{T_1}$ and $\mbf{T_2}$ (by definition, the system is invariant under a translation of any integer number of these vectors). The most common choice for $\mbf{T_1}$ and $\mbf{T_2}$  is
\be
\mbf{T_1} = N_x \mbf{a_1} \ , \ \ \mbf{T_2} = N_y \mbf{a_2}
\ee
where $N_x \times N_y = N_s$ which constrains the aspect ratio to be
\beq
\kappa = \frac{N_x}{N_y \sin (\mbf{a_1}, \mbf{a_2})}
\ee
Some system sizes can thus only be realized with a small aspect ratio; for instance, one can only obtain $N_s = 10$ unit cells with $N_x \times N_y = 5 \times 2$. However, we know that the physics of Fractional Chern Insulators (FCI) is highly influenced by the aspect ratio of the system~\cite{regnault-PhysRevX.1.021014, Lauchli-PhysRevLett.111.126802}. For instance the manybody gap can increase even as the system size increases if the aspect ratio is closer to one. We expect this property to hold true for FTI (it is true at least in the case where pseudospin up and down are coupled neither by the band structure nor by the interaction, as the FTI spectrum can be built exactly from the FCI spectrum). We thus try to perform our calculations on systems that have an aspect ratio close to one. Only three system sizes ($N_s = 8, \ 10, \ 12$) are accessible to exact diagonalization (with a Hilbert space dimension of the order of ten millions for $N = 12$ bosons and $N_s = 12$ in the case where $S_z$ is a good 
quantum 
number). It is then crucial to be able to use all of those system sizes if one is to attempt a gap extrapolation. The geometric aspect ratio can be brought closer to one by using tilted boundary conditions, as mentioned in Ref.~\cite{Lauchli-PhysRevLett.111.126802} in the case of a square Bravais lattice. This method can be generalized to any kind of 2D Bravais lattice, in order to use it with our kagome system (triangular Bravais lattice).
\begin{figure}
\includegraphics[width = 0.95\linewidth]{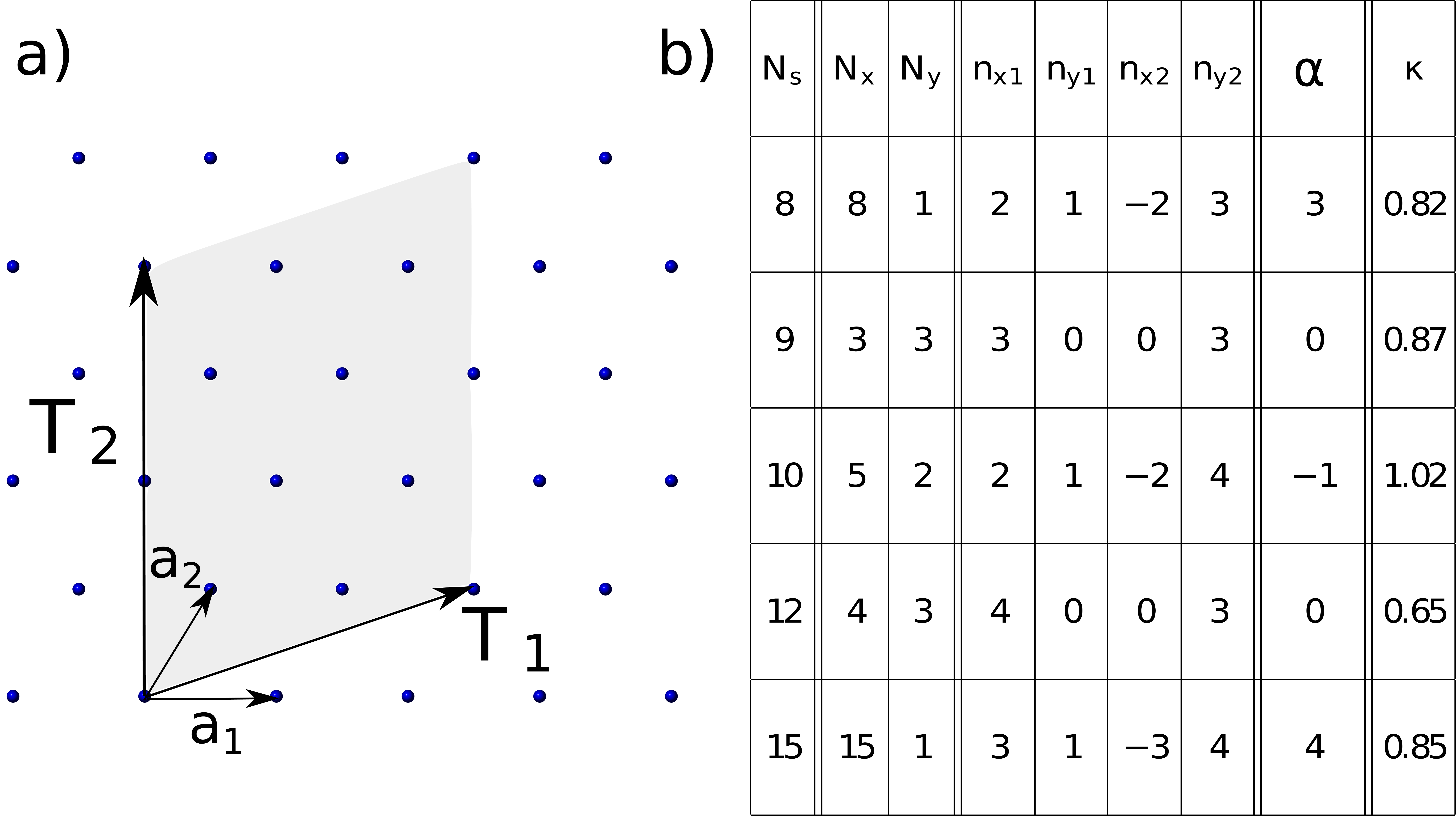}
\caption{\emph{Left panel} Vectors that define the tilted periodic boundary conditions for a triangular Bravais lattice, for a system with $N_s = 10$ unit cells and optimal aspect ratio, as defined in second row of table b). The shaded area represents the finite size system in real space (it contains 10 unit cells). \emph{Right panel} Coordinates of the vectors $\mbf{T_1}$ and $\mbf{T_2}$ that were used in this paper for all numerical calculations, in the $(\mbf{a_1}, \mbf{a_2})$ basis defined in Fig.~\ref{fig:lattice}.}
\label{fig:tiltedLattice}
\end{figure}
We define the boundary condition vectors by their coordinate in the lattice basis
\be
\mbf{T_1} = n_{x1} \mbf a_1 + n_{y1} \mbf a_2\ , \ \ \mbf{T_2} = n_{x2} \mbf a_1 + n_{y2} \mbf a_2
\ee
where $ n_{x1}, \  n_{y1}, \ n_{x2}, \  n_{y2}$ are integers. $\mbf{T_1}$ and $\mbf{T_2}$ define a parallelogram, which encloses a total number of unit cells $N_s$ (see Fig.~\ref{fig:tiltedLattice}a for an example with $N_s = 10$). This leads to the following constraint on the integers
\be
N_s = \frac{||\mbf T_1 \times \mbf T_2||}{||\mbf a_1 \times \mbf a_2||} = |n_{x1} n_{y2} - n_{x2} n_{y1}|
\ee
The invariance of the system wavefunction under a translation of either of these vectors results in a quantization of the momentum $\mbf k$ along the axes defined by $\mbf T_1$ and $\mbf T_2$. Thus any admissible $\mbf k$ should satisfy $\mbf{k}\cdot\mbf{T_1}$ and $\mbf{k}\cdot\mbf{T_2}$ being multiples of $2 \pi$.

We define $\tilde{\mbf T}_1$ and $\tilde{\mbf T}_2$, the reciprocal vectors of $\mbf T_1$ and $\mbf T_2$
\be
\tilde{\mbf{T}}_i\cdot\mbf{T_j} = 2\pi \delta_{ij} \ , \ \ i, \ j = 1, \ 2
\ee
We note $\e = \mbf a_1 \cdot \mbf a_2$. $\e = 1/2$ for a triangular lattice like the kagome lattice.
We can express $\tilde{\mbf T}_1$ and $\tilde{\mbf T}_2$ in the Bravais lattice basis as a function of $\e$.
\ba
\tilde{\mbf{T}}_1 & = & \frac{2\pi}{N_s\left( 1 - \e^2\right)}\left[(n_{y2} + \e n_{x2}) \mbf a_1 - (n_{x2} + \e n_{y2}) \mbf a_2 \right] \\ \nonumber
\tilde{\mbf{T}}_2 & = & \frac{2\pi}{N_s\left( 1 - \e^2\right)}\left[-(n_{y1} + \e n_{x1}) \mbf a_1 + (n_{x1} + \e n_{y1}) \mbf a_2 \right]
\ea

In particular, for a triangular lattice, this gives
\ba
\tilde{\mbf{T}}_1 & = & \frac{4\pi}{3N_s}\left[(2n_{y2} + n_{x2}) \mbf a_1 - (2n_{x2} + n_{y2}) \mbf a_2 \right] \\ \nonumber
\tilde{\mbf{T}}_2 & = & \frac{4\pi}{3N_s}\left[-(2n_{y1} + n_{x1}) \mbf a_1 + (2n_{x1} + n_{y1}) \mbf a_2 \right]
\ea
Now that we have defined the spanning vectors of the momentum space, we would like to define the boundaries of the first Brillouin zone. Two vectors in the reciprocal space are equivalent (the corresponding wavefunctions are equal) if they are equal up to a translation of a vector of the reciprocal lattice $\tilde{\mbf a}_1$ or $\tilde{\mbf a}_2$. The definition of the reciprocal lattice vectors
\be
\tilde{\mbf{a}}_i\cdot\mbf{a_j} = 2\pi \delta_{ij} \ , \ \ i, \ j = 1, \ 2
\ee
allows us to compute their expression in the direct space 
\ba
\tilde{\mbf a}_1 & =  \frac{2\pi}{1 - \e^2}(\mbf a_1 - \e \mbf a_2) \\ \nonumber
\tilde{\mbf a}_2 & = \frac{2\pi}{1 - \e^2}(-\e \mbf a_1 + \mbf a_2)
\ea
For a triangular lattice, this gives:
\ba
\tilde{\mbf a}_1 & = 2\pi \frac{2}{3}(2\mbf a_1 - \mbf a_2) \\ \nonumber
\tilde{\mbf a}_2 & = -2\pi \frac{2}{3}(\mbf a_1 - 2\mbf a_2)
\ea

Their expression in the $\tilde{\mbf T}_1$ and $\tilde{\mbf T}_2$ basis is 
\ba
\label{eq:reciprocalVectors}
\tilde{\mbf a}_1 & = n_{x1}\tilde{\mbf T}_1 + n_{x2}\tilde{\mbf T}_2\\ \nonumber
\tilde{\mbf a}_2 & = n_{y1}\tilde{\mbf T}_1 + n_{y2}\tilde{\mbf T}_2
\ea

The first Brillouin zone lies in the parallelogram defined by $(\tilde{\mbf a}_1, \tilde{\mbf a}_2)$.  We want to label all of the $N_s$ admissible momentum vectors $\mbf k$ within the first Brillouin zone using two integers $p$ and $q$ such that
\be
\label{momentumDefinition}
\mbf k = p \tilde{\mbf{T}}_1' + q \tilde{\mbf{T}}_2'
\ee
We require that $N_x \tilde{\mbf T}_1'$ and $N_y \tilde{\mbf T}_2'$ be vectors of the reciprocal lattice (i.e. they can be written as linear combinations with integer coefficients of $\tilde{\mbf a}_1$ and $\tilde{\mbf a}_2$), with $N_s=N_x \times N_y$. This implies that $p$ (resp. $q$) is in the interval $[0,N_x-1]$ (resp. $[0,N_y-1]$). Note that for $\mbf k$ to be an admissible momentum vector,  $\tilde{\mbf T}_1'$ and $\tilde{\mbf T}_2'$ need to be linear combinations with integer coefficients of $\tilde{\mbf T}_1$ and $\tilde{\mbf T}_2$. We stress that taking both $\tilde{\mbf T}_1'=\tilde{\mbf T}_1$ and $\tilde{\mbf T}_2'=\tilde{\mbf T}_2$ is not a valid solution in general. Still for the sake of simplicity, we set $\tilde{\mbf T}_1=\tilde{\mbf T}_1'$. 

We now want to find the possible values for $N_x$.
Using Eq.~\pref{eq:reciprocalVectors}, we find that
\ba
N_s \tilde{\mbf T}_1=n_{y2}\tilde{\mbf a}_1 - n_{x2}\tilde{\mbf a}_2
\ea
To select $N_x$, we have an additional constraint besides $N_x \tilde{\mbf T}_1$ being a vector of the reciprocal lattice: when $p$ runs from $0$ to $N_x-1$, we should span only inequivalent admissible momentum vectors. In order to satisfy these two constraints, the only possible choice is
\be
\label{nxChoice}
N_x = \frac{N_s}{\text{GCD}(N_s, n_{y2}, n_{x2})}
\ee
$N_y$ is then given by 
\be
\label{nyFromNxChoice}
N_y = \frac{N_s}{N_x} = \text{GCD}(N_s, n_{y2}, n_{x2})
\ee

One now needs to find $\tilde{\mbf T}_2'$. This vector could be written as
\be
\label{T2tildaprimechoice}
\tilde{\mbf T}_2'= \tilde{\mbf T}_2 + \alpha \tilde{\mbf T}_1
\ee
where $\alpha$ is an integer to insure that $k$ is an admissible momentum vector for any $q$. In principle, the coefficient in front of $\tilde{\mbf T}_2$ could be any integer co-prime with $N_y$ to describe all the admissible momentum vectors within the first Brillouin zone. Since this coefficient is also defined modulo $N_y$, we can set it to one. The last step is to find $\alpha$ such that $N_y \tilde{\mbf T}_2'$ is a vector of the reciprocal lattice. While it can be shown that $\alpha N_y$ is an integer using Bezout's identity, there is no guarantee that $\alpha$ is integer. Still we have been able to find possible sets of parameters for all the relevant aspect ratios and system sizes. Fig.~\ref{fig:tiltedLattice}b gives the value of all the lattice parameters ($N_x, \ N_y, n_{x1}, \ n_{y1}, \ n_{x2}, \ n_{y2}, \alpha$) that were used in this article, as well as the resulting aspect ratio $\kappa$ for a triangular Bravais lattice. To examplify the previous analysis, we give the explicit construction of the admissible momentum vectors for 
the tilted lattice for $N_s=10$ shown in Fig.~\ref{fig:tiltedLattice}a.

\begin{figure}
\includegraphics[width = 0.95\linewidth]{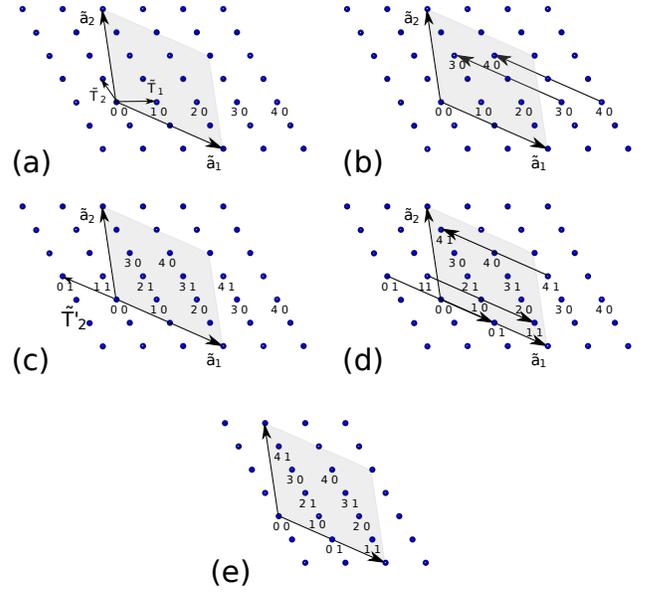}
\caption{Construction of the $(p,q)$  labels for the admissible momentum vectors on the $N_s=10$ tilted lattice of Fig.~\ref{fig:tiltedLattice}a. For this example, we have $N_x=5$ and $N_y=2$ according to the table of Fig.~\ref{fig:tiltedLattice}b. (a) The dots denote the admissible momentum vectors. The shaded area is the first Brillouin zone defined by the two reciprocal vectors $\tilde{\mbf a}_1$ and $\tilde{\mbf a}_2$. Using $\tilde{\mbf T}_1'=\tilde{\mbf T}_1$, we can label the admissible momentum vectors $(p,0)$ with $p$ going from $0$ to $4$. (b) We find the equivalents of $(3,0)$ and $(4,0)$ within the first Brillouin zone using reciprocal vectors. (c) Using $\tilde{\mbf T}_2'=\tilde{\mbf T}_2-\tilde{\mbf T}_1$ (here $\alpha=-1$), we can now label the $(p,1)$ admissible momenta. (d) Once again, we find the equivalents of $(0,1)$, $(1,1)$ and $(4,1)$ within the first Brillouin zone. (e) We show the label of each admissible momentum vector within the first Brillouin zone.}
\label{fig:tiltedLatticeConstruction}
\end{figure}

When we use tilted boundary conditions, the expression of the tight-binding Hamiltonian, which only depends on the structure of the lattice at a local scale, does not change. What changes is the set of admissible values for the momentum, i.e. the set of points where the Bloch Hamiltonian has to be evaluated. More precisely, we only need to evaluate the scalar product of the momentum with the lattice vectors $\mbf k\cdot\mbf a_i , \ i = 1, 2$. We give here its expression
\ba
\mbf k\cdot\mbf a_1 & = & \frac{2\pi}{N_s} \left((p + \alpha q) n_{y2} - q n_{y1}  \right)\\ \nonumber 
\mbf k\cdot\mbf a_2 & = & \frac{2\pi}{N_s} \left(-(p + \alpha q) n_{x2} + q n_{x1}  \right)
\ea
with $(p,q) \in [0, N_x - 1] \times [0, N_y - 1]$.

\section{Additional evidence of the stability of the FTI phase at half filling}
\label{app:HalfFillingSuppMat}
We present here additional data to support the existence and stability of the half filling FTI in the presence of interlayer coupling.

\subsection{Systems with pseudospin conservation}
\begin{figure}
\includegraphics[width = 0.49\linewidth]{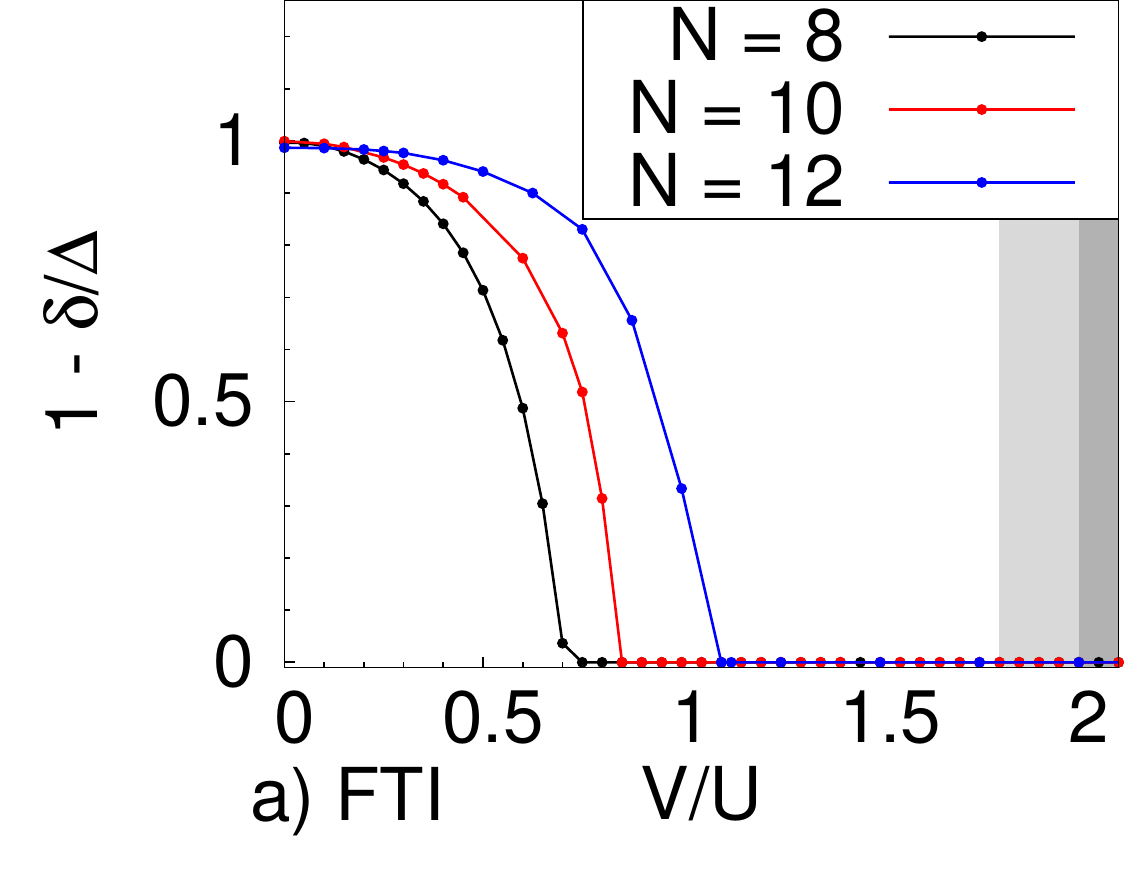}
\includegraphics[width = 0.49\linewidth]{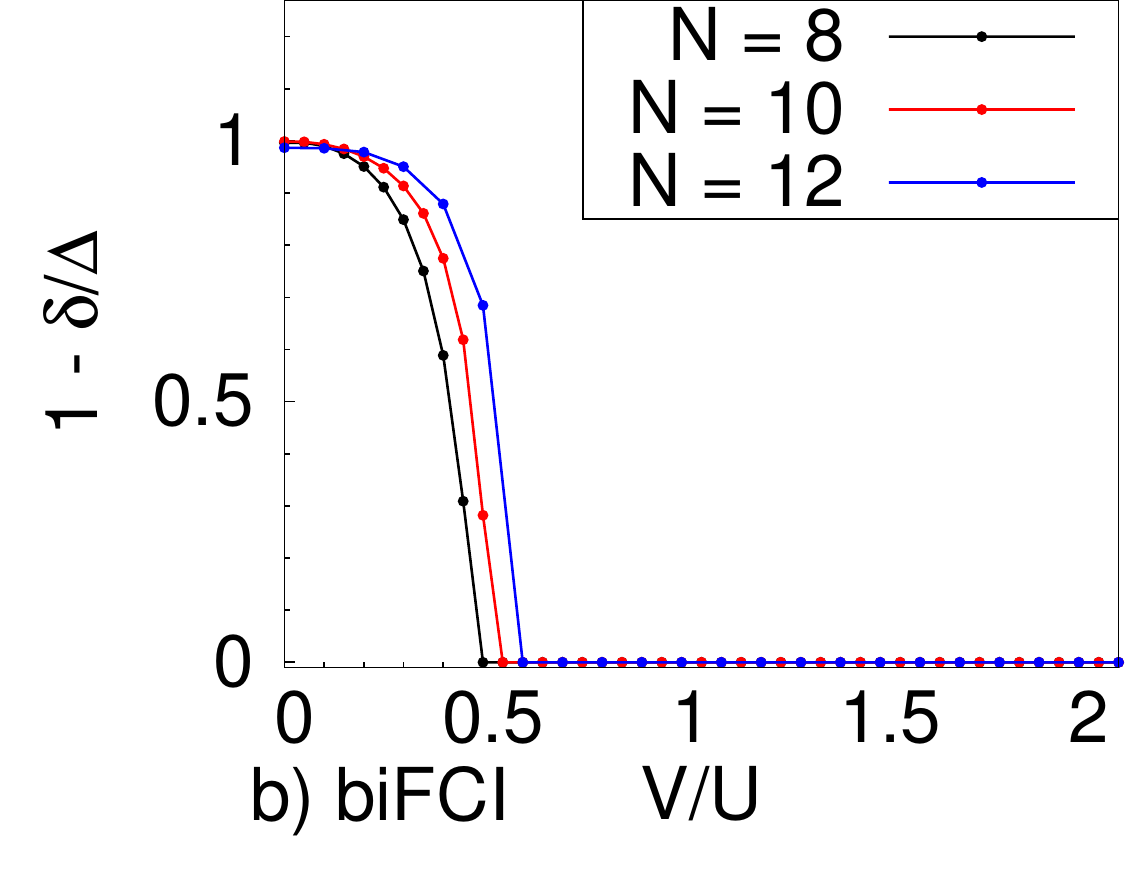}
\caption{Evolution of the quantity $\text{max}(1 - \delta/\Delta, 0)$ for the systems with $N = 8, \ 10, \ 12$ bosons and $N_s = N$ unit cells with respect to the magnitude of the interaction between bosons of opposite pseudospin ($V/U$), for the TRI FTI \emph{left panel} and for the bilayer FCI \emph{right panel}. The shaded area in (a) corresponds to a full polarization of the FTI system, for $N = 10, \ 12$ (light grey) and $N = 8$ (dark grey). The bilayer FCI system only becomes fully polarized for values of $V/U$ that are beyond the scope of this graph (the transition happens for $20 < V/U < 30$ depending on the system size).}
\label{fig:Spread_overlap}
\end{figure}

In the article, we relied on the energetic gap $\Delta$, the PES gap $\Delta_{\xi}(N_A)$ and the overlap to probe the stability of the fractional phase. Another important quantity to characterize the stability of the phase is the energy spread $\delta$. It is defined as the difference between the largest and the smallest energy within the almost degenerate ground state manifold (as depicted Fig.~\ref{fig:EnergySpectrumDecoupled}a). We represent the quantity $\text{max}(1 - \delta / \Delta, 0)$ Fig.~\ref{fig:Spread_overlap}. When $\delta / \Delta > 1$,  the energy separation between the ground state manifold and the excitations cannot be identified. However, we can still extract $\Delta$ and track the overlap. $\text{max}(1 - \delta / \Delta, 0)$  is thus a stricter criterion of the stability of the phase.
In Fig.~\ref{fig:Spread_overlap}a, we show its evolution in the TRI FTI system. There is a well defined fourfold almost degenerate ground state even for relatively large values of $V/U$, for all system sizes. In Fig.~\ref{fig:Spread_overlap}b, we compare this evolution to the bilayer FCI case. The quantity $1 - \delta/\Delta$ decreases faster when time reversal symmetry is broken than when it is preserved, in agreement with the other quantities.

\begin{figure}
\includegraphics[width = 0.49\linewidth]{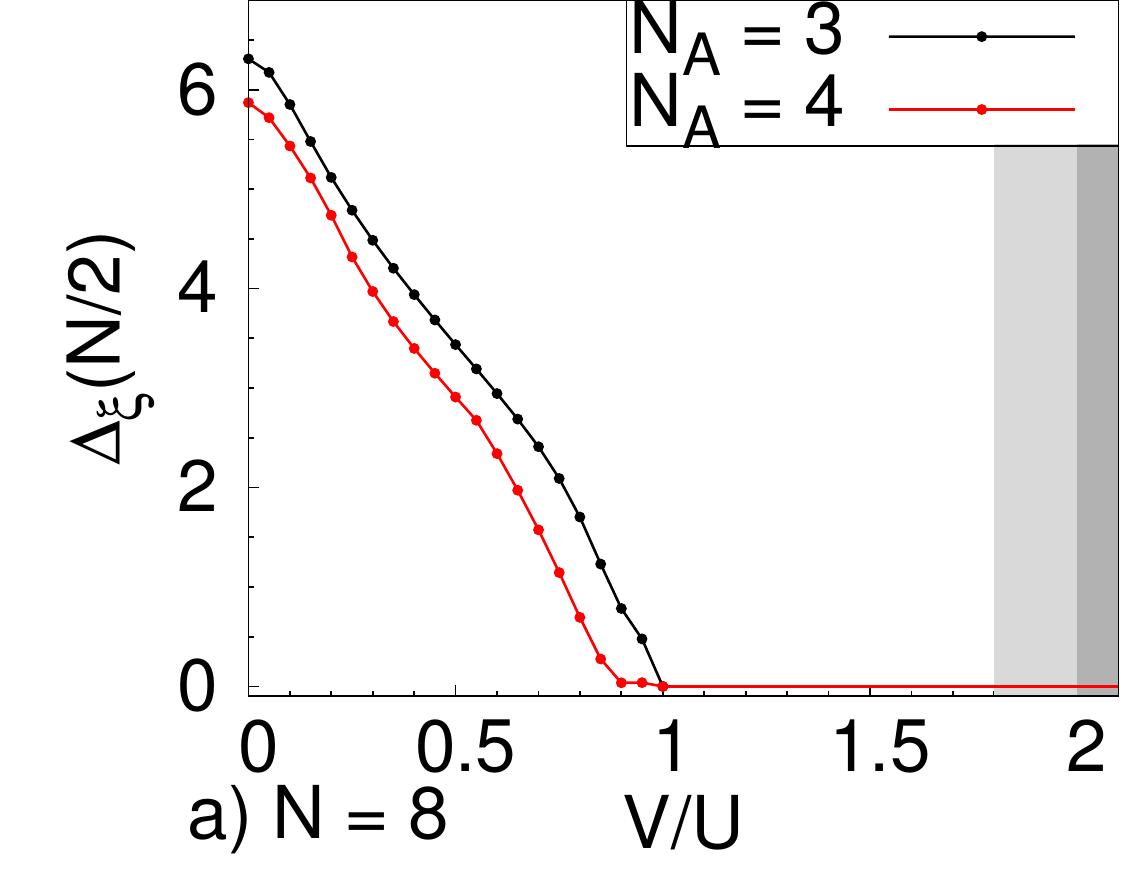}
\includegraphics[width = 0.49\linewidth]{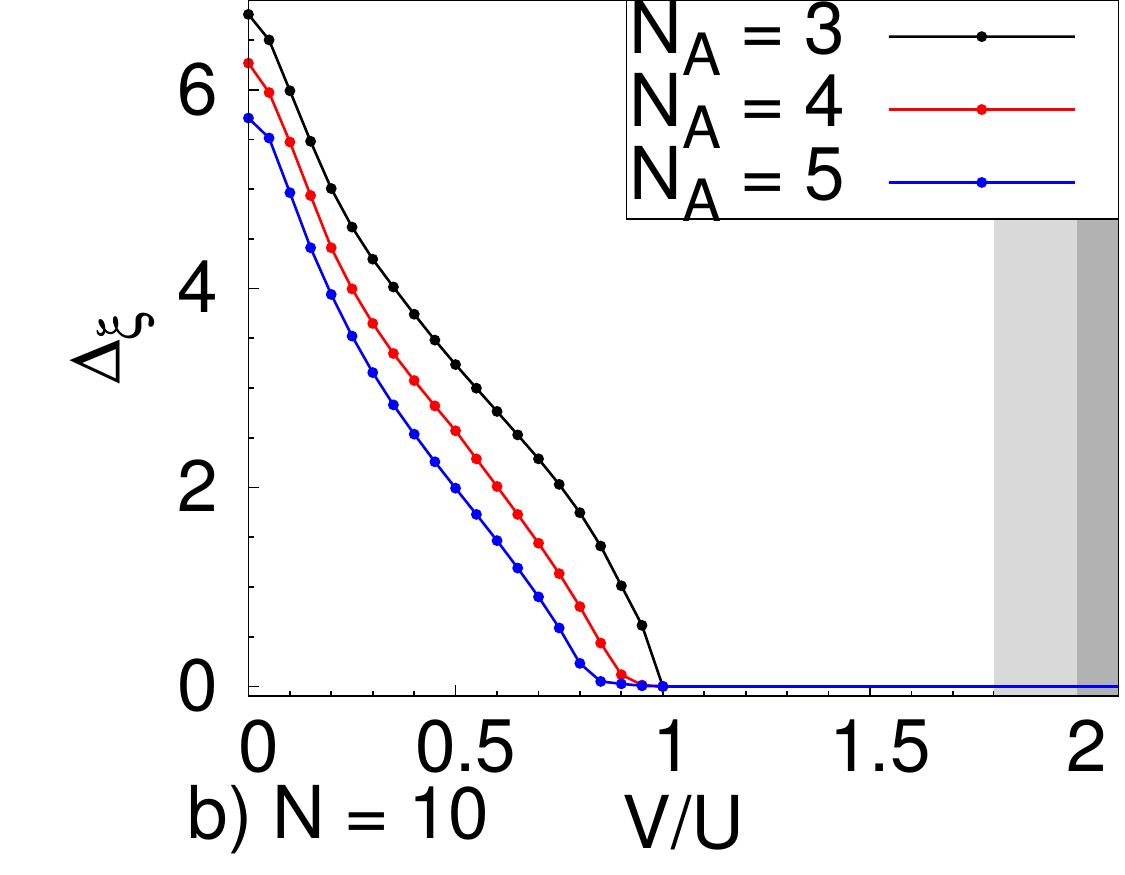}
\includegraphics[width = 0.49\linewidth]{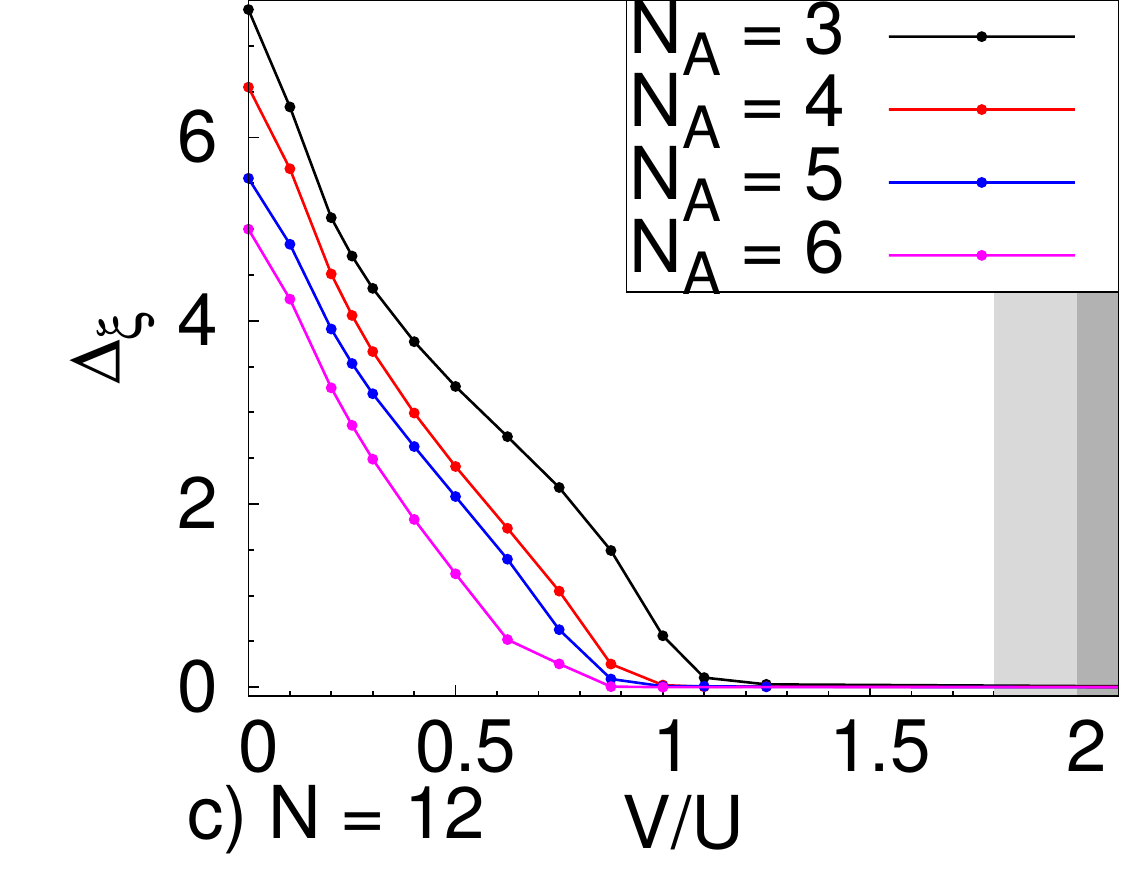}
\caption{Evolution of the entanglement gap with the magnitude of the interaction between bosons of opposite pseudospin ($V$) for the system with $N = 8$ (a), $N = 10$ (b) and $N = 12$ (c), $N_s = N$. The two kagome layers are only coupled by the interaction. The shaded area corresponds to a full polarization of the system, for $N = 10, \ 12$ (light grey) and $N = 8$ (dark grey).}
\label{fig:DecoupledPESN8N10N12}
\end{figure}

We also give some additional results concerning the PES. Fig.~\ref{fig:DecoupledPESN8N10N12}a, b, c give the evolution of the entanglement gap for $N_A \leq N/2$. As mentioned in the article, the sector $N_A = N/2$ has the smallest entanglement gap. In all the cases and all the system sizes, the picture is still consistent with what we have discussed in Sec.~\ref{sec:HalfFilling}.

\subsection{Some properties of the FTI PES}
\label{app:FTIPESProp}
We now derive the number of states below the FTI entanglement gap $\Delta_{\xi}$ starting from the number of states below the FCI gap, and explain the PES structure observed in Fig.~\ref{fig:EnergySpectrumDecoupled}b. We first focus on the states that are separated by an entanglement gap $\Delta_{\xi}$ from the rest of the PES (see Fig.~\ref{fig:EnergySpectrumDecoupled}b). Let us first consider one FCI copy at half filling, say the pseudospin up FCI copy. This system has $N_s$ unit cells and $N_{\uparrow} = N_s/2$ particles. Its ground state has a twofold quasidegeneracy. If we make a cut in the particle space with a number of particles $N_A^{\uparrow}$, we obtain a gapped PES. 
We call ${\cal N}^{\tiny \rm FCI}(N_A^{\uparrow}, \mbf k_{A}^{\uparrow})$ the number of states below the entanglement gap in the momentum sector $\mbf k_A^{\uparrow} = (k_{xA}^{\uparrow}, k_{yA}^{\uparrow})$. We can also define a similar counting for the pseudospin down FCI copy: ${\cal N}^{\tiny \rm FCI}(N_A^{\downarrow}, \mbf k_{A}^{\downarrow})$.
In terms of the FCI quantum numbers, the FTI quantum numbers write:
\ba
\label{eq:additivityNA}
N_A & = & N_A^{\uparrow} + N_A^{\downarrow} \\ 
\label{eq:additivitySzA}
S_{zA} & = & \frac{N_A^{\uparrow} - N_A^{\downarrow}}{2} \\ 
\label{eq:additivitykxA}
k_{xA} & = & (k_{xA}^{\uparrow} - k_{xA}^{\downarrow}) \ \text{mod} \ N_x \\ 
\label{eq:additivitykyA}
k_{yA} & = & (k_{yA}^{\uparrow} - k_{yA}^{\downarrow}) \  \text{mod} \ Ny
\ea
Note that the minus sign in Eqs.~\pref{eq:additivitykxA} and \pref{eq:additivitykyA} comes from the minus sign in the one-body Hamiltonian that acts on the pseudospin down $h^{*}_{\rm CI}(-\boldsymbol k)$.
For given $N_A$ and $S_{zA}$,  Eqs.~\pref{eq:additivityNA} and \pref{eq:additivitySzA} fix $N_A^{\uparrow}$ and $N_A^{\downarrow}$. Naively, the FTI PES counting with a cut $(N_A, S_{zA})$ is thus the following in the $\mbf k_{A}$ sector
\begin{equation}
{\cal N}^{\tiny \rm FTI}_A \left( \mbf k_A\right)= \!\!\!\!\!\!\!\! \sum_{\mbf k_A = \mbf k_A^{\uparrow} - \mbf k_A^{\downarrow}}^{'}  
\!\!\!\!\!\!\!\!
{\cal N}^{\tiny \rm FCI}(N_A^{\uparrow}, \mbf k_{A}^{\uparrow}) \\ \times {\cal N}^{\tiny \rm FCI}(N_A^{\downarrow}, \mbf k_{A}^{\downarrow}) 
\label{eq:momentumCompositionPES}
\end{equation}
where the sum has to be taken over all the FCI momentum sectors that satisfy the constraints of Eq.~\pref{eq:additivitykxA} and \pref{eq:additivitykyA}. The symbol $\sum^{'}$ signals that the momentum constraint has to be taken modulo $(N_x, N_y)$.

As pointed out in Ref.~\cite{sterdyniak-PhysRevB.87.205137}, additional constraints can reduce the number of eigenvalues of the PES  when the ground state possesses an additional symmetry preserved by particle partitioning. Here the ground state has $S_z = 0$, which results in the following constraint
\ba
N_A^{\uparrow} + N_B^{\uparrow} & = & N^{\uparrow} = N / 2 \\ \nonumber
N_A^{\downarrow} + N_B^{\downarrow} & = & N^{\downarrow} = N / 2
\ea
where the the $B$ indices refer to the $B$ partition. Additionally, for a ground state with a degeneracy $d \neq 1$, the PES counting is not the same in partitions A and B
\ba
{\cal N}^{\tiny \rm FCI}(N_A^{\uparrow}, \mbf k_A^{\uparrow}) \neq {\cal N}^{\tiny \rm FCI}(N_B^{\uparrow}, \mbf k_B^{\uparrow})
\ea
As a result, ${\cal N}^{\tiny \rm FTI}_A \left(\mbf k_A \right) \neq {\cal N}^{\tiny \rm FTI}_B\left(\mbf k_B \right)$, and the number of states below the gap in the FTI system reduces to
\begin{equation}
{\cal N}^{\tiny \rm FTI} = \text{min}\left({\cal N}^{\tiny \rm FTI}_A, {\cal N}^{\tiny \rm FTI}_B\right)
\end{equation}

We now focus on the high entanglement energy region of the spectrum. For the FCI, it was shown in Ref.~\cite{Wu-2012PhysRevB.85.075116} that several subgaps appear in the non-universal part of the PES. These gaps are related to the violation of the generalized Pauli principle. In the FTI PES, we observe a similar substructure. We show that the number of states between the first and second gap (see Fig.~\ref{fig:EnergySpectrumDecoupled}b) can be explained by qualitative considerations on the number of states in the FCI PES. We can perform the following qualitative reasoning. Roughly speaking, the ground state of a FCI at half filling is the sum of an ideal FCI Laughlin state $\ket{\Psi_{L}}$ and a perturbative term of amplitude $\e$:
\ba
\ket{\Psi} \simeq \ket{\Psi_{L}} + \e \ket{\Psi_1}
\ea
$\ket{\Psi_{L}}$ alone has an infinite entanglement gap. It contributes ${\cal N}_L$ states to the PES. Meanwhile $\e \ket{\Psi_1}$ contributes ${\cal N}_1$ states to the PES, all in its high entanglement energy part, as shown in Ref.~\cite{Wu-2012arXiv1206.5773W}. In the decoupled case, we can write the following approximation of the FTI wavefunction
\ba
\ket{\Psi_{\rm FTI}} & = & \ket{\Psi^{\uparrow}}\ket{\Psi^{\downarrow}} \\ \nonumber
                 & \simeq & \ket{\Psi_{L}^{\uparrow}}\ket{\Psi_{L}^{\downarrow}} + \e \left(\ket{\Psi_1^{\uparrow}}\ket{\Psi_{L}^{\downarrow}} + \ket{\Psi_{L}^{\uparrow}}\ket{\Psi_1^{\downarrow}}\right)
\ea
The first term is an ideal wavefunction, which contributes ${\cal N}_L^2$ to the low lying part of the PES (the part of the PES that we have studied in the last paragraph). The second term contributes $2{\cal N}_L{\cal N}_1$ to the high entanglement energy part of the PES. Fig.~\ref{fig:EnergySpectrumDecoupled}b, the number of states between the two dotted lines is indeed explained by this simple counting rule, as it is in other system sizes. For a higher number of particles, we observe a substructure in the high entanglement part of the PES. It is related to the FCI subgap hierarchy described in Ref.~\cite{Wu-2012PhysRevB.85.075116}.

Finally, we followed the evolution of the secondary entanglement gap with increasing $V/U$. Results show a much faster decrease of this gap compared to the lower gap $\Delta_{\xi}(N_A)$. This is in agreement with the general observation in FCI that the entanglement gap corresponding to the model state is much more robust to perturbations than the other gaps. 

\subsection{Systems without pseudospin conservation}
\begin{figure}
\includegraphics[width = 0.49\linewidth]{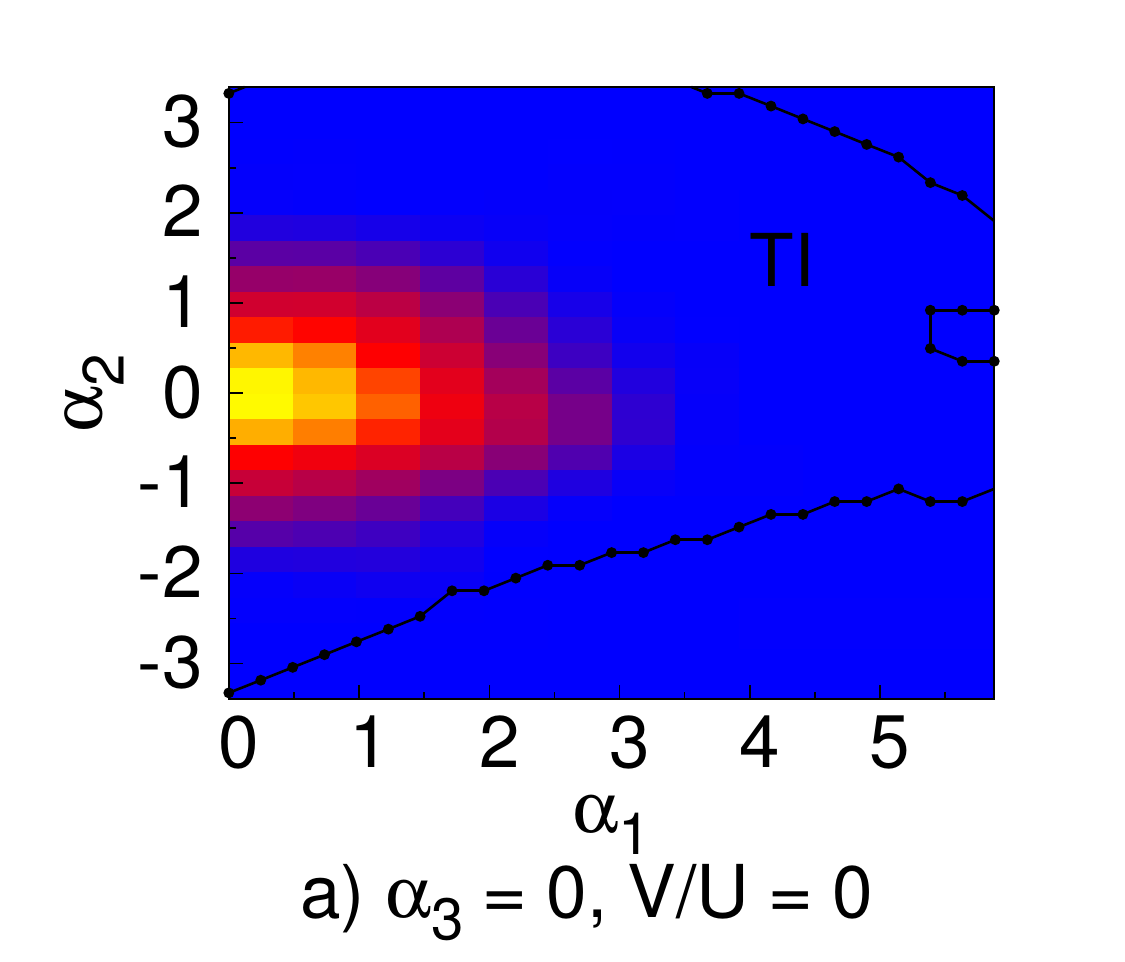}
\includegraphics[width = 0.49\linewidth]{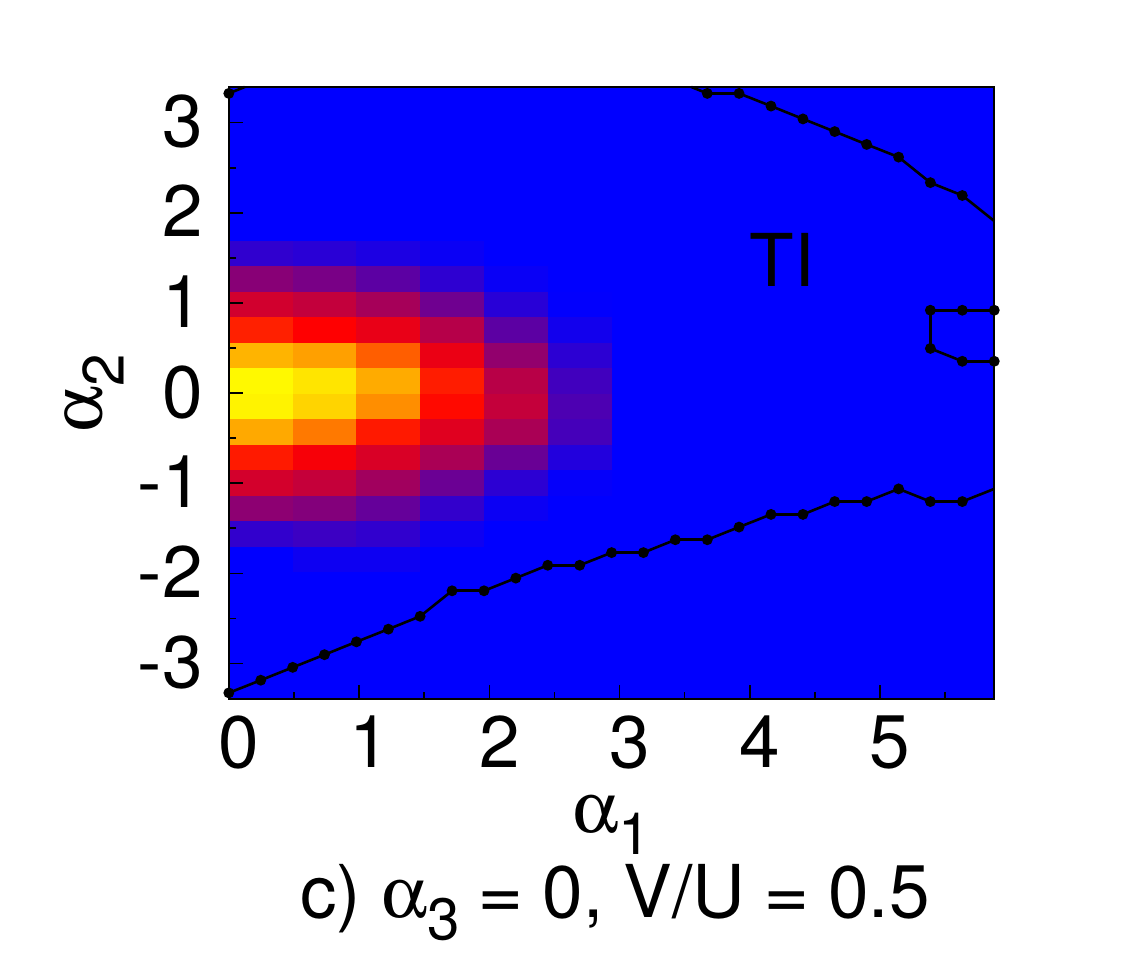}
\includegraphics[width = 0.49\linewidth]{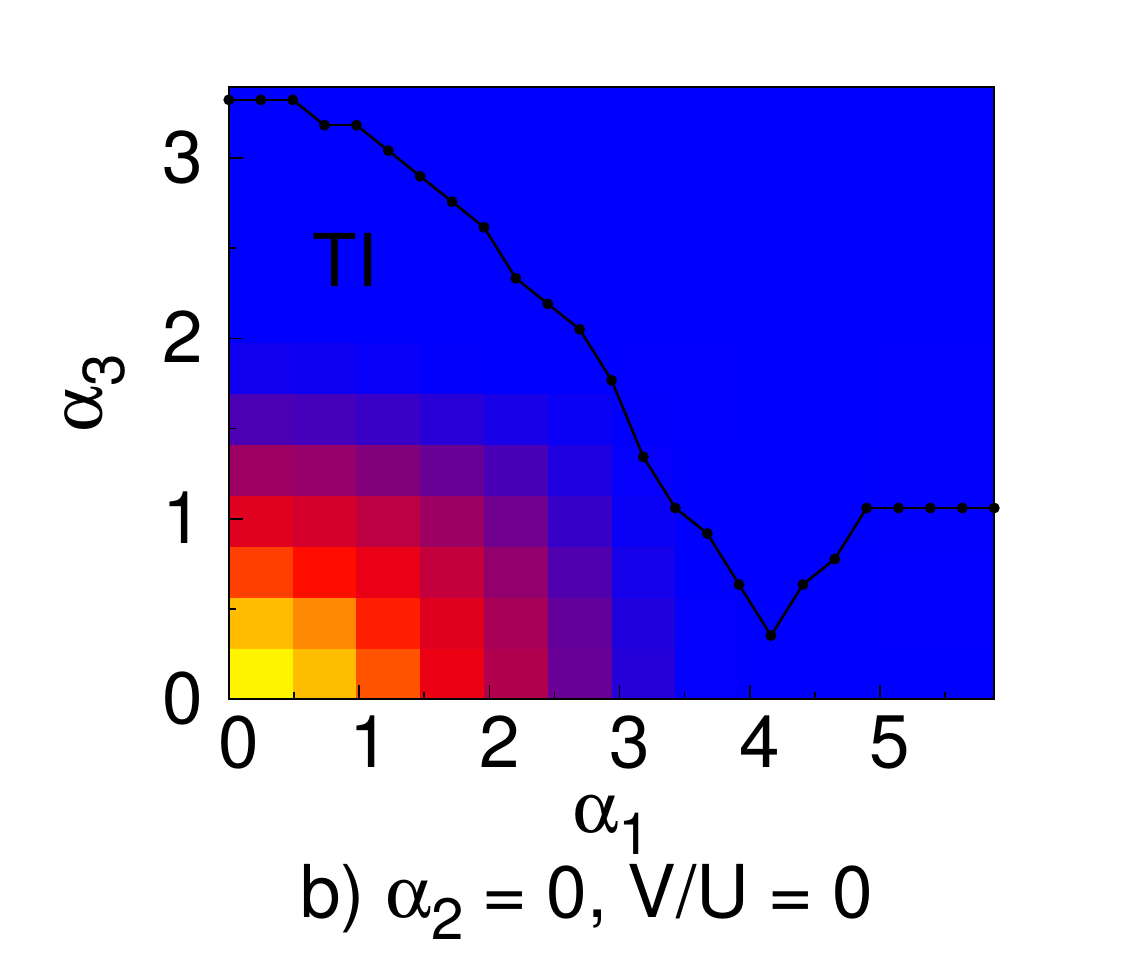}
\includegraphics[width = 0.49\linewidth]{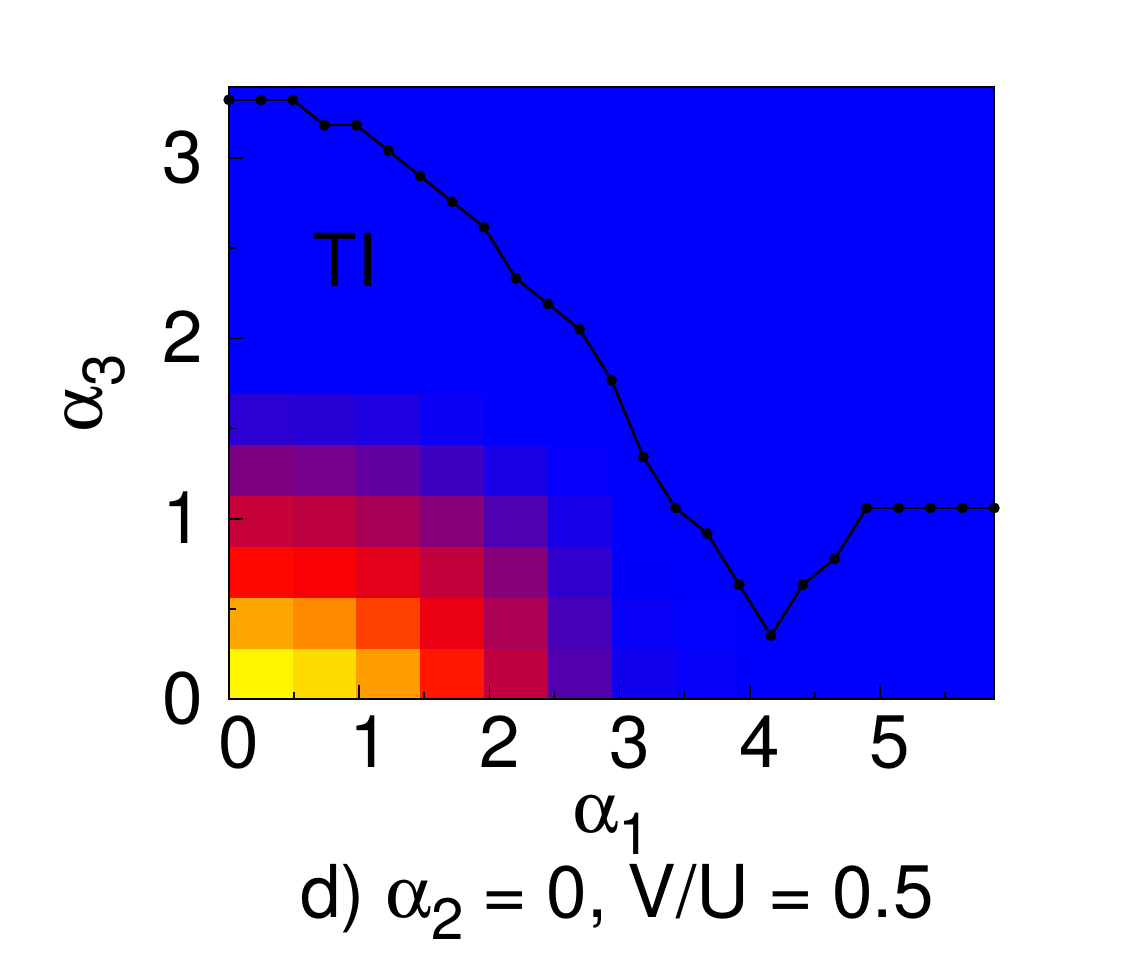}
\includegraphics[width = 0.49\linewidth]{figures/ColorBox.pdf}
\caption{Gap of the kagome lattice model topological insulator at half filling with $N = 8$ bosons, in the plane $\alpha_3 = 0$ (a and b) and in the plane $\alpha_2 = 0$ (c and d), for $V/U = 0$ (a and c) and $V/U = 0.5$ (b and d). $\Delta_\text{max,V}$ is the amplitude of the gap in the case where $R = 0$, with an interlayer interaction $V$. We use the symmetries of the system with respect to the coupling elements $\alpha_i$ and show only the zones $\alpha_1 > 0$ and $\alpha_3 > 0$. The dotted line indicates the boundary between the trivial and topological insulator phases in the non-interacting model.}
\label{fig:PhaseDiagramMixingC3Delta1Delta3N8}
\end{figure}

\begin{figure}
\includegraphics[width = 0.49\linewidth]{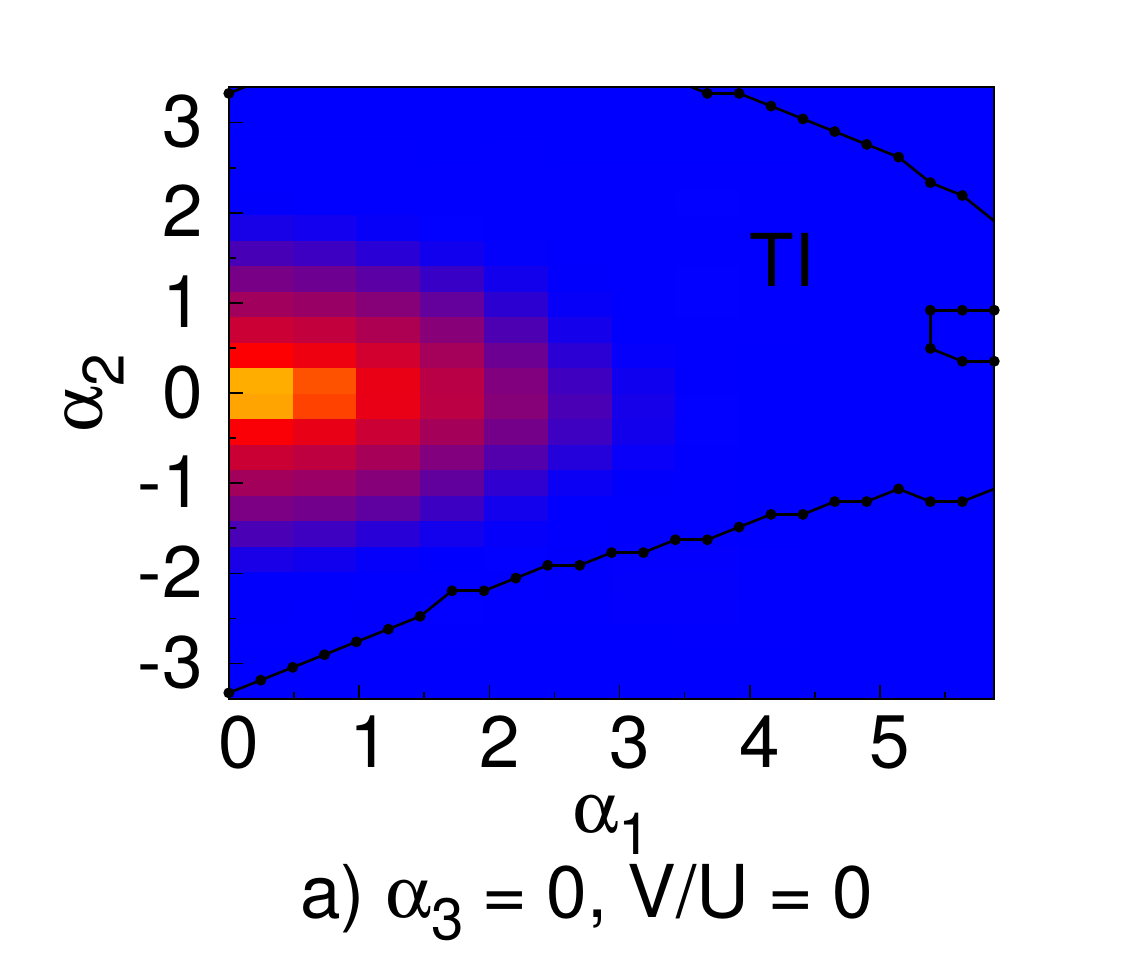}
\includegraphics[width = 0.49\linewidth]{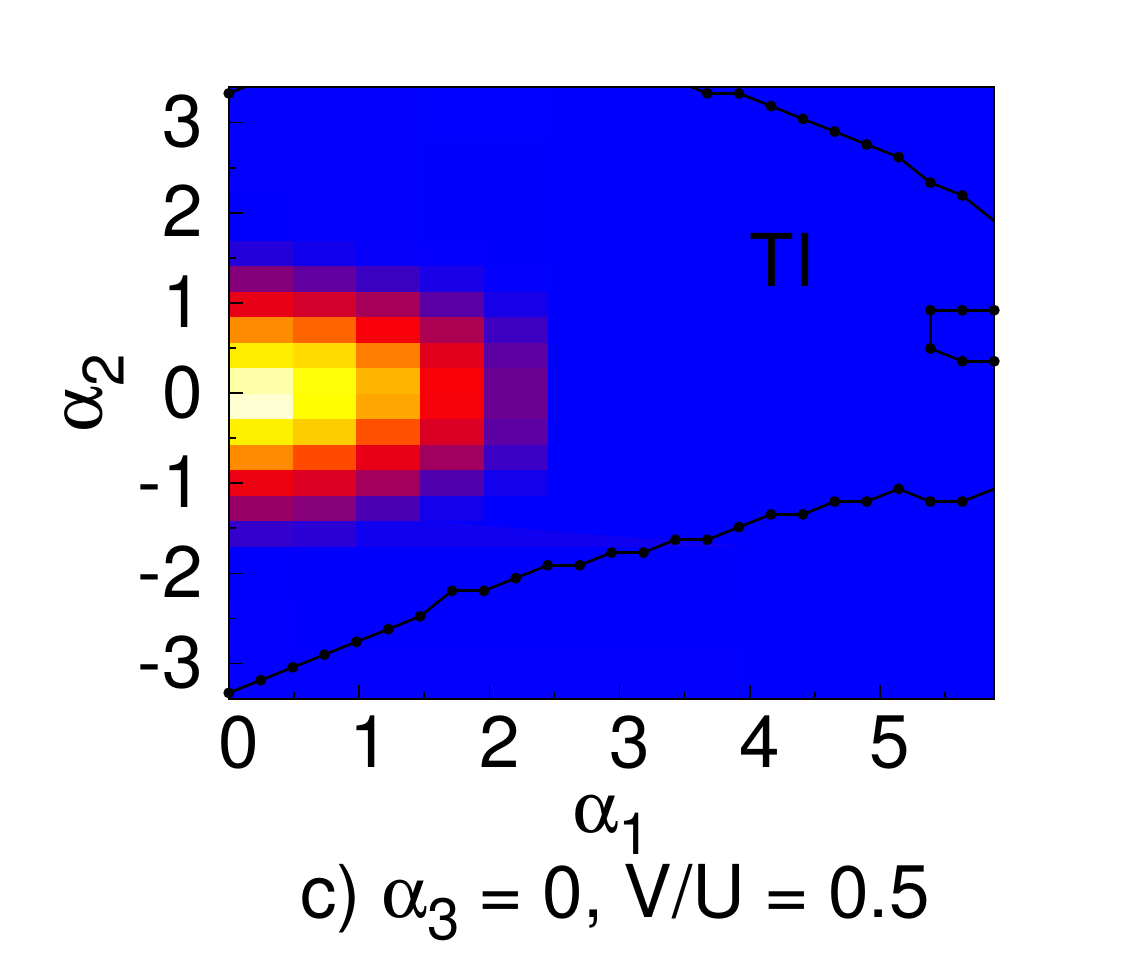}
\includegraphics[width = 0.49\linewidth]{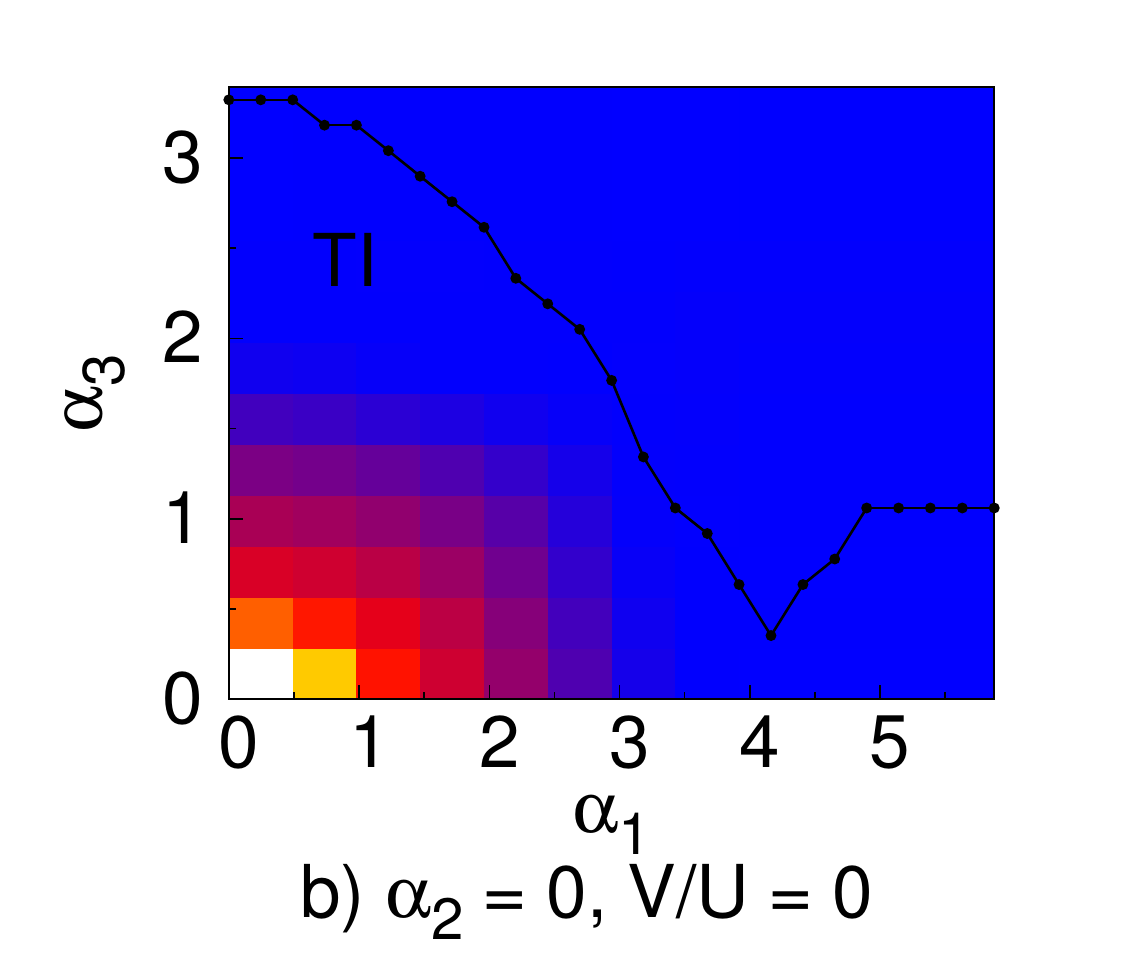}
\includegraphics[width = 0.49\linewidth]{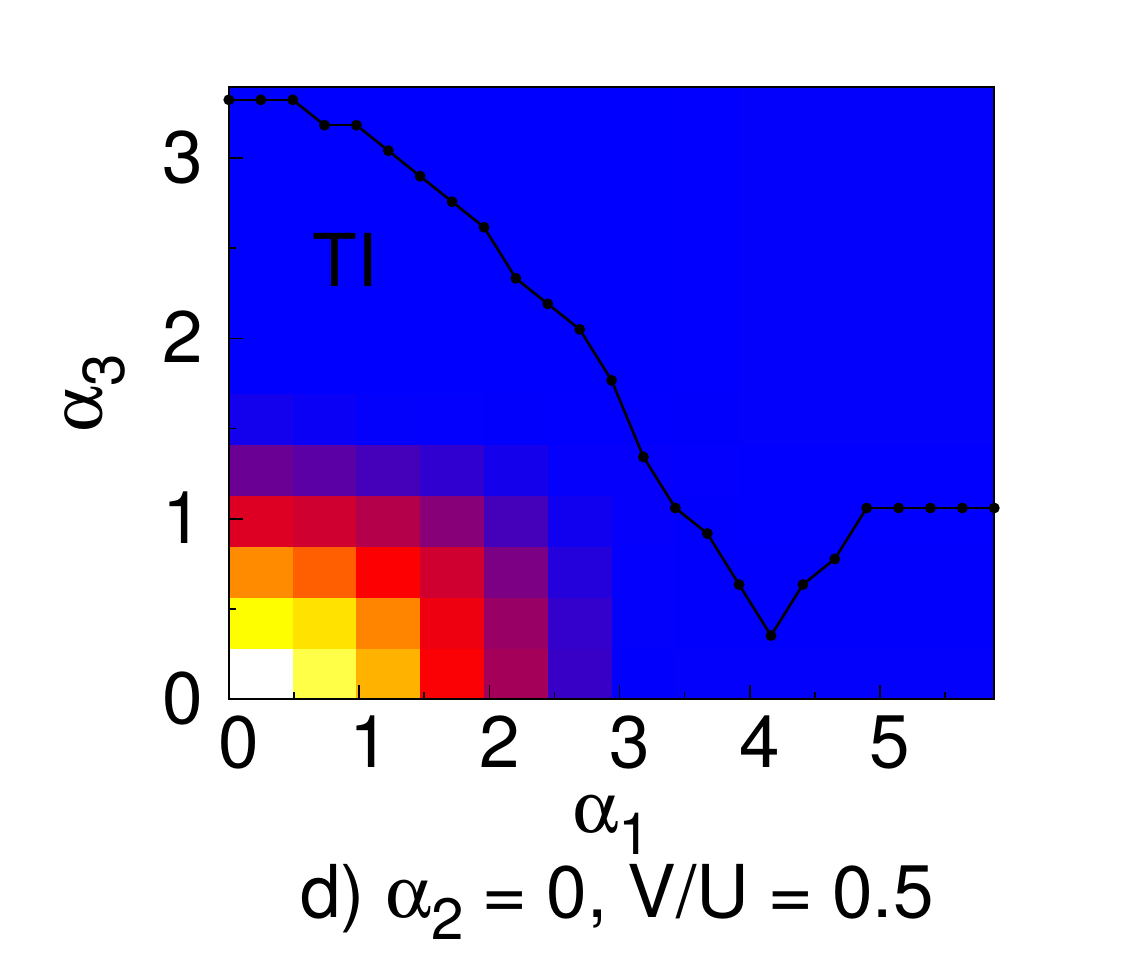}
\includegraphics[width = 0.49\linewidth]{figures/ColorBoxPES.pdf}
\caption{PES gap of the kagome lattice model topological insulator at half filling with $N = 8$ bosons, in the plane $\alpha_3 = 0$ (a and b) and in the plane $\alpha_2 = 0$ (c and d), for $V/U = 0$ (a and c) and $V/U = 0.5$ (b and d). $\Delta_{{\xi}\text{max,V}}(N/2)$ is the amplitude of the PES gap in the case where $R = 0$, with an interlayer interaction $V$. The number of particles in the partition is $N_A = 4$. We use the symmetries of the system with respect to the coupling elements $\alpha_i$ and show only the zones $\alpha_1 > 0$ and $\alpha_3 > 0$. The dotted line indicates the boundary between the trivial and topological insulator phases in the non-interacting model.}
\label{fig:PhaseDiagramMixingC3Delta1Delta2PESN8}
\end{figure}
We give some additional evidence for the stability of the FTI phase upon addition of an inversion symmetry breaking term. In Sec.~\ref{sec:HalfFillingRashba}, we have provided the data for $N = 10$. Here, we give the results for $N = 8$. We look at the gap (see Fig.~\ref{fig:PhaseDiagramMixingC3Delta1Delta3N8}) and the PES gap (see Fig.~\ref{fig:PhaseDiagramMixingC3Delta1Delta2PESN8}) in the planes $\alpha_2 = 0$ and $\alpha_3 = 0$. The results are quantitatively very similar to those obtained for $N = 10$ in the article.

\bibliography{Z2FTI}

\end{document}